\documentclass[10pt]{article}
\usepackage{amsmath}
\usepackage{hyperref}
\usepackage[dvips]{graphicx}
\usepackage{bm}

\renewcommand{\theequation}{\thesection.\arabic{equation}}
\begin{document}

\title{{\Large
Generalization of Faddeev--Popov Rules in Yang--Mills Theories: N=3,4 BRST Symmetries}}
\author{ Alexander Reshetnyak\thanks{%
e-mail address: reshet@ispms.tsc.ru}\\
Institute of Strength Physics and Materials Science of Siberian Branch of  RAS, \\
 634055, Tomsk,
Russia}
\date{}
\maketitle

\begin{abstract}
The Faddeev--Popov rules for a local and Poincare-covariant
procedure of Lagrangian quantization for a gauge theory with gauge
group are generalized to the case of an invariance of the
respective quantum actions, $S_{(N)}$,  with respect to $N$-parametric Abelian SUSY transformations with odd-valued   parameters $\lambda_p$, $p=1,...,N$ and anticommuting generators $s_p$:
$s_p s_q+ s_q s_p =0$, for $N=3,4$, implying the substitution of an $N$-plet of  ghost fields , $C^p$, instead of the  parameter, $\xi$, of  the infinitesimal gauge transformations: $\xi = C^p\lambda_p$.
The total configuration spaces of field variables for a  quantum theory of the same classical model coincide  in the $N=3$ and $N=4$ symmetric cases.
  For the  $N=3$-parametric SUSY transformations the superspace of the irreducible representation includes, in addition to  Yang--Mills fields $\mathcal{A}^\mu$,
 also $3$ ghost odd-valued fields $C^p$, as well as  $3$ new  even-valued  $B^{pq}=-B^{qp}$ and $1$ odd-valued $\widehat{B}$ fields for $p,q=1,2,3$.
It is shown, that in order to construct the quantum action, $S_{(3)}$  a gauge-fixing procedure  achieved by adding to the  classical action of an $N=3$-exact gauge-fixing term (without introduction of non-degenerate odd supermatrix) additionally requires a $1$ antighost field $\overline{C}$, $3$  even-valued  ${B}^{p}$ and $3$ odd-valued $\widehat{B}{}^{pq}$ fields, as well as the Nakanishi--Lautrup field $B$.
  The action of $N=3$ transformations in the space of additional fields, $\overline{\Phi}_{(3)}= (\overline{C}, {B}^{p}, \widehat{B}{}^{pq}, B)$,  not being entangled with  the fields ${\Phi}_{(3)}$ of $N=3$-irreducible representation space is realized as well. These transformations are the  $N=3$ BRST symmetry transformations for the vacuum functional, $Z_3(0)=\int d \Phi_{(3)}d \overline{\Phi}_{(3)} \exp \{(\imath/\hbar )S_{(3)}\}$. It is shown that  the total configuration space of the fields $(\Phi_{(3)}, \overline{\Phi}_{(3)})$, as the space of reducible $N=3$ BRST symmetry transformations, proves to be the space of an irreducible
representation of the fields  $\Phi_{(4)}$ for  $N=4$-parametric SUSY transformations, which contains, in addition to  $\mathcal{A}^\mu$ the $(4+6+4+1)$ ghost-antighost, $C^r=(C^p, \overline{C})$, new  even-valued,  $B^{rs}=-B^{sr}=(B^{pq}, B^{p4}={B}{}^p)$,   odd-valued $\widehat{B}{}^r = (\widehat{B}, \widehat{B}{}^{pq})$ fields and   $B$ for $r,s=1,2,3,4$, $r=(p,4)$. The quantum action  $S_{(4)}$  is constructed by adding to the
classical action an  $N=4$-exact gauge-fixing term with a gauge boson, $F_{(4)}$ as the $s_r$-potential as compared to a gauge fermion $\Psi_{(3)}$ for  $N=3$ case. It is proved that the $N=4$-parametric SUSY transformations are by $N=4$  BRST transformations for the vacuum functional, $Z_4(0)=\int d \Phi_{(4)} \exp \{(\imath/\hbar )S_{(4)}\}$.  The procedures are valid for any admissible gauge. The  equivalence with $N=1$ and $N=2$ BRST-invariant quantization methods are explicitly established.  The finite $N=3,4$ BRST transformations
are derived from the algebraic SUSY transformations. The Jacobians  for a change of
variables related to finite $N=3,4$  SUSY transformations with
field-dependent parameters in the respective path integral are
calculated. The Jacobians imply the presence of a corresponding modified Ward
identity which reduces to a new form of the standard Ward identities in the case of constant
parameters and describe the problem of a gauge-dependence. The gauge-independent Gribov-Zwanziger models with local $N = 3, 4$ BRST
symmetries are proposed  An introduction into diagrammatic Feynman techniques for $N=3, 4$ BRST invariant quantum actions for Yang--Mills theory is suggested. A generalization to the case
of $N=2K-1$ and $N=2K$, $K>2$ BRST transformations is discussed\footnote{The paper is dedicated to the memory of the
outstanding Soviet and Russian theoretical physicist and
mathematician, Academician  Ludwig Dmitrievich  Faddeev (1934-2017)}.
\end{abstract}



\section{Introduction}

\vspace{-1ex}

The problem of Lorentz-covariant quantization for gauge theories
with a non-Abelian gauge group \cite{YangMills} is a long-standing
one, starting with the lecture of R. Feynman   \cite{Feynman}, showing that the naive one-loop
diagram calculation within perturbative techniques with a propagator
constructed, according to quantum electrodynamic, for the photon field $A_\mu$
in the form
  \begin{equation}\label{qedprop}
    G_{\mu\nu}(k)=\frac{1}{k^2+ \imath 0}\left(\eta_{\mu\nu}- \frac{k_\mu k_\nu}{k^2}\right)+ \beta(k)\frac{k_\mu k_\nu}{k^2},
  \end{equation}
turns out to be incorrect\footnote{Let us point out that the elements
of the scattering matrix, among the physical states, do not depend on
the value of $\beta(k)$}.
A modification of calculations for reconstructing the one-loop contribution from
the tree diagrams, using unitarity and analyticity \cite{Feynman}, makes it possible
to interpret the additional contributions as an input from a scalar particle,
which should be, however, considered as a fermion due to the ``-'' sign before this
summand. The solution of this problem was found by L. Faddeev and V. Popov
in their celebrated work \cite{fp} by means of a trick known as the
\emph{insertion of unity}, providing the existence of a path integral for Yang--Mills fields,
$\mathcal{A}_\mu (x)= \mathcal{A}_\mu^m(x)t^m$, given in Minkowski space-time
$\mathbf{R}^{1,3}$ and taking values in a compact Lie group $G$,
with generators $t^m$ for its Lie algebra $\mathcal{G}$, in the form
\begin{eqnarray}\label{PintfpL}
  Z^L_{0} & = & \int  d \mathcal{A} \delta(\partial^\mu \mathcal{A}_\mu)\det M(\mathcal{A}) \exp\Big\{\frac{\imath}{\hbar}S_0(\mathcal{A})\Big\},   \\
  \label{PintfpF} Z^F_{0} & = &  \int  d \mathcal{A} d B \det M(\mathcal{A}) \exp\Big\{\frac{\imath}{\hbar}S_0(\mathcal{A})+ \int d^4x  \big(\partial^\mu \mathcal{A}_\mu + {g^2}B\big)B\Big\}\footnotemark,
\end{eqnarray}\footnotetext{Because of the integration in (\ref{PintfpF}) in powers of $B$ is gaussian, the only way to get after integration the gauge-fixed term: $-   \frac{1}{2g^2} \big(\partial^\mu \mathcal{A}_\mu\big)^2$ to restore coupling constant in the Feynman gauge as it was done above through field $B$.}respectively, for the Landau gauge, $\chi( \mathcal{A})=0$, $\chi( \mathcal{A})=\partial^\mu \mathcal{A}_\mu$,  and then with the use of the proposal of 't~Hooft \cite{t'Hooft} for the Feynman gauge $\chi( \mathcal{A},B)$, $\chi( \mathcal{A},B)=\partial^\mu \mathcal{A}_\mu +{g^2} B$,  with an arbitrary field $B=B^mt^m$ known as Nakanishi-Lautrup field \cite{Nakanishi, Lautrup}. This representation, with a gauge-invariant classical action $S_0$ , in comparison with the case of an Abelian $U(1)$ gauge group, essentially includes a determinant of an non-degenerated operator $M(\mathcal{A}) $:
\begin{eqnarray} \label{operatorFP}
M(\mathcal{A})    & =& \partial^\mu D_{\mu} \ = \    \partial^\mu \big(\partial_\mu -  [\mathcal{A}_\mu,    \ ] \big)\footnotemark
 \end{eqnarray}\footnotetext{Here the notation for $M(\mathcal{A})$ introduced in \cite{fp} was used. In what follows we will use the definition of the covariant derivative $D_{\mu}$ with opposite sign: $D_{\mu}=\partial_\mu +  [\mathcal{A}_\mu,    \ ]$.}known as the Faddeev--Popov operator (having multiple zero-mode eigenfunctions
as compared to the Abelian case, known as \emph{Gribov copies}  \cite{Gribov}).
In \cite{fp-ghosts} (see the review \cite{faddeev}), it was shown, with the use
of F. Berezin \cite{berezin} generalization of the Gaussian integral over Grassmann variables, that the representations (\ref{PintfpL}), (\ref{PintfpF}) may be equivalently presented in a local form by using fictitious scalar Grassman-odd fields $\big(C(x), \overline{C}(x)\big)=\big(C^m(x), \overline{C}{}^m(x)\big)t^m$
\begin{eqnarray}\label{PintfpLloc}
 &\hspace{-1em}&\hspace{-0.5em} Z^L_{0} =  \hspace{-0.3em}\int \hspace{-0.2em} d \mathcal{A}d C d \overline{C} dB \exp \Big\{\hspace{-0.1em}\frac{\imath}{\hbar}S^L_{FP}(\mathcal{A},C, \overline{C}, B)\hspace{-0.1em}\Big\} \ \mathrm{with}  \ S^L_{FP}  =  S_0+ \hspace{-0.2em}\int \hspace{-0.2em}d^4x \Big\{\overline{C}M(\mathcal{A})C+\chi( \mathcal{A})B\Big\},
\end{eqnarray}
and similarly for $Z^F_0$,  where, instead of the quantum action $S^L_{FP}=S^L_{FP}(\mathcal{A},C, \overline{C}, B)$, one should
use the action $S^F_{FP}=S^L_{FP}\big|_{\chi( \mathcal{A})\to\chi( \mathcal{A},B)}$ given
in the Feynman gauge. Independently. the development of the diagrammatic technique without using Grassmann-odd fictitious fields was suggested by B. DeWitt \cite{dewitt}.     The representation (\ref{PintfpLloc}) allows one to replace gauge
transformations for Yang--Mills fields with arbitrary scalar functions $\xi(x)=\xi^m(x)t^m$
by global transformations in the total configuration space
$\mathcal{M}_{tot}$ of fields $\Phi^A = (\mathcal{A},C, \overline{C}, B)$,
with a constant Grassmann-odd parameter $\mu$, $\mu^2=0$ by the rule  $\xi(x)=C(x)\mu$,
being an invariance transformation for the quantum action and for the integral measure
in (\ref{PintfpLloc}), which is known as a BRST symmetry transformation \cite{brs, brs1}.
The BRST symmetry allows one to prove the gauge-invariant renormalizability of a quantum Yang--Mills
theory \cite{Slavnov}, \cite{Slavnov1}, as well as the path integral independence
from a choice of the gauge condition for small variations. This also makes it possible
to obtain the Ward identities for generating functionals of Green's functions \cite{ZinnJustin}.  In \cite{aBRST1, aBRST2} it was shown that the Faddeev--Popov representation (\ref{PintfpL}), (\ref{PintfpF}) admits the form (\ref{PintfpLloc}) for an anti-BRST symmetry transformation with another Grassmann parameter, $\bar{\mu}$:  $\xi(x)=\overline{C}(x)\bar{\mu}$\footnote{For superfield and geometrical interpretation of anti-BRST symmetry see e.g.\cite{Bonora}, \cite{Gregoire}, \cite{Varshovi} and references therein},  which may be considered within the $N=2$ BRST  ( BRSTantiBRST) symmetry \cite{aBRST3} for Yang--Mills theories, describing ghost and antighost fields as an $Sp(2)$-doublet $C^{m{}a}(x)$ of fields: $(C^{m{}1}, C^{m{}2})=(C^{m}, \overline{C}{}^{m})$, as well as the parameters $(\mu, \bar{\mu})= (\mu_1, \mu_2)$, which follows from the substitution $\xi^m(x)= C^{m{}a}(x){\mu}_a$ (with summation over repeated indices).  The lifting of $N=1, 2$ BRST symmetry transformations, given originally in an infinitesimal algebraic form, to a finite group-like form, with finite field-dependent parameters $\mu(\Phi)$, $\mu_a(\Phi)$ has been introduced for $N=1$ case in \cite{sdj}, \cite{ll1} (for gauge theories with a
closed algebra and general gauge theories, see \cite{re}),for $N=2$ case in \cite{re1} (as well as for constrained dynamical systems and general gauge theories in \cite{MR2, MR3, MR4, MR5} with references therein), which allows one to establish that the path integral in different gauges, such as  (\ref{PintfpL}) and  (\ref{PintfpF}), assume the same value.

Recently \cite{UMR}, we have examined special SUSY (distinct from
space-time SUSY)
transformations with $m$ Grassmann-odd generators that form an Abelian superalgebra
$\mathcal{G}_m$ leaving the classical action (in a certain class of field-theoretic models)
invariant and realizing a lifting of  $\mathcal{G}_m$ to an Abelian supergroup ${G}_m$,
with finite parameters and respective group-like elements being functionals of field variables.
We have studied some physical consequences of these transformations at the path integral level.
As a consequence, we are interested in the following question.

Is it possible to find a general solution for the non-local
Faddeev--Popov path integral representations (\ref{PintfpL}),
(\ref{PintfpF}) in a local form which admits an extended $N=k$
global SUSY transformation with $k\geq 3$ Grassmann-odd
parameters, such as those realized by $N=1,2$ BRST symmetries? In the case of a positive solution,  which depends on a possibility to realize on an appropriate $N=k$ SUSY irreducible representation space the $N=k$-invariant gauge-fixing procedure to construct $N=k$-invariant quantum action, $S_{(N)}$,  we are interested in investigating such physical
consequences as gauge-dependence, unitarity,  renormalizability and Ward identities
for the Feynman diagrams in the corresponding path integral with
local $N=3$ and $N=4$-BRST invariant quantum actions.

The paper is devoted to the solution of the problem in question and is organized as follows.
In Section~\ref{GenSUSY}, we expound a generalization of the non-local Faddeev--Popov
path integral to an $N=k$ BRST symmetry realization in Subsection~\ref{N3FPtrick}, starting from the review of $N=1,2$ cases in Subsection~\ref{N12SUSY}.
We derive a local Faddeev--Popov path integral, $Z_{3}$, over fields composing total configuration space, which is the reducible representation superspace of  $N=3$ SUSY transformations being explicitly constructed both for  the fields of $N=3$ irreducible representation superspace  and for auxiliary fields from \emph{non-minimal sector}   in Subsection~\ref{N3FPtrickloc} so as to formulate an $N=3$ BRST
invariant gauge-fixing procedure without a special odd
supermatrix. In Section~\ref{N4}  we consider the fields of  $N=3$ irreducible and additional  representation superspaces on equal footing within explicitly constructed  $N=4$ SUSY   transformations, and formulate   $N=4$ SUSY  invariant gauge-fixing procedure for local path integral, $Z_{4}$, in  Section~\ref{N4gf}, for which these transformations are $N=4$ BRST symmetry transformations.
 In Section~\ref{N=kgauge}, we determine infinitesimal
and finite group-like $N=k$  BRST symmetry transformations, for $k=3,4$, with constant and field-dependent
parameters and compute respective Jacobians for changes of variables in the  path integrals.
In Section~\ref{NkWI}, we apply the results concerning the Jacobians so as to relate the
respective path integral in different gauges, and to obtain new
Ward identities, accompanied by the study of gauge dependence and
gauge-invariant Gribov--Zwanziger formulation both within $N=3$
and $N=4$ BRST local quantum actions for Yang--Mills theories.  The introduction into  Feynman diagrammatic technique in $N=3, N=4$ BRST quantum perturbative  formulations for Yang--Mills theory
is  the basic point of  Section~\ref{diagram34}. The results are summarized in Conclusions.
The proof of an impossibility to realize  $N=3$ BRST invariant gauge-fixing on the configuration space  consisting  of only  the fields of $N=3$ irreducible representation
superspace without an odd nondegenerate supermatrix (based on  an explicit construction of quantum action and $N=3$ BRST transformations) is given in Appendix~\ref{app1}.The details of derivation of $N=4$  BRST invariant  quantum gauge-fixed action
in $R_\xi$-like gauges is considered in Appendix~\ref{AppB}.

We use the DeWitt condensed notation \cite{DeWitt1}. We denote by $\epsilon (F)$
the  value of Grassmann parity  of a quantity $F$ and also
use $\eta_{\mu\nu} = diag (-,+,...,+)$ for the metric tensor of a $d$-dimensional Minkowski
space-time (generalizing the case of $d=4$), with the Lorentz indices $\mu, \nu = 0,1,...,d-1$.
A local orthonormal basis $t^m$ in the semi-simple Lie algebra $\mathcal{G}$ of $G$
is normalized by the Killing metric $\langle t^m,t^n\rangle$ = $-\frac{1}{2}\delta^{mn}$.
Derivatives with respect to the field variables $\Phi ^{A}$ and sources $J_{A}$ are denoted
by $\overleftarrow{\partial }^{A}$ ($\overrightarrow{\partial }^{A}$) for right (left) derivatives  and $\overrightarrow{\partial}_{(J)}^{A}$ for left ones. The symmetrized and antisymmetrized in $p$ and $q$ products of the tensor quantities, $F^{p}$ and $G^q$ are denoted as: $F^{\{p}G^{q\}}$, $F^{\{p}G^{q\}}= F^{p}G^{q}+F^{q}G^{p}$; $F^{[p}G^{q]}$, $F^{[p}G^{q]}=F^{p}G^{q}-F^{q}G^{p}$.
The raising and lowering of $\mathrm{Sp}\left( 2\right)$ indices,
$\big(\overleftarrow{s}^{a},\overleftarrow{s}_{a}\big)=\big(\varepsilon ^{ab}\overleftarrow{s}_{b},\varepsilon _{ab}\overleftarrow{s}^{b}\big)$,
is carried out by a constant antisymmetric tensor
$\varepsilon ^{ab}$, $\varepsilon ^{ac}\varepsilon _{cb}=\delta _{b}^{a}$, $\varepsilon ^{12}=1$.

\section{Generalization of the Faddeev--Popov method}
  \setcounter{equation}{0}

\label{GenSUSY}
Let us consider a configuration space of fields $\mathcal{A}^i = \mathcal{A}^{\mu}(x)=\mathcal{A}^{\mu n}(x)t^n$
in $\mathbf{R}^{1`,d-1}$, taking their values $\mathcal{A}^{\mu n}(x)$ in a Lie algebra
$\mathcal{G}$= $su(\hat{N})$ of a gauge group $G= SU(\hat{N})$ for $n=1,...,\hat{N}{}^2-1$,
with an action $S_0(\mathcal{A})$ invariant under gauge transformations, in the condensed notations
in finite  and infinitesimal form $\delta \mathcal{A}^i = R^i_\alpha(\mathcal{A})\xi^\alpha$, with the generators
$R^i_\alpha(\mathcal{A})$ of the gauge transformations:
\begin{eqnarray}
\hspace{-0.7em} &\hspace{-0.9em}&\ S_{0}(\mathcal{A}) = \frac{1}{2g^2}\int d^{d}x\, tr\, G_{\mu \nu }(x)G^{\mu \nu  }(x) ,
\  G_{\mu \nu }(x)=   \partial _{[
\mu }\mathcal{A}_{\nu ]}(x)   + \big[\mathcal{A}_{\mu }(x),\mathcal{A}_{\nu }(x)\big],  \label{variat} \\
\hspace{-0.7em} &\hspace{-0.9em}&
\mathcal{A}_{\mu }(x) \to \mathcal{A}^\Omega_{\mu }(x)\hspace{-0.1em} = \hspace{-0.1em}\Omega(x) \mathcal{A}_{\mu }(x) \Omega^{-1}(x)\hspace{-0.1em} +  \partial_\mu\Omega(x)\Omega^{-1}(x) \hspace{-0.1em}\Rightarrow   G_{\mu \nu }\to G^\Omega_{\mu \nu }\hspace{-0.1em} =  \Omega G_{\mu \nu }^\Omega  \Omega^{-1}  ,  \Omega \in SU(\hat{N}),
\label{R(A)}\\
\hspace{-0.7em} &\hspace{-0.9em}&\ S_{0}(\mathcal{A})= -\frac{1}{4g^2}\int d^{d}x G_{\mu \nu }^{m}(x)G^{m\mu \nu }(x),
\ G_{\mu \nu }(x) =  G_{\mu \nu }^{m}(x)t^m,  \ G_{\mu \nu }^{m} =\partial _{[\mu }\mathcal{A}_{\nu ]}^{m}+f^{mnl}\mathcal{A}_{\mu }^{n}\mathcal{A}_{\nu }^{l},  \label{variatinf} \\
\hspace{-0.7em} &\hspace{-0.9em}&
\delta \mathcal{A}_{\mu }^{m}(x)=D_{\mu }^{mn}(x)\zeta ^{n}(x)=\int d^{d}y\ R_{\mu
}^{mn}(x;y)\zeta ^{n}(y)\ ,\ \ \mathrm{where}\ \ i=(\mu ,m,x),\alpha =(n,y).
 \label{R(A)inf}
\end{eqnarray}%
Here  $G_{\mu \nu }(x)$, $\Omega(x)$, $g$ and  $D_{\mu }^{mn}(x) = \delta^{mn}\partial_\mu+f^{mon}\mathcal{A}_\mu^o(x)$ are  by the field strength, arbitrary gauge function taking theirs values in  $SU(\hat{N})$, (dimensionless for $d=4$) coupling constant,   covariant derivative with completely antisymmetric structural constants $f^{mno}$: $[t^m,t^n]=gf^{mno}t^o$ of $su(\hat{N})$ and local generators of gauge transformations, $R_{\mu
}^{mn}(x;y) = D_{\mu }^{mn}(x)\delta(x-y)$,  whereas for the infinitesimal gauge transformations (\ref{R(A)inf}) the representation, $\Omega(x) = 1+\zeta ^{m}(x)t^m$ holds.

\subsection{ Review of $N=1, 2$ BRST symmetry}

\label{N12SUSY}

In the case of usual BRST symmetry, the path integral, be it in Landau (\ref{PintfpL}),
Feynman (\ref{PintfpF}), or arbitrary admissible gauges, may be uniquely presented using
a local quantum action, $S_\Psi=S_\Psi(\Phi)$ in the space
$\mathcal{M}^{(N=1)}_{tot}\equiv \mathcal{M}_{tot} $ of fields $\Phi^A$:
\begin{eqnarray}\label{PintfpLlocn}
  &\hspace{-1em}& Z_\Psi =  \int  d \Phi   \exp \Big\{\frac{\imath}{\hbar}S_\Psi(\Phi)\Big\}, \ \mathrm{with}  \ S_\Psi  = S_0+ \Psi(\Phi) \overleftarrow{s}  = S_0+ \Big\{\overline{C}M(\mathcal{A}){C}+\chi( \mathcal{A},B)B\Big\}  ,
\end{eqnarray}
for $ M(\mathcal{A}) = \int d y \big(\partial_{\mathcal{A}^{\mu}(y)}\chi(\mathcal{A},B)\big) D^{\mu }$,
with the help of a \emph{gauge fermion} $\Psi(\Phi)$, encoding the gauge by a gauge function $\chi(\mathcal{A},B)$
linear in the fields $\mathcal{A}_\mu, B$:
\begin{equation}\label{N1gaugef}
  \Psi(\Phi) = \overline{C}\chi(\mathcal{A},B)  +  \widehat{\Psi}(\Phi), \ \epsilon(\Psi) = 1,  \ \mathrm{for} \ deg_{\Phi}\widehat{\Psi}>2 , \ deg_{\Phi}\chi(\mathcal{A},B)=1
\end{equation}
with the use,  first,  the condensed notations in (\ref{PintfpLlocn}) and  (\ref{N1gaugef}), implying the integration over some region in $\mathbf{R}^{1,d-1}$ and trace
over $su(\hat{N})$ indices, second,  of a nilpotent Grassmann-odd ``right-hand'' (left-hand) Slavnov generator
$\overleftarrow{s}$ ($s$), $\overleftarrow{s}{}^2=0$, \cite{Slavnov1} of
$N=1$ BRST transformations acting on the local coordinates of $\mathcal{M}_{tot}$,
as well as on a functional $K(\Phi)$, by the rule \cite{brs, brs1}
\begin{eqnarray}
 \hspace{-0.9em} &\hspace{-0.7em}& \Phi^A \overleftarrow{s}\hspace{-0.1em} = \big(\mathcal{A}_\mu, C, \overline{C}, B \big)\overleftarrow{s}\hspace{-0.1em} =\hspace{-0.1em}  \big(D_{\mu }C, \frac{1}{2}[C,C], B,0 \big) \hspace{-0.1em} \Leftrightarrow \hspace{-0.1em} \big(\mathcal{A}^n_\mu, C^n, \overline{C}{}^n, B^n \big)\overleftarrow{s}\hspace{-0.1em} = \hspace{-0.1em}\big(D_{\mu }^{no}C^o, \frac{1}{2}f^{nop}C^oC^p, B^n, 0 \big)\hspace{-0.1em}, \nonumber \\
\hspace{-0.9em} &\hspace{-0.7em}& sK(\Phi)  =   (s\Phi^A) \overrightarrow{\partial}_AK\  \mathrm{and} \  K(\Phi) \overleftarrow{s} =   K \overleftarrow{\partial}_A(\Phi^A \overleftarrow{s})    \, \Rightarrow  \,   s  \big(\Phi^A,K\big)= -\big( (-1)^{\epsilon_A}\Phi^A, (-1)^{\epsilon(K)}K\big) \overleftarrow{s}.\label{N1BRST}
\end{eqnarray}
The quantum action $S_\Psi$ and the integration measure $d \Phi$
are invariant  under BRST transformations $\Phi^A \to  \Phi^{\prime A}$
with a constant parameter $\mu$,
\begin{equation}\label{brstrans}
  \Phi^{\prime A} = \Phi^A(1+ \overleftarrow{s}\mu): \ \delta_{\mu}\Phi^A=\Phi^A \overleftarrow{s}\mu \Longrightarrow    \delta_{\mu}S_\Psi =0, \ \mathrm{sdet}\|\left( \delta\Phi^{\prime}/  \delta\Phi \right) \| =1,
\end{equation}
providing the invariance of the integrand in $Z_\Psi$ with respect to these transformations.
In turn, for the generating functionals of Green's functions, as well as of correlated
and one-particle irreducible Green's functions (known as well as,  the effective
action $\Gamma(\langle \Phi^A\rangle)$), depending, respectively, on the external sources
$J_A$,  $\epsilon(J_A)=\epsilon_A$ and mean fields, $\langle \Phi^A\rangle$, we have
 \begin{equation}\label{GFGF}
   Z(J)=\int  d \Phi   \exp \Big\{\frac{\imath}{\hbar}S_\Psi(\Phi)+J_A\Phi^A \Big\}= \exp \Big\{\frac{\imath}{\hbar}W(J)\Big\}, \  \Gamma(\langle \Phi^A\rangle) =  W(J) - J_A\langle \Phi^A\rangle
 \end{equation}
by means of a Legendre transformation of $W(J)$ with respect to $J_A$,
for $\langle \Phi^A\rangle = \overrightarrow{\partial}_{(J)}^{A} W $ and
$J_A = - (\delta  \Gamma / \delta \langle \Phi^A\rangle)$.
$N=1$ BRST transformations lead to the presence of respective Ward identities:
\begin{equation}\label{GFGFWIN1}
J_A\langle \Phi^A \overleftarrow{s}\rangle_{\Psi,J}=0, \ \ J_A\langle\langle \Phi^A \overleftarrow{s}\rangle\rangle_{\Psi,J}=0, \ \ \frac{\delta \Gamma}{\delta \langle \Phi^A\rangle}\langle\langle \Phi^A \overleftarrow{s}\rangle\rangle_{\Psi, \langle \Phi\rangle} =0,
 \end{equation}
with respective normalized average expectation values
$\langle L\rangle_{\Psi,J}$, $\langle\langle L\rangle\rangle_{\Psi,J}$,
$\langle\langle L\rangle\rangle_{\Psi, \langle \Phi\rangle}$
for a functional $L=L(\Phi)$ calculated using $Z(J)$, $W(J)$, $\Gamma$
for a given gauge fermion $\Psi$, with the external sources $J_A$ and $\langle \Phi^A\rangle$.

The infinitesimal field-dependent (FD) $N=1$ BRST transformations with a functional parameter
$\mu(\Phi)= (\imath / \hbar) \delta \Psi $ allow one to establish gauge-independence
for the path integral $Z_\Psi$ under an infinitesimal variation of the gauge condition,
$\Psi \to \Psi+\delta\Psi$, due to an input from the superdeterminant of the change
of variables (\ref{brstrans}), $\mathrm{sdet}\|\Phi^{\prime A} \overleftarrow{\partial}_{B} \| = 1-\mu(\Phi)\overleftarrow{s} $,
in the integrand of $Z_{\Psi+\delta\Psi}$:
\begin{equation}\label{N1GI}
  Z_{\Psi+\delta\Psi} =  \int  d \Phi   \mathrm{sdet}\|\Phi^{\prime A} \overleftarrow{\partial}_{B} \| \exp \Big\{\frac{\imath}{\hbar}S_{\Psi+\delta\Psi}(\Phi)\Big\}=   Z_{\Psi}.
\end{equation}
In turn, finite FD $N=1$ BRST transformations, whose set enlarges the Abelian supergroup,
$G(1)$ = $\{g(\mu) : g(\mu)=1 + \overleftarrow{s}\mu\}$, acting in $\mathcal{M}_{tot}$
and providing an non-Abelian supergroup, $\widetilde{G}(1)$= $\{\tilde{g}(\mu): \tilde{g}(\mu)
: \tilde{g}(\mu)=1 + \overleftarrow{s}\mu(\Phi)\}$ with
$\tilde{g}(\mu_1)\tilde{g}(\mu_2)$ = $\tilde{g}(\mu_1\overleftarrow{s}\mu_2)\ne
\tilde{g}(\mu_2\overleftarrow{s}\mu_1)$ = $\tilde{g}(\mu_2)\tilde{g}(\mu_1)$,
introduced for the first time in \cite{sdj}, allow one to obtain a new form
of the Ward identities, depending on an FD parameter, and to establish gauge-independence
for the path integral $Z_\Psi$ under a finite change of the gauge,
$\Psi\to \Psi+\Psi'$: $Z_\Psi = Z_{\Psi+\Psi'}$. In this case, the superdeterminant
of a change of variables (\ref{brstrans}), $\mathrm{sdet}\|\Phi^{\prime A}
\overleftarrow{\partial}_{B} \| =  (1+\mu(\Phi)\overleftarrow{s})^{-1} $,
calculated in \cite{ll1}  -- see also \cite{re} for general gauge theories --
implies a \emph{modified Ward identity}:
\begin{eqnarray}
&& \Big\langle \exp\Big\{\frac{i}{%
\hbar }J_{A}\phi ^{A}
\overleftarrow{s}\mu (\Psi ^{\prime })\Big\} \left( 1+\mu (\Psi ^{\prime })\overleftarrow{s}\right) {}^{-1}%
\Big\rangle _{\Psi ,J} =1, \label{mWI} \\
&& \mathrm{for}%
\ \mu (\Psi ^{\prime })= \frac{i}{\hbar }g(y)\Psi ^{\prime } , \ \Big[y,\,g(y)\Big]=\Big[{%
i}/{\hbar })\Psi ^{\prime }\overleftarrow{s},\,(1-\exp \{y\})/y\Big],  \label{N1mWI}
\end{eqnarray}%
and leads to a solution of the gauge dependence problem for the generating
functional $Z_{\Psi}(J)$:
\begin{eqnarray}
&&Z_{\Psi+\Psi^{\prime }}(J)-Z_{\Psi}(J)=\frac{\imath}{
\hbar }J_{A}\textstyle\left\langle \Phi ^{A} \overleftarrow{s}\mu \left( \Phi |-{\Psi}%
^{\prime }\right) _{\Psi,J}\right\rangle  \Rightarrow \Big(Z_{\Psi+\Psi^{\prime }}(J)-Z_{\Psi}(J)\Big)\vert_{J=0}=0. \label{GDZN1}
\end{eqnarray}

For an $N=2$ BRST symmetry realization for the quantum local action we, once again,
follow the Faddeev--Popov proposal (\ref{PintfpL}), where, instead of the gauge function
$\chi(\mathcal{A})$, a Grassmann-even gauge functional $Y(\mathcal{A})$, $\epsilon(Y)=0$,
is utilized:
  \begin{eqnarray}\label{N2lqact}
  &\hspace{-1em}&\hspace{-0.5em} Z^L_{0}\hspace{-0.3em} =\hspace{-0.2em} \int  d \mathcal{A} \delta\big( \chi^Y(\mathcal{A})\big)\det M(\mathcal{A}) \exp\Big\{\frac{\imath}{\hbar}S_0(\mathcal{A})\Big\}, \, \mathrm{for} \,
  \chi^Y(\mathcal{A})=   \frac{\delta Y}{\delta \mathcal{A}^\mu}D^\mu = \partial_\mu \mathcal{A}^\mu \Leftrightarrow  \chi^Y_\alpha  =   Y_i R^i_\alpha
\end{eqnarray}
(for $Y_i \equiv  \delta Y / \delta A^i $, $A^i=\mathcal{A}^\mu(x)$) which leads to a local representation for the path integral in the same configuration space
$\mathcal{M}_{t ot}^{(N=2)}=\mathcal{M}_{t ot}^{(N=1)}$ of fields $\Phi^A$, arranged into $Sp(2)$-doublet as
$\Phi^A=(\mathcal{A}_\mu, C^a, B)= (\mathcal{A}_\mu^m, C^{m{}a}, B^m)t^m$
\begin{eqnarray}
\label{PintfpLlocN2}
&\hspace{-1em}& Z_Y =  \int  d \Phi   \exp \Big\{\frac{\imath}{\hbar}S_Y(\Phi)\Big\}, \ \mathrm{with}  \ S_Y  = S_0-\textstyle\frac{1}{2}%
Y\overleftarrow{s}{}^a\overleftarrow{s}{}_a \  \mathrm{and} \ -\textstyle\frac{1}{2}%
Y\overleftarrow{s}{}^a\overleftarrow{s}{}_a   =  \Psi(\Phi) \overleftarrow{s}.
\end{eqnarray}
The functional (\ref{PintfpLlocN2}), in the Feynman gauge condition, providing a particular representative
(for $\xi=1$) from the class of $R_\xi$-gauges, $\partial_\mu \mathcal{A}^\mu + {\xi g^2}B$
(Landau gauge for $\xi=0$), takes the form
\begin{eqnarray}
\hspace{-0.5em} &\hspace{-0.5em}&
Z_{Y_\xi} =  \int  d \Phi   \exp \Big\{\frac{\imath}{\hbar}S_{Y_\xi}(\Phi)\Big\} \ \mathrm{for} \  Y_{\xi }(\Phi )= \frac{1}{2}\int
d^{d}x\ tr \big(-\mathcal{ A}_{\mu }\mathcal{A}^{\mu }+{\xi g^2 }\varepsilon
_{ab}C^{a}C^{b}\big)\ ,  \label{Y(A,C)} \\
\hspace{-0.5em} &\hspace{-0.5em}& \label{SY(A,C)}
S_{Y_\xi}(\Phi) = S_0-\textstyle\frac{1}{2}%
Y_\xi\overleftarrow{s}{}^a\overleftarrow{s}{}_a = S_0+S_{\mathrm{gf}} +  S_{\mathrm{gh}}+  S_{\mathrm{add}} ,
\end{eqnarray}
where the gauge-fixing term $S_{\mathrm{gf}}$ and the ghost term $S_{\mathrm{gh}}$
coincide with $N=1$ BRST exact term $\Psi_\xi(\Phi)\overleftarrow{s}$
in the $N=1$ BRST invariant quantum action
$S_{\Psi_\xi}$, for $\xi=1$, whereas the interaction term $S_{\mathrm{add}}$,
quartic in ghosts $C^{sa}$, specific for the $N=2$ BRST symmetry, is given by%
\begin{align}
S_{\mathrm{gf}}+  S_{\mathrm{gh}} &  = \int d^{d}x\ \left[    \partial^{\mu}A_{\mu
}^{m}  + {\xi g^2} B^{m}\right]  B^{m}+ \frac{1}{2}\int d^{d}x\ \left(  \partial^{\mu}%
C^{ma}\right)  D_{\mu}^{mn}C^{nb}\varepsilon_{ab}\ ,\label{Sgh}\\
S_{\mathrm{add}} &  =-\frac{\xi g^2}{ 24}\int d^{d}x\ \ f^{mnl}f^{lrs}C^{sa}%
C^{rc}C^{nb}C^{md}\varepsilon_{ab}\varepsilon_{cd}\footnotemark .  \label{Sadd}
\end{align}\footnotetext{For $g=1$,  the expressions for $S_{\mathrm{gf}}$ (\ref{Sgh}) and $S_{\mathrm{add}}$ (\ref{Sadd}) coincide with ones in \cite{re1} after  rescaling $  \xi\to \frac{1}{2}\xi$.}The quantum action and integration measure are invariant with respect $N=2$ BRST symmetry
transformations at the algebraic level, with right-hand Grassmann-odd generators
$\overleftarrow{s}{}^a$ satisfying the algebra
$\overleftarrow{s}{}^a\overleftarrow{s}{}^b+\overleftarrow{s}{}^b\overleftarrow{s}{}^a=0$, $a,b=1,2$
\begin{eqnarray}\hspace{-0.5em} &\hspace{-0.5em}&
\ \mathrm{for} \  \Phi^A \to \Phi^{\prime A} = \Phi^A(1+\overleftarrow{s}{}^a\mu_a): \nonumber \\
\hspace{-0.5em} &\hspace{-0.5em}&  \big(\mathcal{A}_\mu, C^b, B\big)\overleftarrow{s}{}^a  = \Big(D_\mu C^a, \varepsilon ^{ba}B+%
\textstyle\frac{1}{2}[C^{b}, C^{a}], \,  \textstyle\frac{1}{2}\big(
[B, C^{a}]+\frac{1}{6}[C^{c},[C^{b},C^{a}]]\varepsilon
_{cb}\big)\Big). \label{N2algBRST}
\end{eqnarray}
As in the $N=1$ BRST case, this invariance, for the corresponding generating functionals
of Green's functions, $Z_Y(J) = \exp \{(\imath / \hbar)W_Y(J)\}$, $\Gamma_Y(\langle\Phi\rangle)$
constructed by the rules (\ref{GFGF}) with a given gauge condition $Y(\Phi)$,
leads to the presence of an $Sp(2)$-doublet of Ward identities:
\begin{equation}\label{GFGFWIN2}
J_A\langle \Phi^A \overleftarrow{s}{}^a\rangle_{Y,J}=0, \ \ J_A\langle\langle \Phi^A \overleftarrow{s}{}^a\rangle\rangle_{Y,J}=0, \ \ \frac{\delta \Gamma_Y}{\delta \langle \Phi^A\rangle}\langle\langle \Phi^A \overleftarrow{s}{}^a\rangle\rangle_{Y, \langle \Phi\rangle} =0,
 \end{equation}
with respective normalized average expectation values $\langle L\rangle_{Y,J}$,
$\langle\langle L\rangle\rangle_{Y,J}$, $\langle\langle L\rangle\rangle_{Y, \langle \Phi\rangle}$
for a functional $L=L(\Phi)$ calculated using $Z_Y(J)$, $W_Y(J)$, $\Gamma_Y$
for a given gauge boson $Y$ in the presence of external sources $J_A$ and mean fields
$\langle \Phi^A\rangle$. The gauge independence of the path integral $Z_Y(0)$
under an infinitesimal variation of the gauge condition, $Y \to Y+\delta Y$,
is established using the infinitesimal field-dependent (FD) $N=2$ BRST transformations
\cite{bltsp2, BLT2} with the functional parameters
$\mu_a(\Phi)= (\imath / 2\hbar) \delta Y \overleftarrow{s}_a$
which induce the superdeterminant for the change of variables
(\ref{brstrans}) made in the integrand of $Z_{Y+\delta Y}$,
$\mathrm{sdet}\|\Phi^{\prime A} \overleftarrow{\partial}_{B} \|
= 1- \mu_a(\Phi)\overleftarrow{s}{}^a $, as follows:
\begin{equation}\label{N2GI}
  Z_{Y+\delta Y} =  \int  d \Phi \  \mathrm{sdet}\|\Phi^{\prime A} \overleftarrow{\partial}_{B} \| \exp \Big\{\frac{\imath}{\hbar}S_{Y+\delta Y}(\Phi)\Big\}=   Z_{Y}.
\end{equation}
The finite  $N=2$ BRST transformations acting in $\mathcal{M}_{tot}$,
whose set forms an Abelian  supergroup,
\begin{equation}\label{N2finBRST}
G(2) = \left\{g(\mu_a)
: g(\mu_a)= 1 + \overleftarrow{s}{}^a\mu_a + \frac{1}{4}\overleftarrow{s}{}^a\overleftarrow{s}{}_a \mu_b\mu^b = \exp \left( \hspace{-0.1em}\overleftarrow{s%
}{}^{a}\mu _{a}\hspace{-0.1em}\right)\right\},
\end{equation}
are restored from the algebraic $N=2$ BRST transformations according to \cite{re1}:%
\begin{equation}
\hspace{-0.1em}\left\{ K\left( g(\mu _{a})\Phi \right) =K\left( \Phi
\right) \ \mathrm{and}\ K\overleftarrow{s}{}^{a}=0\right\} \hspace{-0.1em}%
\Rightarrow g\left( \hspace{-0.1em}\mu _{a}\hspace{-0.1em}\right) =\exp \left\{ \hspace{-0.1em}\overleftarrow{s%
}{}^{a}\mu _{a}\hspace{-0.1em}\right\} ,  \label{bab}
\end{equation}%
where $K=K\left( \Gamma \right) $ is an arbitrary regular functional,
and $\overleftarrow{s}{}^{a}$, $%
\overleftarrow{s}{}^{2}\equiv \overleftarrow{s}{}^{a}\overleftarrow{s}{}_{a}$
are the generators of BRST-antiBRST and mixed BRST-antiBRST transformations
in the space of $\Phi^A$. These finite transformations, in a manifest form
\cite{re1}, for $\Delta \Phi^A= \Phi^{\prime A}-\Phi^A$, read as follows:
\begin{eqnarray}\hspace{-0.5em} &\hspace{-0.5em}&\Delta \mathcal{A}_{\mu }=D_{\mu
}C^{a}\mu_{a}-\textstyle\frac{1}{2}\left( D_{\mu }B+%
\frac{1}{2}[C^{a}, D_{\mu }C^{b}]\varepsilon _{ab}\right) \mu
^{2}\ ,  \label{DAmm} \\
\hspace{-0.5em} &\hspace{-0.5em}&\Delta B=\textstyle\frac{1}{2}\big(
[B, C^{a}]+\frac{1}{6}[C^{c},[C^{b},C^{a}]]\varepsilon
_{cb}\big)\mu _{a}\ ,  \label{DBm} \\
\hspace{-0.5em} &\hspace{-0.5em}&\Delta C^{b}=\left( \varepsilon ^{ba}B+%
\textstyle\frac{1}{2}[C^{b},C^{a}]\right) \mu _{a}+\frac{1}{2}%
\left( [B,C^{b}]+\frac{1}{6}[C^{c},[C^{a},C^{b}]]%
\varepsilon _{ca}\right) \mu ^{2}\ ,  \label{DCma}
\end{eqnarray}%
and cannot be presented as group elements (in terms of an $\exp$-like relation)
for an $\mathrm{Sp}(2)$-doublet $\mu _{a}(\Phi)$ which is not closed under BRST-antiBRST
transformations: $\mu _{a}(\Phi)\overleftarrow{s}_{b}\neq 0$.
Once again, the finite FD $N=2$ BRST transformations with functionally-dependent parameters
$\mu_a = \Lambda \overleftarrow{s}_a$ allow one to derive a new form of the Ward identities,
depending on FD parameters, and to study gauge-independence for the generating functionals,
e.g., $Z_Y(J)$ and $Z_Y$, under a finite change of the gauge, $Y\to Y+Y'$, $Z_Y = Z_{Y+Y'}$.
Now, the superdeterminant for a change of variables: $\Phi^A\to \Phi^{\prime A} = \Phi^Ag(\mu_a(\Phi))$,
$\mathrm{sdet}\|\Phi^{\prime A} \overleftarrow{\partial}_{B} \| =
(1- \frac{1}{2}\Lambda(\Phi)\overleftarrow{s}{}^a\overleftarrow{s}{}_a)^{-2} $,
calculated in \cite{re1} --  see also \cite{MR5}   for  general gauge theory and
general form of FD parameters $\mu_a$ -- leads to a \emph{modified Ward identity}
depending on the parameters $\mu
_{a}(Y^{\prime })=\frac{i}{2\hbar }g(y)Y^{\prime }\overleftarrow{s}_{a}$, $%
\Lambda (\Phi|{Y}^{\prime })=\frac{i}{2\hbar }g(y){Y}^{\prime }$, for $%
y\equiv ({i}/{4\hbar })Y^{\prime }\overleftarrow{s}{}^{2}$,  \cite{MR4}, \cite{MR5}
\begin{eqnarray}
&&\textstyle\left\langle \left\{ 1+\frac{i}{\hbar }J_{A}\Phi ^{A}\left[
\overleftarrow{s}^{a}\mu _{a}(\Lambda )+\frac{1}{4}\overleftarrow{s}^{2}\mu
^{2}(\Lambda )\right] -\frac{1}{4}\left( \frac{i}{\hbar }\right)
{}^{2}J_{A}\Phi ^{A}\overleftarrow{s}^{a}J_{B}\Big(\Phi ^{B}\overleftarrow{s}%
_{a}\Big)\mu ^{2}(\Lambda )\right\} \right.  \label{mWIN2} \\
&&\qquad \left. \times \textstyle\left( 1-\frac{1}{2}\Lambda \overleftarrow{s}%
^{2}\right) {}^{-2}\right\rangle _{Y,J}=1,  \nonumber \\
&&Z_{Y+Y^{\prime }}(J)-Z_{Y}(J) =\textstyle\left\langle \frac{i}{%
\hbar }J_{A}\phi ^{A}\left[ \overleftarrow{s}^{a}\mu _{a}\left( \Phi |-{Y}%
^{\prime }\right) +\frac{1}{4}\overleftarrow{s}^{2}\mu ^{2}\left( \Phi |-{Y%
}^{\prime }\right) \right] \right.   \nonumber \\
&&\qquad -\left.  \textstyle(-1)^{\varepsilon _{B}}\left( \frac{i}{2\hbar }%
\right) ^{2}J_{B}J_{A}\left( \Phi ^{A}\overleftarrow{s}{}^{a}\right) \left(
\Phi ^{B}\overleftarrow{s}_{a}\right) \mu ^{2}\left( \Phi |-{Y}^{\prime
}\right) \right\rangle _{Y,J} ,  \label{GDInewN2}
\end{eqnarray}%
vanishing on the mass shell determining by the hypersurface $J_A=0$.

Now, we have all the things prepared to generalize the Faddeev--Popov procedure
in order to realize a more general case of $N=3$  BRST symmetry for an appropriate
local quantum action depending on the entire set of fields, on which the latter symmetry
transformations are defined.

\subsection{Proposal for non-local Faddeev--Popov path integral with $N=3$ BRST symmetry}\label{N3FPtrick}

There are many ways to present the functionals  (\ref{PintfpL}), (\ref{PintfpF}) without using a determinant
and a functional $\delta$-function within perturbation techniques. In the case of Landau and Feynman gauges,
we generalize the path integral (\ref{PintfpL}), \ref{PintfpF}  by the rule
\begin{eqnarray}\label{Pintgen}
 Z^L_{0} & = & \int  d \mathcal{A} \delta\big(\chi(\mathcal{A})\big)\det M(\mathcal{A}) {\det}^{k} M(\mathcal{A}) \,  {\det}^{-k} M(\mathcal{A})\exp\Big\{\frac{\imath}{\hbar}S_0(\mathcal{A})\Big\} , \ k \geq 0,\\
 \label{PintgenF} Z^F_{0} & = &  \int  d \mathcal{A} d B \det M(\mathcal{A}) {\det}^{k} M(\mathcal{A}) \,  {\det}^{-k} M(\mathcal{A})\exp\Big\{\frac{\imath}{\hbar}S_0(\mathcal{A})+ \int d^4x  \big(\partial^\mu \mathcal{A}_\mu + {g^2}B\big)B\Big\}.
\end{eqnarray}
The path integral formulations with local quantum action exist for any $k \in \mathbf{N}_0$ as follows, e.g. for (\ref{Pintgen}):
\begin{eqnarray}\label{PintfpLlock}
 &\hspace{-1em}&\hspace{-0.5em} Z^L_{0} =  \hspace{-0.3em}\int \hspace{-0.2em} d \mathcal{A}dB\prod_{l=0}^{k} d C^{l} d \overline{C}{}^{l} \prod_{l=1}^k d B^{l} d \overline{B}{}^{l} \exp \Big\{\hspace{-0.1em}\frac{\imath}{\hbar}\widetilde{S}{}^L_{(k)}\big(\mathcal{A}, C^0,\overline{C}{}^0, C^{[k]}, \overline{C}{}^{[k]}, B^{[k]}, \overline{B}{}^{[k]}, B\big)\Big\} \\
   &\hspace{-1em}&\hspace{-0.5em}\ \mathrm{with}  \ \widetilde{S}{}^L_{(k)}  =  S_0+ \int \hspace{-0.2em}d^dx \Big\{\sum_{l=1}^k\big(\overline{C}{}^lM(\mathcal{A}){C}^l+\overline{B}{}^lM(\mathcal{A}){B}{}^{l}\big)+\overline{C}{}^0M(\mathcal{A}){C}^0+\chi( \mathcal{A})B\Big\}, \label{SintfpLlock}\\     &\hspace{-1em}&\hspace{-0.5em}\ \mathrm{for} \ D^{[k]}=(D^1,...,D^k), \ D\in\{C,\overline{C},B, \overline{B}\},  \ (C^0, \overline{C}{}^0) \equiv (C, \overline{C}{}),
\end{eqnarray}
where   odd-valued fields $C^{[k]}, \overline{C}{}^{[k]}$ and even-valued fields $B^{[k]}, \overline{B}{}^{[k]} $  taking values in Lie group $G$, whose numbers coincide.

However, it is not for any $k$  there exists a local representation for the path integral (\ref{PintfpLlock}) such that the total set of  fields, $\widetilde{\Phi}_{(k)}$, $\widetilde{\Phi}_{(k)}=(\mathcal{A}, C^0, \overline{C}{}^0$,  $C^{[k]}, \overline{C}{}^{[k]}, B^{[k]}, \overline{B}{}^{[k]}, B)$ forms the representation space of  Abelian group  of SUSY transformations, like $N=1,2$ BRST symmetry, for $k=0$, but with larger numbers of $N\geq 3$, so that the Grassmann-odd: with $C^l,\overline{C}{}^l$; Grassmann-even  : with $B^l,\overline{B}{}^l$ ghost actions with Faddeev-Popov operator and gauge-fixed term with $\chi( \mathcal{A})B$ would be generated as the  exact terms  with respect to the action of being searched  $N$-parametric generators of BRST symmetry transformations.

More exactly, the fact  is  that

\noindent
\textbf{Statement 1:} In order  the action functional $\widetilde{S}{}^L_{(k)}$, (\ref{SintfpLlock}) to be given on the configuration space of fields $\widetilde{\Phi}_{(k)}=(\mathcal{A}, C^0, \overline{C}{}^0$,  $C^{[k]}, \overline{C}{}^{[k]}, B^{[k]}, \overline{B}{}^{[k]}, B)$ permitting the local presentation for the path integral, $Z^L(0)$, (\ref{Pintgen}) in the form  (\ref{PintfpLlock})  will be invariant with respect to $N=N(k)$-parametric Abelian SUSY transformations with Grassmann-odd  generators $\overleftarrow{s}{}^{p_k}$: $\overleftarrow{s}{}^{p_k}\overleftarrow{s}{}^{q_k}+\overleftarrow{s}{}^{q_k}\overleftarrow{s}{}^{p_k}=0$,   $q_k,p_k=1,...,N$, and will be presented in the form:
\begin{eqnarray}
   &\hspace{-0.8em}& \hspace{-0.7em} S^L_{(N(k))}(\Phi_{(N(k))}) = S_0(\mathcal{A})- \frac{(-1)^N}{N!} F_{(N(k))}\big(\Phi_{(N(k))}\big)\prod_{e=1}^N\overleftarrow{s}{}^{p^e_k}\varepsilon_{p^1_kp^2_k\ldots p^N_k}, \ \epsilon( F_{(N(k))})=N \label{qe}, \end{eqnarray}
with completely antisymmetric $N(k)$-rank (Levi-Civita)  tensors $\varepsilon_{p^1_kp^2_k\ldots p^N_k}$, $\varepsilon^{p^1_kp^2_k\ldots p^N_k}$ normalized as,
  \begin{equation}\label{lchevn}
 \varepsilon_{p^1_kp^2_k\ldots p^N_k}\varepsilon^{p^1_kp^2_k\ldots p^N_k} = N! \ \mathrm{for}  \ k>2,
\end{equation}
with some gauge-fixing functional,  $F_{(N(k))}$ corresponding to the Landau gauge, so that the fields $\Phi_{(N(k))}$\footnote{When the exponential index $k$ in the representations (\ref{Pintgen}), (\ref{PintgenF}) is related to $N=N(k)$-parametric SUSY transformations we will  denote  the fields parameterizing configuration space, the quantum action and gauge-fixing functional as, $\Phi_{(N(k))}$, $S^L_{(N(k))}$, $F_{(N(k))}$ in opposite case we add "tilde" over it: $\widetilde{\Phi}_{(k)}$, $\widetilde{S}^L_{(k)}$, $\widetilde{F}_{(k)}$ so that for $N(k)=k$ in general: $\Phi_{(N(k))}\ne  \widetilde{\Phi}_{(k)}$, $S^L_{(N(k))} \ne \widetilde{S}^L_{(k)}$.} should  parameterize   the  irreducible representation superspace of   the Abelian superalgebra $\mathcal{G}(N(k))$ of $N(k)$-parametric  SUSY transformations,
the spectra of integer $k=k(N)$ should be found as:
\begin{eqnarray}\label{kN}
1) \qquad k(1) =0,\   k(N) = 2^{N-2}-1, \ \mathrm{for} \ N\geq 2\,.
\end{eqnarray}
If in addition, the gauge-fixing functional $F_{(N(k))}$ should be determined without introducing auxiliary  Grassmann-odd scalar or supermatrix the spectra of integer $k=k_u(N)$  is determined by the relation:
\begin{equation}\label{kNu}
2) \qquad k_u(1) =0,\   k_u(N) = 2^{2\left[\frac{N-1}{2}\right]}-1, \ \mathrm{for} \ N\geq 2,
\end{equation}
 for integer part, $[x]$, of real $x$.

 \vspace{1ex}
 \noindent
  Note, the  local path integral $Z^L_{F_{(N(k))}}(0) = \int  d \Phi_{(N(k))}\exp\{(\imath/\hbar) S^L_{(N(k))}(\Phi_{(N(k))})\}$ $\ne $ $Z^L_0$ for $N(k)>2$   due to the presence of possible  additional vertexes in fictitious fields in $S^L_{(N(k))}$. In addition, in the second case the requirement of the irreducibility of the $\mathcal{G}(N(k))$ (finite-dimensional) representations for each $N(k)$ is weakened. The irreducibility will be hold only for even $N$: $N=2K$, $K\in \mathbf{N}$.

  Indeed, this leads, for $N=1, k(1)=0$, to the standard Faddeev--Popov representation
(\ref{PintfpLloc}) with the BRST symmetry, whereas, for $k(2)=0 $, this leads
to the $N=2$ BRST symmetry with a local path integral (\ref{PintfpLlocN2}).

For $N=3$, $k(3) = 1$, there arises a first non-trivial case for the case $1$ (\ref{kN}) and $k_u(3)=3$ for the case $2$ (\ref{kNu}) of the Statement 1.
For $N=4$ for both cases we have from (\ref{kN}), (\ref{kNu}): $k(4) = k_u(4) = 3$.

The  validity of  the first part (\ref{kN}) follows from the simple fact that any field finite-dimensional irreducible tensor representation superspace  of the Abelian superalgebra
$\mathcal{G}(N)$ with Grassmann-odd generators $\overleftarrow{s}{}^p$: $\prod_{l=1}^{N+1}\overleftarrow{s}{}^{p_l}=0$,  contains in addition to the gauge fields $\mathcal{A}^\mu$, whose infinitesimal gauge transformations are changed on the global transformations with constant Grassmann-odd parameters, $\lambda_p$, $\epsilon(\lambda_p)=1$:
\begin{equation}\label{gtbt}
  \delta \mathcal{A}^\mu(x) = D^\mu \xi (x) =  D^\mu C^p(x)\lambda_p = \delta_\lambda  \mathcal{A}^\mu(x) = \mathcal{A}^\mu(x) \overleftarrow{s}{}^p\lambda_p  ;
\end{equation}
(where the summation with respect to repeating indices, $p$, is implied)
the $N$-plet of Grassmann-odd fields,
$C^p$, $\frac{1}{2}N(N-1)$ new Grassmann-even fields, $B_1^{p_1p_2}$, and so on up to $N$-plet of new  fields, $B^{p_1p_2...p_{N-1}}$, ($\epsilon(B^{p_1p_2...p_{N-1}})=N-1$) and new single field, $B_{(N)}$, ($\epsilon(B_{(N)}) = N$). All the new fields take theirs values in  $su(\hat{N})$ and appear from the  chain:
\begin{align}
  & \mathcal{A}^\mu \overleftarrow{s}{}^p= D^\mu C^p, &&  C^{p_1}\overleftarrow{s}{}^{p_2}=  B^{p_1p_2}+\mathcal{O}(CB), \ldots,  \label{chain1} \\
  & \ldots\ldots\ldots\ldots, && \ldots\ldots\ldots\ldots, \nonumber\\
  & B^{p_1p_2...p_{N-2}}\overleftarrow{s}{}^{p_{N-1}} =B^{p_1p_2...p_{N-1}}+\mathcal{O}(CB),
 & & B^{p_1p_2...p_{N-1}}\overleftarrow{s}{}^{p_{N-1}} =B^{p_1p_2...p_{N-1}}+\mathcal{O}(CB), \\
  &  B^{p_1p_2...p_{N-1}}\overleftarrow{s}{}^{p_{N}} =\epsilon^{p_1p_2...p_{N}}B_{(N)}+\mathcal{O}(CB), &&  B_{(N)}\overleftarrow{s}{}^p=\mathcal{O}(CB), \label{chain2}
\end{align}
generated by a nilpotent of the order $(N+1)$ differential-like element, $\overleftarrow{d}:$ $\overleftarrow{d} = \sum_p \overleftarrow{s}{}^p$, the such that  $(\overleftarrow{d})^{N+1}=0$.
The length of the chain, $l$, is equal to, $l=(N+1)$, whereas its non-vanishing linear part  in fields $C^p$, $B^{p_1...p_l}$, $B$, for $l=2,...,N-1$,  due to the last equation in (\ref{chain2}) has the length, $l_{\mathrm{lin}} = N$.
The Grassmann-odd and Grassmann-even numbers of new degrees of freedom  for additional to $ \mathcal{A}^\mu$ fields in the   multiplet $\Phi_{(N(k))}$ of the irreducible representation of the superalgebra $\mathcal{G}(N)$ without decomposition in $su(\hat{N})$ generators $t^m$
are equal to, $(2^{N-1}, 2^{N-1}-1) $.  Indeed, for $N=1,$ the only ghost field $C(x)$ contains in $N=1$ irreducible multiplet. For $N=2$, two ghost-antighost $C^p\equiv C^a$, $a=1,2$  and Nakanishi-Lautrup, $B_2\equiv B$, fields. Then, first, extracting the   degrees of freedom relating to the  usual ghost and antighost $C, \overline{C}$ and $B$ fields, second,  dividing any subset on pairs  of Grassmann-odd (and Grassmann-even) fields as it is given  in (\ref{SintfpLlock}), we get to the  value of $k=k(N)$ for the respective exponent of the determinants in (\ref{Pintgen}):
 \begin{equation}\label{kNp}
   (2^{N-1}, 2^{N-1}-1)  \to  \big(2^{N-1}-2, 2^{N-1}-2\big) \to  \frac{1}{2}\big(2^{N-1}-2, 2^{N-1}-2\big) \stackrel{\ref{kN}}{=} \big(2^{N-2}-1, 2^{N-2}-1\big).
 \end{equation}
 However, we meet the problem when going to construct the action functional (quantum action),  $S^L_{(N(k))}$, by the rule   (\ref{qe}) for odd $N=2K-1$, in particular, for $N=3$ SUSY transformations on the $\mathcal{G}(N)$-irreducible representation superspace.
Indeed, the  respective gauge-fixing  functional,  $F_{(3)}\big(\Phi_{(3)}\big)$  due to the linear part of the $N=3$ SUSY transformations (\ref{chain1}), (\ref{chain2})  should be, at least, quadratic in the  fields $\mathcal{A}^\mu$. The fact that, the Grassmann parity of $F_{(3)}$ determines it as the fermion, $\epsilon(F_{(3)})=1$, means  the necessity to introduce some additional Grassmann odd non-degenerate supermatrix  in order to realize the prescription (\ref{qe}) for the quantum action.
The details of  using of such kind of odd supermatrix, which should both to determine the required  Grassmann parity of $F_{(3)}$ and to change the basis of additional fields $(C,B)$ in the configuration space parameterized by $\Phi_{(3)}$ to construct $N=3$ SUSY invariant action  $S^L_{(3)}(\Phi_{(3)})$ for $k(3)=1$ are considered in the Appendix~\ref{app1}.

For $k(N)\big|_{N=5;6,...}=7;15,\ldots $, etc, the situation is more involved,
and we leave its detailed consideration out of the paper scope (see as well comments in the Conclusions).

The validity of the second part (\ref{kNu}) of the Statement we consider here only for $N=3$  case, whereas for even,  $N=2K$,  case its both parts (\ref{kN}) and (\ref{kNu}) coincide.

To do so we should  determine the total configuration space, $\mathcal{M}^{(N=3)}_{tot}\equiv \mathcal{M}^{(3)}_{tot}$, of fields parameterizing it, $\big(\Phi_{(3)}, \overline{\Phi}_{(3)}\big)$=$\widetilde{\Phi}_{(3)} $ being sufficient to construct the (bare) quantum action, ${S}^L_{(3)}$, which must form a finite-dimensional field completely reducible representation of Abelian $\mathcal{G}(3)$ superalgebra. That means, that on the fields $\overline{\Phi}_{(3)}$ it will be realized the another irreducible representation  of $N=3$-parametric $\mathcal{G}(3)$ superalgebra not being entangled with the irreducible  $\mathcal{G}(3)$-representation acting on the fields $\Phi_{(3)}$.

First of all, let us find exactly the action of  the generators $\overleftarrow{s}{}^{p}$ of  $\mathcal{G}(N)$-representation for  $N=3$ on the fields $\Phi_{(3)}$, $\Phi_{(3)}=\big(\mathcal{A}^\mu, C^{p_1}, B^{p_1p_2}, B_{(3)}= \widehat{B}\big)$ parameterizing irreducible representation superspace from (\ref{chain1}) --(\ref{chain2}).

\noindent
\textbf{Lemma 1}: The action of the generators  $\overleftarrow{s}{}^{p}$ of the Abelian superalgebra $\mathcal{G}(3)$ on the fields $\Phi_{(3)}$ is given by the relations:
 \begin{equation}\label{minN3}
\begin{array}{|c|c |}\hline
  \downarrow \  \leftarrow \!&\!  \overleftarrow{s}{}^{p}\!\! \!\! \\
   \hline
    \! \! \mathcal{A}^\mu &\! D^\mu(\mathcal{A}) C^p \!\!   \\

 \!\! C^{p_1}    &\! B^{p_1p}+%
\textstyle\frac{1}{2}\big[C^{p_1}, C^{p}\big] \!\!   \\
 \!\! B^{p_1p_2}    &\! \varepsilon^{p_1p_2p}\widehat{B}+ \textstyle\frac{1}{2}\Big(
\big[B^{p_1p_2}, C^{p}\big]-\frac{1}{6}\big[C^{[p_1},\big[C^{p_2]},C^{p}\big]\big]\Big) \!\!   \\
    \!\! \widehat{B}    &\! \textstyle\frac{1}{2}\big[\widehat{B},\,C^p\big] - \Big\{\frac{1}{8}\big[\big[B^{p_1p_2},\,C^{p_3}\big],\,C^p\big] +\frac{1}{6} \big[\big[B^{p_1p},\,C^{p_2}\big],\,C^{p_3}\big]\Big\} \varepsilon_{p_1p_2p_3}  \!\!   \\
   \cline{1-2}
\end{array}. \end{equation}
The respective $N=3$ SUSY transformations with triplet of anticommuting parameters, $\lambda_p$, on the fields $\Phi_{(3)}$  are determined  as: $\delta_\lambda \Phi_{(3)} = \Phi_{(3)} \overleftarrow{s}{}^{p} \lambda_p $.

To prove the representation (\ref{minN3})  we start from the boundary condition for such transformation inherited from the gauge transformations for $\mathcal{A}^\mu$ (\ref{gtbt}) and present the  realization for  the sought-for generators as series:
\begin{equation}\label{sersp}
\overleftarrow{s}^p = \sum_{e\geq 0}\overleftarrow{s}^p_e: \qquad  \mathcal{A}^\mu\overleftarrow{s}{}^{p} =  \mathcal{A}^\mu\overleftarrow{s}{}^{p}_0 = D^\mu(\mathcal{A})C^{p} \ \mathrm{and }\  C^{p_1}\overleftarrow{s}{}^{p}_0\equiv 0.
\end{equation}
Since, first,
\begin{equation}\label{s0ne0}
\mathcal{A}_\mu\Big(\overleftarrow{s}{}_0^p\overleftarrow{s}{}_0^r
+\overleftarrow{s}{}^r_0\overleftarrow{s}{}^p_0\Big)  \ne 0,
\end{equation}
 we must add to
$\overleftarrow{s}{}_0^p$  the nontrivial action of new  $\overleftarrow{s}{}^p_1$ on $C^{p_1}$ (vanishing when acting on $\mathcal{A}_\mu$: $\mathcal{A}_\mu\overleftarrow{s}{}^p_1 \equiv 0$),
starting from the Grassmann-even  triplet  of the fields $B^{p_1p_2}= B^{p_1p_2{}m}t^m$ (BRST-like variation of $C^{p_1}$) (\ref{chain1})
\begin{eqnarray}\label{3brstC}
  C^{p_1} \overleftarrow{s}{}^{p_2}_1 \ = \   B^{p_1p_2} + (\kappa_{C1})^{p_1p_2}_{r_1r_2}  \big[C^{r_1},\,C^{r_2}\big]  ,   \ \mathrm{for}  \ B^{p_1p_2} = - B^{p_2p_1} = \big(B^{ 12}, B^{13}, B^{ 23}\big), \ \epsilon(B^{p_1p_2})=0
\end{eqnarray}
(where the summation with respect to repeated indices is assumed)
with unknown real numbers: $ (\kappa_{C1})^{p_1p_2}_{r_1r_2}$  $=(\kappa_{C1})^{p_1p_2}_{r_2r_1}$,
to be determined from the consistency of $3\times 3$ equations:
\begin{equation}\label{consistA}
\mathcal{A}_{\mu}\big(\overleftarrow{s}{}^{p_1}_{[1]}\overleftarrow{s}{}^{p_2}_{[1]}+\overleftarrow{s}{}^{p_1}_{[1]}\overleftarrow{s}{}^{p_2}_{[1]}\big) =0 ,  \  \ \mathrm{where} \ \overleftarrow{s}^p_{[l]}\equiv \sum_{n\geq0}^l\overleftarrow{s}^p_{n}, \ \mathrm{and} \  C^{p_1} \overleftarrow{s}{}^{p_2}_0 \equiv 0,
\end{equation}
from which, in fact,  follows the property of antisymmetry for $B^{p_1p_2}$ in the indices $p_1,p_2$.
The solution for (\ref{consistA})  determines:
\begin{equation}\label{solconstA}
 (\kappa_{C1})^{p_1p_2}_{r_1r_2} = \frac{1}{4} \delta_{\{r_1}^{p_1}\delta_{r_2\}}^{p_2},
\end{equation}
providing the validity of the $2$-nd row in the table (\ref{minN3}).
Having in mind, that any completely antisymmetric tensor, $\sigma^{p_1...p_n}$ of the $n$-th rank,  is vanishing
for $n>3$, there are only the third-rank independent completely antisymmetric
constant tensor with upper, $\varepsilon^{p_1p_2p_3}=-\varepsilon^{p_1p_3p_2}=-\varepsilon^{p_2p_1p_3}$,
and lower, $\varepsilon_{p_1p_2p_3}$, indices, which are normalized
by the conditions (according with (\ref{lchevn}))
\begin{equation}\label{eprt}
  \varepsilon^{123}= 1, \, \varepsilon^{p_1p_2p_3}\varepsilon_{r_1r_2p_3}= \big(\delta^{p_1}_{r_1}\delta_{r_2}^{p_2} -\delta^{p_2}_{r_1}\delta_{r_2}^{p_1} \big), \, \varepsilon^{p_1p_2p_3}\varepsilon_{r_1p_2p_3}=2\delta^{p_1}_{r_1}.
\end{equation}
Second, because of
\begin{equation}\label{consistC1}
C^{p}\big(\overleftarrow{s}^{p_1}_{[1]}
\overleftarrow{s}^{p_2}_{[1]}+\overleftarrow{s}^{p_2}_{[1]}
\overleftarrow{s}^{p_1}_{[1]}\big) \ne 0,
\end{equation}
we should determine, for a nontrivial action of $\overleftarrow{s}{}^{p}_2$
on $B^{ p_1p_2}$ (vanishing when acting on $\mathcal{A}_\mu, C^p$: $\big(\mathcal{A}_\mu$, $C^{p_1}\big)\overleftarrow{s}{}^p_2$ $\equiv$ $0$), in the form of a general anzatz, starting from the new Grassmann-odd
field variables $\widehat{B}= \widehat{B}{}^{m}t^m$ (BRST-like variation of $B^{p_1p_2}$) (\ref{chain2})
up to the third power in $C^p$ with a preservation of Grassmann homogeneity
in each summand, as in the (\ref{3brstC}),
\begin{eqnarray}\label{3brstB}
\hspace{-0.5em}  &\hspace{-0.5em}&\hspace{-0.5em}  B^{p_1p_2} \overleftarrow{s}{}^{p_3}_2 \ = \ \epsilon^{p_1p_2p_3}   \widehat{B}  +  (\kappa_{B1})^{p_1p_2p_3}_{r_1r_2r_3}  \big[B^{r_1r_2},\,C^{r_3}\big]+ (\kappa_{B2})^{p_1p_2p_3}_{r_1r_2r_3}   \big[C^{r_1},\,\big[C^{r_2},\,C^{r_3}\big]\big]  ,   \,  \epsilon( \widehat{B}) = 1.
\end{eqnarray}
with unknown real numbers: $(\kappa_{Bj})^{p_1p_2p_3}_{r_1r_2r_3}$ = $\hspace{-0.1em}-\hspace{-0.1em}(\kappa_{Bj})^{p_2p_1p_3}_{r_1r_2r_3}$, $j=1,2$;  $(\kappa_{B1})^{p_1p_2p_3}_{r_1r_2r_3}$  $=\hspace{-0.1em}-\hspace{-0.1em}(\kappa_{B1})^{p_1p_2p_3}_{r_2r_1r_3}$; $\hspace{-0.1em}(\kappa_{B2})^{p_1p_2p_3}_{r_1r_2r_3}$  $\hspace{-0.1em}=(\kappa_{B2})^{p_1p_2p_3}_{r_1r_3r_2}$,  to be determined from the fulfillment of
the $3\times3\times 3$ equations
\begin{eqnarray}\label{consistC}
  && C^{p_1}\big(\overleftarrow{s}{}^{p_2}_{[2]}\overleftarrow{s}{}^{p_3}_{[2]}+\overleftarrow{s}{}^{p_3}_{[2]}\overleftarrow{s}{}^{p_2}_{[2]}\big) =0, \ \mathrm{where}  \ B^{p_1p_2}\overleftarrow{s}{}^{p_3}_{l}\equiv 0,\, l=0,1.
\end{eqnarray}
Its general  solution has the form:
\begin{eqnarray}\label{solconstC}
\hspace{-0.5em}  &\hspace{-0.5em}&\hspace{-0.5em} (\kappa_{B1})^{p_1p_2p_3}_{r_1r_2r_3} = \frac{1}{4} \delta^{[p_1}_{r_1}\delta^{p_2]}_{r_2} \delta^{p_3}_{{r_3}}:\  \ (\kappa_{B1})^{p_1p_2p_3}_{r_1r_2r_3} \big[B^{r_1r_2},\,C^{r_3}\big] = \frac{1}{2} \big[B^{p_1p_2},\,C^{p_3}\big], \\
  \hspace{-0.5em}  &\hspace{-0.5em}&\hspace{-0.5em}  (\kappa_{B2})^{p_1p_2p_3}_{r_1r_2r_3} = -\frac{1}{12} \delta^{[p_1}_{r_1}\delta^{p_2]}_{r_2} \delta^{p_3}_{r_3}:\ \    (\kappa_{B2})^{p_1p_2p_3}_{r_1r_2r_3}  \big[C^{r_1},\,\big[C^{r_2},\,C^{r_3}\big]\big] = - \frac{1}{12} \big[C^{[p_1},\,\big[C^{p_2]},\,C^{p_3}\big]\big].
\end{eqnarray}
providing the validity of the $3$-rd row in the table (\ref{minN3}).

Third, due to
\begin{equation}\label{consistB}
B^{p_1p_2}\big(\overleftarrow{s}{}^{p_3}_{[2]}\overleftarrow{s}{}^{p_4}_{[2]}
+\overleftarrow{s}{}^{p_4}_{[2]}\overleftarrow{s}{}^{p_3}_{[2]}\big) \ne 0,
\end{equation}
we should determine for a nontrivial action of $\overleftarrow{s}{}^{p}_3$ on $\widehat{B}$, (vanishing when acting on $\mathcal{A}_\mu, C^p, B^{p_1p_2}$: $\big(\mathcal{A}_\mu,C^p, B^{p_1p_2}\big)\overleftarrow{s}{}^{p_3}_{3} \equiv 0$)
 a general ansatz without using the new field variables
(due to of the $4$-th order nilpotency of $\overleftarrow{s}{}^{p}$:  $\prod_{l=1}^4\overleftarrow{s}{}^{p_l}\equiv 0$)
up to the fourth order in $C^p$ with a preservation of Grassmann homogeneity
in each summand, as in the case of (\ref{3brstC}) and (\ref{3brstB}),
\begin{eqnarray}
 \widehat{B} \overleftarrow{s}{}^p_3 & = &  (\sigma_{\widehat{B}1}) \big[\widehat{B},\,C^p\big]+ (\sigma_{\widehat{B}2})^{p}_{r_1r_2r_3r_4}  \big[\big[B^{r_1r_2},\,C^{r_3}\big],\,C^{r_4}\big]\big]+(\sigma_{\widehat{B}3})^{p}_{r_1r_2r_3r_4}  \big[B^{r_1r_2},\,B^{r_3r_4}\big] \nonumber \\
  &&  +(\sigma_{\widehat{B}4})^{p}_{r_1r_2r_3r_4}  \big[C^{r_1},\,\big[C^{r_2},\,\big[C^{r_3},\,C^{r_4}\big]\big]\big]  .\label{3brsthB}
\end{eqnarray}
Here   unknown real numbers $\sigma_{\widehat{B}1}$,
$ (\sigma_{\widehat{B}2})^{p}_{r_1r_2r_3r_4}$ = $-(\sigma_{\widehat{B}2})^{p}_{r_2r_1r_3r_4} $,
$(\sigma_{\widehat{B}3})^{p}_{r_1r_2r_3r_4}$ = $-(\sigma_{\widehat{B}3})^{p}_{r_2r_1r_3r_4}=
-(\sigma_{\widehat{B}3})^{p}_{r_1r_2r_4r_3}$= $- (\sigma_{\widehat{B}3})^{p}_{r_3r_4r_1r_2} $,
$(\sigma_{\widehat{B}4})^{p}_{r_1r_2r_3r_4}$ = $(\sigma_{\widehat{B}4})^{p}_{r_1r_2r_4r_3}$,
should be determined from the $3\times3\times 3$ equations:
\begin{eqnarray}\label{consistBa}
&&  B^{p_1p_2}\big(\overleftarrow{s}{}^{p_3}_{[3]}\overleftarrow{s}{}^{p_4}_{[3]}+\overleftarrow{s}{}^{p_4}_{[3]}\overleftarrow{s}{}^{p_3}_{[3]}\big) =0, \ \mathrm{where} \ \widehat{B}\overleftarrow{s}{}^{p_3}_{l}\equiv 0,\,  l=0,1,2 \end{eqnarray}
Its general  solution looks as
\begin{equation}\label{solconstB}
  (\sigma_{\widehat{B}1}) = \frac{1}{2}, \  (\sigma_{\widehat{B}2})^{p}_{r_1r_2r_3r_4} = -\frac{1}{8}\varepsilon_{r_1r_2r_3}\delta^p_{r_4} -  \frac{1}{12}\varepsilon_{[r_1r_3r_4}\delta^p_{r_2]}, \ (\sigma_{\widehat{B}3})^{p}_{r_1r_2r_3r_4} = (\sigma_{\widehat{B}4})^{p}_{r_1r_2r_3r_4} =0,
\end{equation}
providing the validity of the last  row in the table (\ref{minN3}). In deriving (\ref{solconstB})
the use has been made of the symmetry for the commutator $[C^p,C^r]=[C^r,C^p]$,
and the following relations
\begin{eqnarray}
&& \big[\big[B^{p[p_1},\, C^{p_2]}\big],\,C^{p_3}\big] +  \big[\big[B^{p_1p_2},\, C^{p}\big],\,C^{p_3}\big]   =    \varepsilon^{pp_1p_2} P^{p_3}  ,     \label{ident1} \\
 && \big[\big[B^{[pp_3},\, C^{p_1}\big],\,C^{p_2]}\big] - \big[\big[B^{[pp_3},\, C^{p_2]}\big],\,C^{p_1}\big] + \big[\big[C^{[p},\,B^{p_1p_3}\big],\,C^{p_2]}\big]  =   \varepsilon^{pp_1p_2} Q^{p_3},   \label{ident2}\\
 && \mathrm{for }\  \ P^{p_3}  = \frac{1}{2} \big[\big[B^{pp_1},\, C^{p_2}\big],\,C^{p_3}\big] \varepsilon_{pp_1p_2}, \  \mathrm{and} \ Q^{p_3}  \hspace{-0.1cm} =  \hspace{-0.1cm} \big[\big[B^{pp_3},\, C^{p_1}\big],\,C^{p_2}\big] \varepsilon_{pp_1p_2} ,\label{aident3}
\end{eqnarray}
as well as the Jacobi identities, which establish the absence of the $4$-th power in the fields
$C^p$ in the transformation for $\widehat{B}$ (\ref{3brsthB}):
\begin{eqnarray}
& \hspace{-1em}& \hspace{-1em} B^{p_1p_2}\big(\overleftarrow{s}{}^{p_3}_{[3]}\overleftarrow{s}{}^{p_4}_{[3]}+\overleftarrow{s}{}^{p_4}_{[3]}\overleftarrow{s}{}^{p_3}_{[3]}\big)\big|_{(\widehat{B}=B^{pq}=0)} =  - \frac{1}{12} \Big[C^{[p_1}, \big[C^{p_4}, \big[C^{p_2]}, C^{p_3}\big] \big]+ \big[C^{p_3}, \big[ C^{p_4}, C^{p_2]}\big] \big]\nonumber \\
 & \hspace{-1em}&\hspace{-1em} \ +\big[C^{p_2]}, [C^{p_3}, C^{p_4}] \big]\Big] +  \varepsilon^{p_1p_2\{p_3}\widehat{B}  \overleftarrow{s}{}^{p_4\}}\big|_{(\widehat{B}=B^{pq}=0)}\, =\, \varepsilon^{p_1p_2\{p_3}\widehat{B}  \overleftarrow{s}{}^{p_4\}}\big|_{(\widehat{B}=B^{pq}=0)}=0 , \label{Jacobiid}
\end{eqnarray}
 meaning that we may put $(\sigma_{\widehat{B}4}) =0$. One can easily see that the $3\times 3$ equations (\ref{consistB}) considered
for $\widehat{B}$ are fulfilled as well:
\begin{equation}\label{consisttot}
  \widehat{B}\big(\overleftarrow{s}{}^{p_1}_{[3]}\overleftarrow{s}{}^{p_2}_{[3]}+\overleftarrow{s}{}^{p_2}_{[3]}\overleftarrow{s}{}^{p_1}_{[3]}\big) =0 \ \Leftrightarrow \  \overleftarrow{s}{}^{\{p_1}_{[3]}\overleftarrow{s}{}^{p_2\}}_{[3]} =0.
\end{equation}
Therefore, $\overleftarrow{s}{}^{p} = \overleftarrow{s}{}^{p}_{[3]}$ are the generators of the irreducible representation of  $\mathcal{G}(3)$ superalgebra
of $N=3$-parametric transformations
in the field superspace, $\mathcal{M}^{(3)}_{min}$, parameterized by the fields, $\Phi^{A_3}_{(3)}$.  That fact completes the proof of the Lemma~1.

 Thus, in order to have  the superspace of irreducible representation being closed with respect to the action of  abelian Lie superalgebra $\mathcal{G}(3)$   with Grassmann odd scalar generators $\overleftarrow{s}{}^p$ this superspace should  parameterized by the set of fields:
 \begin{equation}\label{N3space}
\{\Phi^{A_3}_{(3)}\}  = \big\{\mathcal{A}_\mu, C^{p}, {B}{}^{p_1p_2}, \widehat{B}\big\} = \big\{\mathcal{A}^n_\mu, C^{p{}n}, {B}{}^{p_1p_2{}{}n}, \widehat{B}{}^n\big\}t^n
  \end{equation}
used as local coordinates in the  configuration space $\mathcal{M}^{(3)}_{min}$ with dimension:
$\dim \mathcal{M}^{(3)}_{min}= (\hat{N}{}^2-1)\big(d+3,3+1 \big) $,
for an irreducible gauge theory of the fields $\mathcal{A}^\mu$ with a non-Abelian gauge group $SU(\hat{N})$.
It is obvious that $\mathcal{M}^{(3)}_{min} \supset \mathcal{M}^{(i)}_{tot}$ for  $i=1,2$.
We will call $\mathcal{M}^{(3)}_{min}$ as the \emph{minimal configuration space}.

Now, due to insufficiency  of the  $\mathcal{M}^{(3)}_{min}$ to provide gauge-fixing procedure without using of  additional  odd supermatrix or Grassmann-odd parameter  let us extend the $\mathcal{M}^{(3)}_{min}$ by  the fields $\overline{\Phi}{}_{(3)}$  of so-called \emph{non-minimal sector}, starting from a new  antighost field, $\overline{C}(x)=\overline{C}{}^m(x)t^m$, to provide a determination of the gauge fermion $F_{(3)}\equiv \Psi_{(3)}$ as the quadratic functional for the Landau gauge, $\chi (\mathcal{A})=0$:
\begin{equation}\label{gfn3}
\Psi^L_{(3)} (\overline{C}, \mathcal{A}) = \int d^dx\, tr\, \overline{C}\chi (\mathcal{A}).
\end{equation}
Properly the fields $\overline{\Phi}{}_{(3)}$ contain the Nakanishi-Lautrup fields,  $B$, and  have the contents
 \begin{equation}\label{fieldsnm3}
   \overline{\Phi}{}_{(3)} = \big( \overline{C}, B^{p}, \widehat{B}{}^{p_1p_2}, B \big), \ \  \epsilon\big(\overline{C},\widehat{B}{}^{p_1p_2}\big)+(1,1)=\epsilon\big(B^{p_1},B\big)=(0,0)
 \end{equation}
with even and odd degrees of freedom, $(3+1, 1+3)$ (modulo general factor with $\dim SU(\hat{N})$) and  determine  the  action of generators $\overleftarrow{s}{}^p_{(n)}$ of the representation of  the Abelian superalgebra $\mathcal{G}(3)$ in the superspace, $\mathcal{M}^{(3)}_{\mathrm{nm}}$,  with the local coordinates $\overline{\Phi}{}_{(3)}$.

\noindent
\textbf{Lemma 2}: The action of the generators  $\overleftarrow{s}{}^p_{(n)}$ of the Abelian superalgebra $\mathcal{G}(3)$ on the fields $\overline{\Phi}_{(3)}$ is determined by the relations:
 \begin{equation}\label{nminN3}
\overline{C}\overleftarrow{s}{}^p_{(n)} =  B^p, \ \  B^{p_1}\overleftarrow{s}{}^p_{(n)} =
  \widehat{B}{}^{p_1p},\ \  \widehat{B}{}^{p_1p_2} \overleftarrow{s}{}^p_{(n)} = \varepsilon^{p_1p_2p}B,\ \ {B}\overleftarrow{s}{}^p_{(n)} =0. \end{equation}
The respective $N=3$ SUSY transformations with triplet of anticommuting parameters, $\lambda_p$, on the fields $\overline{\Phi}_{(3)}$  are given by the rule: $\delta_\lambda \overline{\Phi}_{(3)} = \overline{\Phi}_{(3)} \overleftarrow{s}{}^{p}_{(n)} \lambda_p $.

\vspace{1ex}
\noindent
Indeed, the relations (\ref{nminN3})  repeat  by its form linearized chain (\ref{chain1}) --(\ref{chain2}) without non-linear terms.
It easy to check, that the generators $\overleftarrow{s}{}^{p}_{(n)}$ satisfy to the defining relations:
\begin{equation}\label{defnm}
  \overleftarrow{s}{}^{p_1}_{(n)}\overleftarrow{s}{}^{p_2}_{(n)}+\overleftarrow{s}{}^{p_2}_{(n)}\overleftarrow{s}{}^{p_1}_{(n)}=0,\qquad \prod_{l=1}^4\overleftarrow{s}{}^{p_l}_{(n)}=0.
\end{equation}
In particular, we have the exact sequence
\begin{equation}\label{exseqnm}
  \big( \overline{C}, B^{p}, \widehat{B}{}^{p_1p_2}, B \big) \hspace{-0.15em}\stackrel{\overleftarrow{s}{}^{p_3}_{(n)}}{\to} \hspace{-0.15em}\big(  B^{p_3}, \widehat{B}{}^{pp_3}, \varepsilon^{p_1p_2p_3}B,0 \big)\hspace{-0.15em}\stackrel{\overleftarrow{s}{}^{p_4}_{(n)}}{\to} \hspace{-0.15em}\big(   \widehat{B}{}^{p_3p_4}, \varepsilon^{pp_3p_4}B,0,0 \big)\hspace{-0.15em}\stackrel{\overleftarrow{s}{}^{p_5}_{(n)}}{\to} \hspace{-0.15em}\big(  \varepsilon^{p_3p_4p_5}B, 0,0,0 \big)\hspace{-0.15em}\stackrel{\overleftarrow{s}{}^{p_6}_{(n)}}{\to} \hspace{-0.15em}0
\end{equation}
of the length, equal to $4$.

We will call the representation (\ref{nminN3}) as the $N=3$ \emph{trivial representation}  of the  superalgebra $\mathcal{G}(3)$.

Finally. we construct the reducible representation of the superalgebra $\mathcal{G}(3)$ in the  total configuration space, $\mathcal{M}^{(3)}_{tot}$, parameterized by the fields, $\big({\Phi}_{(3)}, \overline{\Phi}_{(3)}\big) = \big(\widetilde{\Phi}_{(3)}\big)$,  with dimension in each space-time point $x\in \mathbf{R}^{1,d-1}$,
\begin{equation}\label{dimtot}
  \dim \mathcal{M}^{(3)}_{tot} =(\hat{N}{}^2-1) \Big(d +2^3-1,  2^3\Big).
\end{equation}
The generators of this representation we will denote as, $\overleftarrow{s}{}^{p_3}_{tot} = \overleftarrow{s}{}^{p_3} +\overleftarrow{s}{}^{p_3}_{(n)}$, (and then we will omit index $"tot"$ in it as it done  for the generally-adopted  notations in $N=1,2$ BRST symmetry cases).  The action of  $\overleftarrow{s}{}^{p_3}_{tot} $ is completely determined by (\ref{minN3}) and (\ref{nminN3}).

Now, let us turn to the gauge-fixing procedure, construction of the quantum action and path integral, whose integrand will be invariant with respect to derived $N=3$ SUSY transformations.

 \subsection{$N=3$ BRST-invariant path integral and quantum action}\label{N3FPtrickloc}

Let us determine the local   path integral, ${Z}_{3}$, and generating functionals of Green functions in any admissible gauge, turning to the non-degenerate Faddeev-Popov matrix, for Yang-Mills theory underlying above constructed explicit  $N=3$ SUSY invariance (\ref{minN3}), (\ref{nminN3}) in the total configuration space $\mathcal{M}^{(3)}_{tot}$, with triplet of anticommuting parameters $\lambda_p$  and the local  quantum action $S_{\Psi_{(3)}}(\Phi_{(3)}, \overline{\Phi}_{(3)})$ given by the prescription (\ref{qe})  as follows:
\begin{eqnarray}
\label{Pintlocn3}
  &\hspace{-1em}& \hspace{-0.5em}{Z}_{3|\Psi}(0) =  \hspace{-0.15em}\int \hspace{-0.15em}  d \Phi_{(3)} d \overline{\Phi}_{(3)}  \hspace{-0.15em}\exp \hspace{-0.15em}\Big\{\frac{\imath}{\hbar}S_{\Psi_{(3)}}\big(\Phi_{(3)}, \overline{\Phi}_{(3)}\big)\Big\} \ \mathrm{with}  \ S_{\Psi_{(3)}}\hspace{-0.15em}  = S_0(\mathcal{A}) \hspace{-0.15em}+\frac{1}{3!}%
\Psi_{(3)}\overleftarrow{s}{}^p\overleftarrow{s}{}^q\overleftarrow{s}{}^r \varepsilon_{pqr},\\
\label{GFlocn3}
  &\hspace{-1em}& \hspace{-0.5em} {Z}_{3|\Psi}(\widetilde{J}) =  \int  d \Phi_{(3)} d \overline{\Phi}_{(3)}  \exp \Big\{\frac{\imath}{\hbar}S_{\Psi_{(3)}}\big(\Phi_{(3)}, \overline{\Phi}_{(3)}\big) + J\Phi_{(3)}+\overline{J} \overline{\Phi}_{(3)}\Big\} \ =\ \exp \Big\{\frac{\imath}{\hbar}W_{3|\Psi}(\widetilde{J})\Big\},   \end{eqnarray}
with gauge fermion functional, $\Psi_{(3)}=\Psi_{(3)}\big(\widetilde{\Phi}_{(3)}\big)$, depending on the fields $\widetilde{\Phi}_{(3)}$ as follows (confer with (\ref{N1gaugef})):
\begin{equation}\label{N3gaugef}
  \Psi_{(3)}(\widetilde{\Phi}_{(3)}) = \overline{C}\chi_{(3)}(\mathcal{A},B)  +  \widehat{\Psi}_{(3)}(\widetilde{\Phi}_{(3)}), \   \ \mathrm{for} \ deg_{\widetilde{\Phi}}\widehat{\Psi}_{(3)}>2, \ \ deg_{\Phi}{} \chi_{(3)}(\mathcal{A},B) = 1,
\end{equation}
and external sources $ \widetilde{J}_{A^t_3} = \big(J_{A_3}$, $\overline{J}_{A^n_3}\big)$  to the respective Green functions related to the fields $\Phi^{A_3}_{(3)}, \overline{\Phi}{}^{A^n_3}_{(3)}$ with the same Grassmann parities: $\epsilon(J_{A_3}) = \epsilon(\Phi^{A_3}_{(3)})$,  $\epsilon(\overline{J}_{A^n_3}) = \epsilon(\overline{\Phi}{}^{A^n_3}_{(3)})$.

 It is easy  to check that both the' functional measure, $d \Phi_{(3)} d \overline{\Phi}_{(3)}= d\widetilde{\Phi}_{(3)}$, as well as the quantum action,
 $S_{\Psi_{(3)}}$, are invariant with respect to the change of variables, $\widetilde{\Phi}{}^{A^t}_{(3)}\to  \widetilde{\Phi}{}^{\prime  A^t}_{(3)}$ generated by  $N=3$ SUSY transformations (\ref{minN3}), (\ref{nminN3}), with accuracy   up to the first order in constant $\lambda_p$  (equally with infinitesimal  $\lambda_p$):
\begin{equation}\label{N3inv}
 \widetilde{\Phi}{}^{\prime  A^t}_{(3)}=\widetilde{\Phi}{}^{A^t}_{(3)}(1+\overleftarrow{s}{}^p\lambda_p): \ \delta_{\lambda}\widetilde{\Phi}{}^{  A^t}_{(3)}=\widetilde{\Phi}{}^{  A^t}_{(3)}\overleftarrow{s}{}^p\lambda_p \Longrightarrow    \delta_{\lambda}S_{\Psi_{(3)}} =o(\lambda), \ \mathrm{sdet}\left\| \delta\widetilde{\Phi}{}^{\prime}_{(3)}/  \delta\widetilde{\Phi}_{(3)}  \right\| =1+o(\lambda),
\end{equation}
 We will call, therefore, the transformations:
 \begin{equation}\label{n3brst}
   \delta_{\lambda}\widetilde{\Phi}{}^{  A^t}_{(3)}\ = \ \widetilde{\Phi}{}^{\prime  A^t}_{(3)}-\widetilde{\Phi}{}^{  A^t}_{(3)}\ = \ \widetilde{\Phi}{}^{  A^t}_{(3)}\overleftarrow{s}{}^p\lambda_p,\
 \end{equation}
 with the explicit action of the generators $\overleftarrow{s}{}^p$ (\ref{minN3}), (\ref{nminN3}) on  the component fields as $N=3$-\emph{parametric}  \emph{BRST transformations}.

 The particular representations for the path integrals (\ref{Pintlocn3}), (\ref{GFlocn3}) in  the Landau and Feynman gauges are easily obtained within the same $R_\xi$-family of the gauges as for the $N=1$ BRST invariant case (\ref{PintfpLlocn}) due to obvious coinciding choice of the gauge functions, $\chi_{(3)}(\mathcal{A},B)$, for  $\widehat{\Psi}_{(3)}=0$, in (\ref{N3gaugef}) with  one, $\chi(\mathcal{A},B)=(\partial^\mu A_{\mu}+\xi g^2 B=0) $, in (\ref{N1gaugef}). The quantum action, $S_{\Psi_{(3)\xi}}$, has the representation:
\begin{align}\label{qexi3}
  S_{\Psi_{(3)\xi}}\big(\widetilde{\Phi}_{(3)}\big) & = S_0+\frac{1}{3!}%
\Psi_{(3)\xi}\overleftarrow{s}{}^p\overleftarrow{s}{}^q\overleftarrow{s}{}^r \varepsilon_{pqr}= S_0+S_{\mathrm{gf}(3)} +  S_{\mathrm{gh}(3)}+  S_{\mathrm{add}(3)} ,\\
 S_{\mathrm{gf}(3)} &  =  \int d^{d}x\ tr\, \Big[    \partial^{\mu}A_{\mu
}  + {\xi g^2} B\Big]  B,
\label{gfxi3}\\
   S_{\mathrm{gh}(3)}&  = \int d^{d}x\ tr\, \Big[ %
 \overline{C}M(\mathcal{A}) \widehat{B}   +\frac{1}{2} \big\{B^pM(\mathcal{A}) {B}{}^{qr}  +\widehat{B}{}^{pq}M(\mathcal{A}) {C}{}^{r}\big\}\varepsilon_{pqr}\Big], \label{ghxi3} \\
 S_{\mathrm{add}(3)} &= \frac{1 }{ 6}\int d^{d}x \ tr\,  \Big(-3(\partial_\mu B^p) \big[D^\mu(\mathcal{A}) C^q,\, C^r\big] -  (\partial_\mu \overline{C}) \Big\{ 2 \big[D^\mu(\mathcal{A}) C^r,\, B^{pq}\big]      \nonumber \\
 \phantom{ S_{add(3)}} & \ \  +  \big[D^\mu(\mathcal{A}) B^{pq},\,C^r \big] +\big[\big[D^\mu(\mathcal{A}) C^p,\,  C^q\big] ,\,C^r\big] \Big\} \Big)   \varepsilon_{pqr} , \label{addxi3}
\end{align}
where we have used the identities,
\begin{eqnarray}
&& \label{auxsaK} M(\mathcal{A}) \overleftarrow{s}{}^{p} \ =\  \big[M(\mathcal{A}),\, C^p\big]  \Longleftrightarrow M^{mn}(\mathcal{A};x,y) \overleftarrow{s}{}^{p} =  f^{mrn}M^{rs}(\mathcal{A};x,y)C^{sp}(y),\\
&& \label{auxsaK1} \big(M(\mathcal{A})C^{q}\big)\overleftarrow{s}{}^{p} \ =\  -  \partial_\mu \big[D^\mu(\mathcal{A}) C^p,\,C^q\big] + M(\mathcal{A})\big(C^q\overleftarrow{s}{}^{p}\big) \\
 && \phantom{\big(M(\mathcal{A})C^{q}\big)\overleftarrow{s}{}^{p}}\ = \ -\big[M(\mathcal{A}) C^p,\, C^q\big] -\big[D^\mu(\mathcal{A}) C^p,\, \partial_\mu C^q\big] + M(\mathcal{A})\Big\{B^{qp}+\frac{1}{2}\big[C^{q},\,C^{p} \big]\Big\}  , \nonumber\\
&&  \left(  AB\right)\overleftarrow{s}{}^{p}  =\left(  A\overleftarrow{s}{}^{p}\right)  B\left(  -1\right)^{\epsilon(B)}+A\left(  B\overleftarrow{s}{}^{p}\right),\label{sAB}\\
& &  \left(  AB\right) \overleftarrow{s}{}^{p}\overleftarrow{s}{}^{q}\varepsilon_{pqr} =\left[ A\overleftarrow{s}{}^{p}\overleftarrow{s}{}^{q}  B + 2  A\overleftarrow{s}{}^{p}
\left(  B\overleftarrow{s}{}^{q}\right)  \left(  -1\right) ^{\epsilon(B)}+A\left(
B\overleftarrow{s}{}^{p}\overleftarrow{s}{}^{q}\right) \right]\varepsilon_{pqr} , \label{s2(AB)} \\%
&& \left(  AB\right) \overleftarrow{s}{}^{p}\overleftarrow{s}{}^{q}\overleftarrow{s}{}^{r}\varepsilon_{pqr} =\Big[ A \overleftarrow{s}{}^{p}\overleftarrow{s}{}^{q}\overleftarrow{s}{}^{r} B(-1)^{\epsilon(B)}+3A\overleftarrow{s}{}^{p}
\left(  B\overleftarrow{s}{}^{q}\overleftarrow{s}{}^{r}\right)  \left(  -1\right) ^{\epsilon(B)} \label{s3(AB)}\\
&& \phantom{\left(  AB\right) \overleftarrow{s}{}^{p}\overleftarrow{s}{}^{q}\overleftarrow{s}{}^{r}\varepsilon_{pqr}} +3A\overleftarrow{s}{}^{p}\overleftarrow{s}{}^{q}
\left(  B\overleftarrow{s}{}^{r}\right) +A\left(
B\overleftarrow{s}{}^{p}\overleftarrow{s}{}^{q}\overleftarrow{s}{}^{r}\right)\Big]\varepsilon_{pqr} ,  \nonumber%
\end{eqnarray}
where the latter relations (\ref{sAB})--(\ref{s3(AB)}) appear by readily established Leibnitz-like properties of the generators  of
$N=3$ BRST transformations, $\overleftarrow{s}{}^{p}$ acting on the product of
any functions $A$, $B$ with definite Grassmann parities depending on the fields $\widetilde{\Phi}{}^{A^t}_{(3)}$. Indeed, e.g. the validity of (\ref{sAB}) follows from the calculation of variations:
\begin{eqnarray}\label{varAs}
  &\hspace{-1em}&\hspace{-1em} \delta_{\lambda}A \hspace{-0.15em}= A \partial_{\widetilde{\Phi}{}^{A^t}_{(3)}} \left(\widetilde{\Phi}{}^{A^t}_{(3)} \overleftarrow{s}{}^{p}\right) \lambda_p \  \Longrightarrow \  A \partial_{\widetilde{\Phi}{}^{A^t}_{(3)}} \left(\widetilde{\Phi}{}^{A^t}_{(3)} \overleftarrow{s}{}^{p}\right) \equiv A \overleftarrow{s}{}^{p}, \\
  &\hspace{-1em}&\hspace{-1em} \delta_{\lambda}(AB)\hspace{-0.15em}=(\delta_{\lambda}A)B\hspace{-0.15em}+A(\delta_{\lambda}B) \hspace{-0.15em}= (A \overleftarrow{s}{}^{p}\lambda_p )  B\hspace{-0.15em} + A ( B\overleftarrow{s}{}^{p}\lambda_p ) \hspace{-0.15em}= \left\{(A \overleftarrow{s}{}^{p})  B(  -1)^{\epsilon(B)}\hspace{-0.15em}+ A ( B\overleftarrow{s}{}^{p} )\right\}\lambda_p,\label{varABs}
\end{eqnarray}
and the same for the second:  $\delta_{\lambda_1}\delta_{\lambda_2}(AB)$, and third: $\delta_{\lambda_1}\delta_{\lambda_2}\delta_{\lambda_3}(AB)$ variations.

For instance, the ghost-dependent  functional, $S_{\mathrm{add}(3)}$, with cubic and quartic in fictitious fields  terms is derived from the expression:
\begin{eqnarray}
   &&  S_{\mathrm{add}(3)} =  \frac{1 }{ 6}\int d^{d}x\ tr\,   \Big(3B^p \Big\{\big[M(\mathcal{A}) C^q,\, C^r\big]  + \big[D^\mu(\mathcal{A}) C^q,\,  \partial_\mu C^r\big] \Big\}   \label{eaddxi3} \\
 &&  \phantom{ S_{add(3)}} +  \overline{C} \Big\{2\big[M(\mathcal{A}) C^p,\, B^{qr}\big]  + 2\big[D^\mu(\mathcal{A}) C^p,\,  \partial_\mu B^{qr}\big]
 - \big[M(\mathcal{A}) B^{pr},\, C^q \big]  -\big[D^\mu(\mathcal{A}) B^{pr},\,  \partial_\mu C^q\big]
\nonumber\\
&&\phantom{ S_{add(3)}}  +     \big[\big[M(\mathcal{A}) C^r,\,C^p\big] ,\,C^q\big]  +     \big[\big[D^\mu(\mathcal{A}) C^r,\,\partial_\mu C^p\big] ,\,C^q\big]+  \big[\big[D^\mu(\mathcal{A}) C^r,\, C^p\big] ,\,\partial_\mu C^q\big]\Big\}   \Big)\varepsilon_{pqr},\nonumber
\end{eqnarray}
where we have omitted vanishing  terms, $\big[C^{p}, C^{q}\big]\varepsilon_{pqr} \equiv 0$,  and have used of the antisymmetry in $p,q,r$ as well as the integration by parts. The representation (\ref{addxi3}) immediately follows from (\ref{eaddxi3}).
Note, the each term in $S_{\mathrm{add}(3)}$, which determine the interaction vertexes from the sector of fictitious fields,  contains  the space-time differential operator for any gauge from $R_\xi$-gauges, that looks as more  nontrivial analog of  $S_{\mathrm{add}}$ (\ref{Sadd}) for $N=2$ BRST symmetry.

Let us study some consequences of the suggested $N=3$ BRST transformations.
 As in the $N=1,2$ BRST case, the $N=3$  invariance, for the corresponding generating functionals
of Green's functions, ${Z}_{3|\Psi}(\widetilde{J})$ , ${W}_{3|\Psi}(\widetilde{J})$ and effective action, $\Gamma_{3|\Psi}(\langle\widetilde{\Phi}_{(3)}\rangle)$:
\begin{equation}\label{GFGF3}
\Gamma_{3|\Psi}(\langle\widetilde{\Phi}_{(3)}\rangle) =  {W}_{3|\Psi}(\widetilde{J})- \widetilde{J}_{A^t}\langle \widetilde{\Phi}^{A^t}_{(3)}\rangle, \  \widetilde{J}_{A^t} = - (\delta  \Gamma_{3|\Psi} / \delta \langle \widetilde{\Phi}^{A^t}_{(3)}\rangle),\ \langle \widetilde{\Phi}^{A^t}_{(3)}\rangle = \overrightarrow{\partial}_{(\widetilde{J})}^{A^t} W_{3|\Psi}(\widetilde{J}),
 \end{equation}
with a given gauge condition $\Psi_{(3)}(\widetilde{\Phi}_{(3)})$,
leads to the presence of an $\mathcal{G}(3)$-triplet of Ward identities:
\begin{equation}\label{GFGFWIN3}
\widetilde{J}_{A^t}\langle \widetilde{\Phi}{}^{A^t}_{(3)}\overleftarrow{s}{}^p\rangle_{\Psi_{(3)},\widetilde{J}}=0, \ \ \widetilde{J}_{A^t}\langle\langle \widetilde{\Phi}^{A^t}_{(3)} \overleftarrow{s}{}^p\rangle\rangle_{\Psi_{(3)},\widetilde{J}}=0, \ \ \frac{\delta \Gamma_{3|\Psi}}{\delta \langle \widetilde{\Phi}^{A^t}_{(3)}\rangle}\langle\langle \widetilde{\Phi}{}^{A^t}_{(3)} \overleftarrow{s}{}^p\rangle\rangle_{\Psi_{(3)} \langle \widetilde{\Phi}\rangle} =0,
 \end{equation}
with respective normalized  average expectation values $\langle M \rangle_{\Psi_{(3)},\widetilde{J}}$,
$\langle\langle M\rangle\rangle_{\Psi_{(3)},\widetilde{J}}$, $\langle\langle M\rangle\rangle_{\Psi_{(3)}, \langle \widetilde{\Phi}\rangle}$, so that $\langle 1 \rangle_{\Psi_{(3)},\widetilde{J}}$ = $1$,
for a functional $M=M(\widetilde{\Phi}_{(3)})$ calculated using ${Z}_{3|\Psi}(\widetilde{J})$, ${W}_{3|\Psi}(\widetilde{J})$, $\Gamma_{3|\Psi}$
for a given gauge fermion $\Psi_{(3)}$ in the presence of external sources $,\widetilde{J}_{A^t}$ and mean fields
$\langle \widetilde{\Phi}{}^{A^t}_{(3)}\rangle$. The gauge independence of the path integral $Z_{3|\Psi}(0)\equiv Z_{3|\Psi_{(3)}}(0)$
under an infinitesimal variation of the gauge condition, $\Psi_{(3)} \to \Psi_{(3)}+\delta \Psi_{(3)}$:
\begin{equation}\label{N3GIinf}
  Z_{3|\Psi_{(3)}+\delta \Psi_{(3)}}(0) =     Z_{3|\Psi_{(3)}}(0).
\end{equation}
is established using the \emph{infinitesimal FD} $N=3$ BRST transformations
 with the functional parameters,
  \begin{equation}\label{N3FDP}
 \lambda_p(\widetilde{\Phi})= \frac{1}{3!}\big(\imath / \hbar\big) \delta \Psi_{(3)}(\widetilde{\Phi}) \overleftarrow{s}{}^q\overleftarrow{s}{}^r \varepsilon_{pqr},
.\end{equation}
 which we consider in details   in the Section~\ref{N=kgauge}.

The equivalence of $N=3$ and $N=1$
 BRST invariant path integrals ${Z}_{3|\Psi}(0)$  (\ref{Pintlocn3}), $
Z_\Psi $  (\ref{PintfpLlocn}),. e.g in $R_\xi$-like gauges  immediately follows from  the structure of the quantum action $S_{\Psi_{(3)\xi}}$,   (\ref{qexi3}). Indeed, integrating by the fields $\widehat{B}{}^{pq}$, second,  with respect to $C^{p}$, then trivially  with respect  to $B^p$  and $B^{pq}$ we get:
\begin{eqnarray}
  {Z}_{3|\Psi(\xi)}(0) \hspace{-0.25em}&=\hspace{-0.45em}& \hspace{-0.25em}\int \hspace{-0.15em}  d {\Phi}_{(3)}d \overline{C} d B^p  dB^{pq} {\det}^{3} M(\mathcal{A})  \delta(C^p) \exp \hspace{-0.15em}\Big\{\frac{\imath}{\hbar}\Big(S_{\Psi_{(3)\xi}}\big(\widetilde{\Phi}_{(3)} \big) -\frac{1}{2} \int \hspace{-0.15em} d^{d}x\ tr\, \widehat{B}{}^{pq}M(\mathcal{A}) {C}{}^{r}\varepsilon_{pqr}
  \Big)\Big\} \nonumber \\
  \hspace{-0.25em}&=\hspace{-0.45em}& \hspace{-0.25em}\int \hspace{-0.15em}  d {\Phi}d B^{pq}  {\det}^{3} M(\mathcal{A}) {\det}^{-3} M(\mathcal{A})  \delta(B^{pq})  \exp \hspace{-0.15em}\Big\{\frac{\imath}{\hbar}S_{\Psi{\xi}}\big({\Phi} \big ) \Big\} = {Z}_{\Psi},\label{N3N1eqv}
\end{eqnarray}
where, e.g. $ \delta(C^p)= \prod_x\prod_{k=1}^3\delta(C^k(x)) $ appears by the functional $\delta$-function and $S_{\Psi{\xi}}\big({\Phi} \big )$ is the $N=1$ BRST invariant  quantum action (\ref{PintfpLlocn}) given in the $R_\xi$- gauges.
The functional ${Z}_{\Psi}$ coincides with one given in (\ref{PintfpLlocn}) after identification for the field $\widehat{B}$ as $\widehat{B}=C$ which plays now the role of the ghost field.

The crucial point of the found  $N=3$ BRST symmetry transformations  in $\mathcal{M}^{(3)}_{tot}$ that  the whole  fields $\widetilde{\Phi}_{(3)}$ due to the relations (\ref{kN}), (\ref{kNu}) of the  Statement leading to: $k_u(3)= k(4)=k_u(4)$,  maybe organized in the respective multiplet  of $N=4$ field irreducible SUSY transformations with constant $4$ Grassmann-odd parameters, $\lambda_r$, $r=1,2,3,4$.   The  construction of the respective $N=4$ SUSY transformations will be  the main aim of the next Section.

\section{$N=4$ global SUSY  transformations}\label{N4}

 \setcounter{equation}{0}

Before introducing the  $N=4$ SUSY  transformations we consider  additional $N=1$-parametric SUSY transformations in $\mathcal{M}^{(3)}_{tot}$ with new Grassmann-odd nilpotent generator, $\overleftarrow{\bar{s}}$,  parameter, $\bar{\lambda}$: $\big(\bar{\lambda}{}^2, \overleftarrow{\bar{s}}{}^2\big)=0$, anticommuting with triplet of $\lambda^p$: $\bar{\lambda}\lambda^p +\lambda^p\bar{\lambda}=0$, where as for $N=1$ antiBRST transformations \cite{aBRST1}, \cite{aBRST2}
the role of the antighost field $\overline{C}$, as well as the rest multiplet (\ref{fieldsnm3})
   $\overline{\Phi}{}_{(3)}$ from the non-minimal sector  should be considered in opposite way ac compared to the multiplet $\Phi_{(3)}=\big(\mathcal{A}^\mu, C^{p_1}, B^{p_1p_2}, \widehat{B}\big)$ from the $\mathcal{G}(3)$ irreducible (minimal) representation.

\subsection{Additional $N=1$ BRST  transformations on the fields of $N=3$ representation}\label{N31}

It is valid  the following

 \noindent
\textbf{Lemma 3}: The action of the generator  $\overleftarrow{\bar{s}}$ of the Abelian superalgebra $\mathcal{G}(1)$ on the fields $\big({\Phi}_{(3)}, \overline{\Phi}_{(3)}\big)$ parameterizing $\mathcal{M}^{(3)}_{tot}$ is determined by the relations:
 \begin{equation}\label{aN1}
\begin{array}{|c|c |}\hline
  \downarrow \  \leftarrow \!&\!  \overleftarrow{\bar{s}}{}\!\! \!\! \\
   \hline
    \! \! \mathcal{A}^\mu &\! D^\mu(\mathcal{A}) \overline{C} \!\!   \\
\!\! \overline{C}    &\! 
\textstyle\frac{1}{2}\big[\overline{C}, \overline{C}\big] \!\!   \\
 \!\! B^{p_1}    &\!
\big[B^{p_1}, \overline{C}\big] \!\!   \\
    \!\! \widehat{B}{}^{p_1p_2}    &\!
     \big[\widehat{B}{}^{p_1p_2},\,\overline{C}\big]      \!\!   \\
 \!\! C^{p_1}    &\! B^{p_1}+%
\big[C^{p_1}, \overline{C}\big] \!\!   \\
 \!\! B^{p_1p_2}    &\!  \widehat{B}{}^{p_1p_2}+ \big[B^{p_1p_2}, \overline{C}\big] \!\!   \\
    \!\! \widehat{B}    &\! B  \!\!   \\
    \!\! B    &\!  0 \!\!   \\
   \cline{1-2}
\end{array}. \end{equation}
The respective $N=1$ SUSY transformations with  anticommuting parameter, $\lambda$, on the fields $\widetilde{\Phi}_{(3)}$  are given by the rule: $\delta_{\bar{\lambda}} \widetilde{\Phi}_{(3)} = \widetilde{\Phi}_{(3)} \overleftarrow{\bar{s}} \bar{\lambda} $.

\vspace{1ex}
Note,  the transformations (\ref{aN1}) reflects the fact that only the field $\overline{C}(x)$ appears by the active (as compared to $C^p$) connection.

\noindent
To prove the correctness of (\ref{aN1}) it is sufficient to   check,  the nilpotency of  $\overleftarrow{\bar{s}}$  on each field from the multiplet, because of the homogeneity in Grassmann grading is obvious.  The nilpotency calculated on the gauge field $\mathcal{A}^\mu$: $\mathcal{A}^\mu  \overleftarrow{\bar{s}}{}^2 =0$,  means that the set of local generators of the gauge transformations, $R^i_\alpha(\mathcal{A}) = R_{\mu
}^{mn}(x;y)$ (\ref{R(A)inf}), for $\epsilon_{\alpha}=0$,  forms  the local algebra Lie (as well as for the case of Lemma 1 (\ref{consistA}) but for $\overleftarrow{s}{}^p$):
\begin{equation}\label{clalg}
  R^i_\alpha(\mathcal{A})\overleftarrow{\partial}_jR^j_\beta(\mathcal{A})-(-1)^{\epsilon_{\alpha}\epsilon_{\beta}}R^i_\beta(\mathcal{A})\overleftarrow{\partial}_jR^j_\alpha(\mathcal{A})= - F^\gamma_{\alpha\beta} R^i_\gamma(\mathcal{A}), \ \mathrm{for} \ F^\gamma_{\alpha\beta} =f^{mnl}\delta(x-z)\delta(x-y) .
\end{equation}
The nilpotency on any other fields follows,  first, from the Leibnitz rule of acting of $\overleftarrow{\bar{s}}$ on the commutator of any functions $A, B$ with definite Grassmann parities:
\begin{equation}  \big[  A,\,B\big]\overleftarrow{\bar{s}}  =\left[ A\overleftarrow{\bar{s}},\,  B\right](  -1)^{\epsilon(B)}+\left[A,\,  \big(B\overleftarrow{\bar{s}}\big)\right],\label{sABcom}
\end{equation}
 second,  from the Jacobi identity:
\begin{equation}  \left[ \left[ A,\, \overline{C}\right],\,\overline{C}\right] (-1)^{\epsilon(A)} - \left[ \left[ \overline{C},\,A\right],\,\overline{C}\right]+ \left[ \left[ \overline{C},\, \overline{C}\right],\,A\right] (-1)^{\epsilon(A)}    =0,\label{JidAC}
\end{equation}
  for any $A \in \big\{\overline{C}, B^{p_1}, B^{p_1p_2}, C^{p_1}, \widehat{B}{}^{p_1p_2}, \widehat{B}, B\big\}$.  E.g. for Grassmann-even  $A = B^{p_1p_2}$ we have,
\begin{eqnarray}
&& B^{p_1p_2}\overleftarrow{\bar{s}}{}^2  = \big(\widehat{B}{}^{p_1p_2}+ \big[B^{p_1p_2}, \overline{C}\big]\big)\overleftarrow{\bar{s}}  \nonumber\\
&&  \phantom{ B^{p_1p_2}\overleftarrow{\bar{s}}{}^2}
= \big[\widehat{B}{}^{p_1p_2}, \overline{C}\big]  - \big[\widehat{B}{}^{p_1p_2}+ \big[B^{p_1p_2}, \overline{C}\big],\, \overline{C}\big] + \textstyle\frac{1}{2}\big[B^{p_1p_2}, \big[\overline{C},\,\overline{C}\big]\big] \nonumber \\
&& \phantom{ B^{p_1p_2}\overleftarrow{\bar{s}}{}^2} = - \frac{1}{2}\Big(\big[B^{p_1p_2}, \overline{C}\big],\, \overline{C}\big] - \big[\overline{C},\,B^{p_1p_2} \big],\, \overline{C}\big] + \big[\big[\overline{C},\,\overline{C}\big],\,B^{p_1p_2} \big] \Big) =0,
\label{Jidex}
\end{eqnarray}
 where we have used the relations (\ref{aN1}), linearity and Leibnitz rule (\ref{sABcom}) for $\overleftarrow{\bar{s}}$, Jacobi identity (\ref{JidAC}) and generalized antisymmetry for the (super)commutator.

The  transformations,
\begin{equation}\label{abrst3}
\widetilde{\Phi}_{(3)} \to \widetilde{\Phi}{}^{\prime}_{(3)} = \widetilde{\Phi}_{(3)}(1+\overleftarrow{\bar{s}} \bar{\lambda}) =\widetilde{\Phi}_{(3)}+\delta_{\bar{\lambda}} \widetilde{\Phi}_{(3)},
\end{equation}
  appear by the  invariance transformations of  following path integral and quantum action:
 \begin{eqnarray}
\label{Pintlocn1}
  &\hspace{-1em}& {Z}_{1|\Psi}(0) =  \int  d \Phi_{(3)} d \overline{\Phi}_{(3)}  \exp \Big\{\frac{\imath}{\hbar}S_{\Psi_{(1)}}\big(\Phi_{(3)}, \overline{\Phi}_{(3)}\big)\Big\}, \ \mathrm{with}  \ S_{\Psi_{(1)}}  = S_0(\mathcal{A}) +%
\Psi_{(1)}\overleftarrow{\bar{s}},
   \end{eqnarray}
with a new gauge fermion functional, $\Psi_{(1)}=\Psi_{(1)}\big(\widetilde{\Phi}_{(3)}\big)$, which should determine a non-degenerate  quantum action $S_{\Psi_{(1)}}$ on the $\mathcal{M}_{tot}^{(3)}$, i.e. with non-degenerate supermatrix of the second derivatives  in $\widetilde{\Phi}{}^{A_t}_{(3)}, \widetilde{\Phi}{}^{B_t}_{(3)}$ of $S_{\Psi_{(1)}}$ evaluated on a some vicinity of the solutions, $\widetilde{\Phi}{}^{A_t}_{0(3)} = (\mathcal{A}^\mu_0,0,...,0)$  of the respective  equations of motions: $S_0 \overleftarrow{\partial}_j=0$:
\begin{equation}\label{aN1gaugef}
  \Psi_{(1)} = \widehat{B}\chi_{(1)}(\mathcal{A},B)  +  C^pB^{qr}\varepsilon_{pqr}  + \widehat{\Psi}_{(1)}(\widetilde{\Phi}_{(3)}), \   \ \mathrm{for} \ deg_{\widetilde{\Phi}}\widehat{\Psi}_{(1)}>2, \ \ deg_{\Phi}{} \chi_{(1)}(\mathcal{A},B) = 1.
\end{equation}
Indeed,  from the invariance of the integration measure,  $d \widetilde{\Phi}_{(3)}$, and  quantum action, $S_{\Psi_{(1)}}$,   due to the same reason as for the standard $N=1$ BRST realization in $\mathcal{M}_{tot}$ (\ref{brstrans}):
 \begin{equation}\label{abrstrans}
   \delta_{\bar{\lambda}}S_{\Psi_{(1)}} =0, \quad  d \widetilde{\Phi}{}^{\prime}_{(3)}=d \widetilde{\Phi}_{(3)}\mathrm{sdet}\left \| \delta\widetilde{\Phi}{}^{\prime}_{(3)}/  \delta\widetilde{\Phi}_{(3)} \right \| =d \widetilde{\Phi}_{(3)},
\end{equation}
it follows  the invariance of the integrand in $Z_{1|\Psi}(0) $ with respect to these transformations.
It justifies a definition of  the transformations (\ref{abrst3}) as $N=1$ \emph{antiBRST symmetry transformations} in $\mathcal{M}^{(3)}_{tot}$.

Choosing, $\widehat{\Psi}_{(1)}(\widetilde{\Phi}_{(3)})=0$ in (\ref{aN1gaugef}) for the quadratic gauge functional,  $ \Psi_{(1)}$, (in particular, for $R_\xi$-gauges: $\chi_{(1)}(\mathcal{A},B) =\chi(\mathcal{A},B) $) we find for the quantum action,
 $S_{\Psi_{(1)}}$, the representation:
 \begin{eqnarray}
  &\hspace{-1em}& \hspace{-0.5em} S_{\Psi_{(1)}}  = S_0(\mathcal{A}) +  \hspace{-0.3em}  \int \hspace{-0.2em}d^{d}x\ tr\,\Big(\Big[    \partial^{\mu}A_{\mu
}  + {\xi g^2} B\Big]  B + \widehat{B} M(\mathcal{A}) \overline{C}   + \big\{B^p  {B}{}^{qr}  + C^p\widehat{B}{}^{qr}\big\}\varepsilon_{pqr} \Big)+S_{\mathrm{add}(1)}, \label{Sqen1}\\
  &\hspace{-1em}&  \hspace{-0.5em} S_{\mathrm{add}(1)}  = \int d^{d}x\ tr\,\Big(C^p  \big[B^{qr}, \overline{C}\big] + \big[C^p, \overline{C}\big] {B}{}^{qr}\Big)\varepsilon_{pqr} . \label{Sadd1}
   \end{eqnarray}
   Integrating out of $B^p$, $\widehat{B}{}^{qr}$ fields    we get   for the path integral:
   \begin{eqnarray}
\label{Pintlocn1fin}
  &\hspace{-1em}&\hspace{-0.5em} {Z}_{1|\Psi}(0) =  \int  d \mathcal{A} d\widehat{B}d \overline{C}dB dC^p dB^{qr}\delta({B}{}^{q_1r_1}) \delta({C}{}^{p_1}) \exp\Big\{\frac{\imath}{\hbar}\Big(S_{\Psi}\big(\Phi_{(1)}\big) +S_{\mathrm{add}(1)}\Big)\Big\}  \\
  &\hspace{-1.0em}& \hspace{-0.5em}   =\int \hspace{-0.15em} d \mathcal{A} d\widehat{B}d \overline{C}dB \hspace{-0.15em}\exp \Big\{\frac{\imath}{\hbar}S_{\Psi}\big(\Phi_{(1)}\big)\Big\}  \ \mathrm{with} \   S_{\Psi}  = S_0(\mathcal{A}) +%
\Psi \big(\Phi_{(1)}\big) \overleftarrow{\bar{s}}, \ \Psi \big(\Phi_{(1)}\big)= \widehat{B}\chi_{(1)}(\mathcal{A},B),  \label{eqZpsi0}
   \end{eqnarray}
where the resulting (after integration) fields $\Phi^A_{(1)}$, in fact, coincide with the fields given by the local formulation for the path integral (\ref{PintfpLlocn}) within Faddeev-Popov rules with $N=1$ BRST symmetry, in particular, for the Landau gauge (\ref{PintfpLloc}) under identification:
\begin{equation}\label{identN1n1}
  \Phi^A_{(1)} = \big(\mathcal{A}, \widehat{B} ,\overline{C}, B\big) \to \Phi^A = \big(\mathcal{A}, C ,\overline{C}, B\big).
\end{equation}
The only difference consists in the realization $N=1$ antiBRST symmetry for ${Z}_{1|\Psi}(0)$ given in $\mathcal{M}^{(3)}_{tot}$
and of $N=1$  BRST symmetry for ${Z}_{\Psi}$ (\ref{PintfpLlocn}) determine  over $\mathcal{M}_{tot}$.  After replacing $(\widehat{B} ,\overline{C})\to (\overline{C}, {C})$ the above path integral will coincide exactly

 Thus, we  reached  the validity of the

 \noindent
 {\textbf{Statement 2}}: The path integral, ${Z}_{1|\Psi}(0)$,  (\ref{Pintlocn1}) with the quantum action,  $S_{\Psi_{(1)}}$, (\ref{Sqen1}) at least, for the special quadratic gauge fermion, $ \Psi_{(1)}$, (\ref{aN1gaugef})  with $\widehat{\Psi}_{(1)}=0$ determined in $N=3$ reducible representation space, $\mathcal{M}^{(3)}_{tot}$,  of $\mathcal{G}(3)$ superalgebra,  but with realization of the additional $N=1$ antiBRST symmetry (\ref{aN1}), (\ref{abrstrans})  coincide with respective path integral (\ref{Pintlocn1fin}),  with the quantum action, $S_{\Psi}$, (\ref{eqZpsi0}) obtained with use of $N=1$ antiBRST symmetry transformations acting in the standard  configuration space, $\mathcal{M}_{tot}$.

 Now, we may reveal the physical contents of the fields spectrum for the $  {Z}_{3|\Psi}(0)$ (\ref{Pintlocn3}),  $
  S_{\Psi_{(3)\xi}}\big(\widetilde{\Phi}_{(3)}\big) $ (\ref{qexi3})--(\ref{addxi3}) being invariant with respect to $N=3$ BRST symmetry transformations (\ref{minN3}), (\ref{nminN3}).
  Namely, the fields $\widehat{B}$,   $\overline{C}$ from $\mathcal{M}_{tot}^{(3)}$ space  correspond respectively to  the pair of ghost field $C$ inheriting the gauge symmetry    and antighost field, $\overline{C}$, introducing the gauge condition in the gauge fermion for $N=1$ BRST symmetry realization of the standard Faddeev-Popov path integral. The triplet of the ghost fields $C^p$ and triplet of dual to $\widehat{B}{}^{p_1p_2}$ fields: $\widehat{B}_{p_3}=\frac{1}{2}\varepsilon_{p_1p_2p_3}\widehat{B}{}^{p_1p_2}$ = $\big(\widehat{B}{}^{23},\widehat{B}{}^{31},\widehat{B}{}^{12}\big)$ are organized into the pairs of $N=3$ triplet of Grassmann-odd ghost-antighost pairs: $\big(C^p,\,\widehat{B}_p\big)$. The triplet of the Grassmann-even fictitious fields
  $B^p$ and triplet of dual to ${B}{}^{p_1p_2}$ fields: $\overline{B}_{p_3}=\frac{1}{2}\varepsilon_{p_1p_2p_3}{B}{}^{p_1p_2}$ = $\big({B}{}^{23},{B}{}^{31},{B}{}^{12}\big)$ forms the pairs of $N=3$ triplet  of Grassmann-even ghost-antighost pairs: $\big({B}^p,  \overline{B}_p\big)$.
  The role of the Nakanishi-Lautrup field $B$  remains the same as in case of standard $N=1$ BRST symmetry formulation, i.e. as the Lagrangian multiplier (at least for Landau gauge) introducing the gauge into the quantum action.

  Because of, the term  in the ghost part, $S_{\mathrm{gh}(3)}$, (\ref{ghxi3}) with Grassmann-even triplet of ghost-antighost pairs  maybe presented as follows,
    \begin{equation}\label{Gegh}
  \frac{1}{2}\int d^dx\,  tr\, B^pM(\mathcal{A}) {B}{}^{qr}\varepsilon_{pqr} \equiv \int d^dx\,  tr\, B^pM(\mathcal{A}) \overline{B}_p
 \end{equation}
 we can immediately identify the fields, $(C^0, \overline{C}{}^0; C^{[3]}, \overline{C}{}^{[3]}; B^{[3]}, \overline{B}{}^{[3]})$ in the quantum action  (\ref{SintfpLlock}) for the local representation (\ref{PintfpLlock}) of the generalized  path integral (\ref{Pintgen}) ,  for $k_u(3)=3$,
 with singlet and Grassmann-odd and Grassmann-even triplets of ghost pairs as follows:
 \begin{equation}\label{identifgh}
   \big(C^0, \overline{C}{}^0; C^{[3]}, \overline{C}{}^{[3]}; B^{[3]}, \overline{B}{}^{[3]}\big) =  \big(\widehat{B}, \overline{C}; C^{p}, \widehat{B}_p;   \overline{B}_{p}, B^{p}\big).
 \end{equation}
 Note, first, that for  $N=1$ antiBRST symmetry realization in the  configuration space $\mathcal{M}^{(3)}_{tot}$  it is possible in addition to the path integral formulation (\ref{Pintlocn1}) introduce all necessary for diagrammatic Feynman technique generating functionals of Green functions  as it was done for $   N=1$ and $N=3$ BRST symmetry case in the Subsections~\ref{N12SUSY}, \ref{N3FPtrickloc} and study theirs respective properties (Ward identities, gauge-independence problem).
 Second, as for the above developed $N=1$ antiBRST symmetry concept in  $\mathcal{M}^{(3)}_{tot}$ it is possible to construct  a so-called $N=3$ \emph{antiBRST symmetry} transformations as the $N=3$ SUSY transformations of $\mathcal{G}(3)$ superalgebra  with the triplets of the anticommuting  generators $\overleftarrow{\bar{s}}{}_p$ with lower indices $p=1,2,3$ and Grassmann-odd parameters, $\bar{\lambda}{}^p$. Doing so we should, to change all the Grassmann-odd and Grassmann-even ghosts on its antighosts in the $N=3$ SUSY transformations described by Lemmas 1, 2, starting from the change for the gauge parameters $\xi$:
 $\xi=\widehat{B}_p\bar{\lambda}{}^p$ and the first relations in a chain of these transformation
 \begin{equation}\label{aN3brst}
   \mathcal{A}^{\mu}\overleftarrow{\bar{s}}{}_p = D^{\mu}(\mathcal{A})\widehat{B}_p, \ \widehat{B}_q\overleftarrow{\bar{s}}{}_p= \varepsilon_{pqr}B^r +\frac{1}{2}\big[\widehat{B}_q,\,\widehat{B}_p\big] , \ldots,
    \end{equation}
    and finishing with the construction of the respective path integral, whose action and functional measure should be invariant with respect to these transformations.
   We leave the details of this interesting concept  out of the paper scope.

\subsection{$N=4 =3+1$  SUSY  transformations}\label{N4SUSY}

Now, we are able to consider the triplet of the Grassmann-odd ghost fields $C^p$ and singlet $\overline{C}$, triplets of the Grassmann-even ghost fields $B^{pq}$ and $B^p$, triplet of new Grassmann-odd ghost fields $\widehat{B}{}^{pq}$ and singlet $\widehat{B}$ on the equal footing within corresponding Grassmann-odd quartet, $C^r$, Grassmann-even sextet, $B^{r_1r_2}$, and Grassmann-odd quartet, $B^{r_1r_2r_3}$ for $r,r_1,r_2,r_3=1,2,3,4$ as the elements (with the fields $ \mathcal{A}^\mu, B$)
of the irreducible tensor representation of the Abelian  $\mathcal{G}(4)$ superalgebra. In fact, the $N=3$ and $N=1$ representations  of $\mathcal{G}(3)$ and $\mathcal{G}(1)$ superalgebra in the same $\mathcal{G}(3)$-representation space of the fields $\widetilde{\Phi}_{(3)}$  are nontrivially entangled  in unique $N=4$ irreducible representation in the same representation space  $\mathcal{M}^{(3)}_{tot}=\mathcal{M}^{(4)}_{tot}$  whose local coordinates (fields) are organized into $\mathcal{G}(4)$-irreducible antisymmetric tensors, as well as the parameters  and generators have the structures:
\begin{eqnarray}
  &&   \Big(\stackrel{\mathcal{G}(4)}{\overbrace{C^r, B^{r_1r_2}, B^{r_1r_2r_3},B}}\Big) =  \Big(\big(\stackrel{\mathcal{G}(3)}{\overbrace{C^p,\overline{C}}}\big), \big(\stackrel{\mathcal{G}(3)}{\overbrace{{B}{}^{p_1p_2},{B}^{p_1}}}\big)  , \big(\stackrel{\mathcal{G}(3)}{\overbrace{\widehat{B},\widehat{B}{}^{p_1p_2}}}\big), B\Big) , \label{corg3g4}\\
   && \lambda_r=\big(\lambda_p, \bar{\lambda}\big);\ \ \overleftarrow{s}{}^r =\big(\overleftarrow{s}{}^p, \overleftarrow{\bar{s}}\big);\ \ \ r=(p,4)=(1,2,3,4). \label{corg3g42}
\end{eqnarray}

\noindent
\textbf{Lemma  4}: The action of the generators  $\overleftarrow{s}{}^{r}$ of $N=4$-parametric  Abelian superalgebra $\mathcal{G}(4)$ on the fields $\Phi_{(4)}=\big(\mathcal{A}^\mu,C^r, B^{r_1r_2}, B^{r_1r_2r_3},B\big)$ is given by the relations:
 \begin{eqnarray}
&\hspace{-0.5em}&\hspace{-0.5em}\begin{array}{|c|c |}\hline
  \downarrow \  \leftarrow \!&\!  \overleftarrow{s}{}^{r}\!\! \!\! \\
   \hline
    \! \! \mathcal{A}^\mu &\! D^\mu(\mathcal{A}) C^r \!\!   \\
 \!\! C^{r_1}    &\! B^{r_1r}+%
\textstyle\frac{1}{2}\big[C^{r_1}, C^{r}\big] \!\!   \\
 \!\! B^{r_1r_2}    &\! {B}^{r_1r_2r}+ \textstyle\frac{1}{2}\Big(
\big[B^{r_1r_2}, C^{r}\big]-\frac{1}{6}\big[C^{[r_1},\big[C^{r_2]},C^{r}\big]\big]\Big) \!\!   \\
    \!\! B^{r_1r_2r_3}    &\! \hspace{-0.2em}\varepsilon^{r_1r_2r_3r}{B}+\textstyle\frac{1}{2}\hspace{-0.1em}\big[B^{r_1r_2r_3},C^r\big] - \displaystyle\sum_{P}\hspace{-0.15em}(-1)^{P(r_1,r_2,r_3)}\hspace{-0.1em}\Big\{\hspace{-0.1em}\textstyle\frac{1}{8}\big[\big[B^{r_1r_2},C^{r_3}\big],C^r\big] +\frac{1}{6} \big[\big[B^{r_1r},C^{r_2}\big],C^{r_3}\big] \hspace{-0.1em} \Big\} \hspace{-0.1em} \!\!   \\
 \!\! B   &\! \textstyle\frac{1}{2}\big[{B},\,C^r\big] - \frac{1}{4!}\big[\big[B^{r_1r_2r_3},\,C^{r_4}\big],\,C^r\big] \varepsilon_{r_1r_2r_3r_4} \!\!   \\
     \cline{1-2}
\end{array} \nonumber \\
&\hspace{-0.5em}&\hspace{-0.5em} \ \mathrm{with} \ \sum_{P}(-1)^{P(r_1,r_2,r_3)} X^{r_1r_2r_3r} =X^{r_1r_2r_3r}-X^{r_2r_1r_3r}- X^{r_1r_3r_2r} + \ldots, \label{minN4}\end{eqnarray}
where the sign, $\sum_{P}(-1)^{P(r_1,r_2,r_3)} X^{r_1r_2r_3r}$ means the summation over all (odd with sign $"-"$ and even with $"+"$) $3!$  permutations  of the indices  $(r_1,r_2,r_3)$.
The respective $N=4$ SUSY transformations with quartet of anticommuting parameters, $\lambda_r$, on the fields $\Phi_{(4)}$  are determined  as: $\delta_\lambda \Phi_{(4)} = \Phi_{(4)} \overleftarrow{s}{}^{r} \lambda_r $.

\vspace{1em}
The form of the transformations (\ref{minN4}) follows from the chain (\ref{chain1}), (\ref{chain2}) for $N=4$. To prove the Lemma we will follow the algorithm elaborated when the Lemma 1 was proved.
We start from the boundary condition for the  transformations (\ref{minN4}) inherited from the gauge transformations for $\mathcal{A}^\mu$ (\ref{gtbt}) and present the  realization for  the sought-for generators as series:
\begin{equation}\label{sersr}
\overleftarrow{s}^r = \sum_{e\geq 0}\overleftarrow{s}^r_e: \qquad  \mathcal{A}^\mu\overleftarrow{s}{}^{r} =  \mathcal{A}^\mu\overleftarrow{s}{}^{r}_0 = D^\mu(\mathcal{A})C^{r} \ \mathrm{and }\  C^{r_1}\overleftarrow{s}{}^{r}_0\equiv 0.
\end{equation}
Then, because of,
\begin{equation}\label{s0ne04}
\mathcal{A}_\mu\Big(\overleftarrow{s}{}_0^{r_1}\overleftarrow{s}{}_0^{r_2}
+\overleftarrow{s}{}^{r_2}_0\overleftarrow{s}{}^{r_1}_0\Big)  \ne 0,
\end{equation}
 we should add to
$\overleftarrow{s}{}_0^r$  the nontrivial action of new part  $\overleftarrow{s}{}^r_1$ on $C^{r_1}$ (vanishing when acting on $\mathcal{A}_\mu$: $\mathcal{A}_\mu\overleftarrow{s}{}^r_1 \equiv 0$),
starting from the Grassmann-even  sextet  of the fields $B^{r_1r_2}=B^{r_1r_2{}m}t^m$ (BRST-like variation of $C^{r_1}$) (\ref{chain1})
\begin{eqnarray}\label{4brstC}
  C^{r_1} \overleftarrow{s}{}^{r_2}_1 \ = \   B^{r_1r_2} + (\eta_{C1})^{r_1r_2}_{s_1s_2}  \big[C^{s_1},\,C^{s_2}\big]  ,   \ \mathrm{for}  \ B^{r_1r_2} = - B^{r_2r_1} , \ \epsilon(B^{r_1r_2})=0
\end{eqnarray}
with unknown real numbers: $ (\eta_{C1})^{r_1r_2}_{s_1s_2}$= $(\eta_{C1})_{s_2s_1}^{r_1r_2}$,
to be determined from the consistency of $4\times 4$ equations:
\begin{equation}\label{consistA4}
\mathcal{A}_{\mu}\Big(\overleftarrow{s}{}^{r_1}_{[1]}\overleftarrow{s}{}^{r_2}_{[1]}+\overleftarrow{s}{}^{r_1}_{[1]}\overleftarrow{s}{}^{r_2}_{[1]}\Big) =0 ,  \  \ \mathrm{where} \ \overleftarrow{s}^r_{[l]}\equiv \sum_{n\geq0}^l\overleftarrow{s}^r_{n}, \ \mathrm{and} \  C^{r_1} \overleftarrow{s}{}^{r_2}_0 \equiv 0,
\end{equation}
from which  follows the  antisymmetry for $B^{r_1r_2}$ in the indices $r_1,r_2$.
The solution for (\ref{consistA4})  looks as:
\begin{equation}\label{solconstA4}
 (\eta_{C1})_{s_1s_2}^{r_1r_2} = \frac{1}{4} \delta^{r_1}_{\{s_1}\delta^{r_2}_{s_2\}} , \ \mathrm{for} \  \delta^{r_1}_{\{s_1}\delta^{r_2}_{s_2\}}  \equiv \delta^{r_1}_{s_1}\delta^{r_2}_{s_2}+\delta^{r_1}_{s_2}\delta^{r_2}_{s_1},
\end{equation}
that proves  the validity of the $2$-nd row in the table (\ref{minN4}).

Second, in view of
\begin{equation}\label{consistC4a}
C^{r}\big(\overleftarrow{s}^{r_1}_{[1]}
\overleftarrow{s}^{r_2}_{[1]}+\overleftarrow{s}^{r_2}_{[1]}
\overleftarrow{s}^{r_1}_{[1]}\big) \ne 0,
\end{equation}
we should determine, for a nontrivial action of $\overleftarrow{s}{}^{r}_2$
on $B^{ r_1r_2}$ (vanishing when acting on $\mathcal{A}_\mu, C^r$: $\big(\mathcal{A}_\mu$, $C^{r_1}\big)\overleftarrow{s}{}^r_2$ $\equiv$ $0$), in the form of a general anzatz, starting from the  Grassmann-odd
field variables ${B}^{ r_1r_2r_3}= {B}{}^{{ r_1r_2r_3}{}m}t^m$ (BRST-like variation of $B^{r_1r_2}$) (\ref{chain2})
up to the third power in $C^r$ with a preservation of Grassmann homogeneity
in each summand, as in the (\ref{4brstC}),
\begin{eqnarray}\label{4brstB}
\hspace{-0.7em}&\hspace{-0.7em}& \hspace{-0.7em}  B^{r_1r_2} \overleftarrow{s}{}^{r_3}_2 \, = \,   {B}^{r_1r_2r_3} \hspace{-0.15em}  +  (\eta_{B1})^{r_1r_2r_3}_{s_1s_2s_3}  \big[B^{s_1s_2},\,C^{s_3}\big]\hspace{-0.15em} + (\eta_{B2})^{r_1r_2r_3}_{s_1s_2s_3}   \big[C^{s_1},\,\big[C^{s_2},\,C^{s_3}\big]\big]  ,   \,  \epsilon( {B}^{r_1r_2r_3}) = 1.
\end{eqnarray}
with unknown real numbers: $(\eta_{Bj})^{r_1r_2r_3}_{s_1s_2s_3}$, $j=1,2$;  satisfying the same antisymmetry properties as for $(\kappa_{Bj})^{p_1p_2p_3}_{r_1r_2r_3}$  in (\ref{3brstB}) and  to be determined from the solution of
the $4\times4\times 4$ equations
\begin{eqnarray}\label{consistC4}
  && C^{r_1}\big(\overleftarrow{s}{}^{r_2}_{[2]}\overleftarrow{s}{}^{r_3}_{[2]}+\overleftarrow{s}{}^{r_3}_{[2]}\overleftarrow{s}{}^{r_2}_{[2]}\big) =0, \ \mathrm{where}  \ B^{r_1r_2}\overleftarrow{s}{}^{r_3}_{l}\equiv 0,\, l=0,1.
\end{eqnarray}
Its general  solution has the form:
\begin{eqnarray}\label{solconstC4}
 && (\eta_{B1})^{r_1r_2r_3}_{s_1s_2s_3} = \frac{1}{4} \delta^{[r_1}_{s_1}\delta^{r_2]}_{s_2} \delta^{r_3}_{{s_3}}:\  \ (\eta_{B1})^{r_1r_2r_3}_{s_1s_2s_3} \big[B^{s_1s_2},\,C^{s_3}\big] = \frac{1}{2} \big[B^{r_1r_2},\,C^{r_3}\big], \\
  &&   (\eta_{B2})^{r_1r_2r_3}_{s_1s_2s_3} = -\frac{1}{12} \delta^{[r_1}_{s_1}\delta^{r_2]}_{s_2} \delta^{r_3}_{s_3}:\ \    (\eta_{B2})^{r_1r_2r_3}_{s_1s_2s_3}  \big[C^{s_1},\,\big[C^{s_2},\,C^{s_3}\big]\big] = - \frac{1}{12} \big[C^{[r_1},\,\big[C^{r_2]},\,C^{r_3}\big]\big].
\end{eqnarray}
providing the validity of the $3$-rd row in the table (\ref{minN4}).

Third,  there are only the fourth-rank independent completely antisymmetric
constant tensors with upper, $\varepsilon^{r_1r_2r_3r_4}$,
and lower, $\varepsilon_{r_1r_2r_3r_4}$, indices, which are normalized
by the conditions (according with (\ref{lchevn}))
\begin{eqnarray}\label{eprts}
  && \varepsilon^{1234}= 1, \, \varepsilon^{r_1r_2r_3r_4}\varepsilon_{s_1s_2s_3r_4}=  \det \|\delta^{r_i}_{s_j}\|, \, i,j=1,2,3€; \\
  && \varepsilon^{r_1r_2r_3r_4}\varepsilon_{s_1s_2r_3r_4}=2\big(\delta^{r_1}_{s_1}\delta_{s_2}^{r_2} -\delta^{r_2}_{s_1}\delta_{s_2}^{r_1} \big) ;  \varepsilon^{r_1r_2r_3r_4}\varepsilon_{s_1r_2r_3r_4}= 6\delta^{r_1}_{s_1}. \nonumber
\end{eqnarray}
 Due to
\begin{equation}\label{consistB4a}
B^{r_1r_2}\big(\overleftarrow{s}{}^{r_3}_{[2]}\overleftarrow{s}{}^{r_4}_{[2]}
+\overleftarrow{s}{}^{r_4}_{[2]}\overleftarrow{s}{}^{r_3}_{[2]}\big) \ne 0,
\end{equation}
we should determine for a nontrivial action of $\overleftarrow{s}{}^{r}_3$ on ${B}^{r_1r_2r_3}$, (vanishing when acting on $\mathcal{A}_\mu, C^r, B^{r_1r_2}$: $\big(\mathcal{A}_\mu,C^r, B^{r_1r_2}\big)\overleftarrow{s}{}^{r_3}_{3} \equiv 0$)
 a general ansatz with use of the new Grassman-even field variable, $B$,
\begin{eqnarray}
B^{r_1r_2r_3} \overleftarrow{s}{}^r_3 & = & \varepsilon^{r_1r_2r_3r}B+ (\sigma_{{B}1})_{s_1s_2s_3s}^{r_1r_2r_3r}  \big[B^{s_1s_2s_3},\,C^s\big]+ (\sigma_{{B}2})_{s_1s_2s_3s}^{r_1r_2r_3r}  \big[\big[B^{s_1s_2},\,C^{s_3}\big],\,C^{s}\big] \nonumber \\
&+& (\sigma_{{B}3})^{r_1r_2r_3r}_{s_1s_2s_3s}  \big[B^{s_1s_2},\,B^{s_3s}\big] +(\sigma_{{B}4})^{r_1r_2r_3r}_{s_1s_2s_3s}  \big[C^{s_1},\,\big[C^{s_2},\,\big[C^{s_3},\,C^{s}\big]\big]\big]. \label{4brsthB}
\end{eqnarray}
Here ,  the unknown real numbers $(\sigma_{{B}i})^{r_1r_2r_3r}_{s_1s_2s_3s}$, $i=1,2,3,4$,  obey the analogous properties of (anti)sym\-met\-ry as for the coefficients $(\sigma_{\widehat{B}2})^{p}_{r_1r_2r_3r_4}$ (\ref{3brsthB}) in the respective lower and upper indices that is now dictated  by antisymmetry for $B^{r_1r_2r_3}$, $B^{s_1s_2}$ and symmetry for  $\big[C^{s_3},\,C^{s}\big] $ in $\mathcal{G}(4)$-indices.
They should be determined from the $6\times 4\times 4$ equations:
\begin{eqnarray}\label{consistB4}
&&  B^{r_1r_2}\big(\overleftarrow{s}{}^{r_3}_{[3]}\overleftarrow{s}{}^{r_4}_{[3]}+\overleftarrow{s}{}^{r_4}_{[3]}\overleftarrow{s}{}^{r_3}_{[3]}\big) =0, \ \mathrm{where} \ B^{r_1r_2r_3} \overleftarrow{s}{}^{r}_{l}\equiv 0,\,  l=0,1,2 \end{eqnarray}
Its general  solution looks as
\begin{eqnarray}\label{solconstB41}
\hspace{-0.9em}  &\hspace{-0.9em}&\hspace{-0.9em} (\sigma_{B1})^{r_1r_2r_3r}_{s_1s_2s_3s} \big[B^{s_1s_2s_3},\,C^s\big] = \frac{1}{2} \big[B^{r_1r_2r_3},\,C^r\big], \\
\hspace{-0.9em}  &\hspace{-0.9em}&\hspace{-0.9em}
 (\sigma_{{B}2})^{r_1r_2r_3r}_{s_1s_2s_3s} \hspace{-0.1em}\big[\big[B^{s_1s_2},C^{s_3}\big],C^{s}\big]\big] \hspace{-0.1em} = \hspace{-0.1em} - \hspace{-0.15em} \displaystyle\sum_{P}\hspace{-0.15em}(-1)^{P(r_1,r_2,r_3)}\hspace{-0.1em}\Big\{\hspace{-0.1em}\textstyle\frac{1}{8}\big[\big[B^{r_1r_2},C^{r_3}\big],C^r\big] \hspace{-0.15em}+\hspace{-0.1em}\frac{1}{6} \big[\big[B^{r_1r},C^{r_2}\big],C^{r_3}\big]\hspace{-0.1em} \hspace{-0.1em}\Big\}\hspace{-0.1em},\label{solconstB423}\\
\hspace{-0.9em}  &\hspace{-0.9em}&\hspace{-0.9em} (\sigma_{{B}3})^{r_1r_2r_3r}_{s_1s_2s_3s} = (\sigma_{{B}4})^{r_1r_2r_3r}_{s_1s_2s_3s} =0 \label{solconstB42},
\end{eqnarray}
providing the validity of the $4$-th  row in the table (\ref{minN4}). In deriving (\ref{solconstB41})--(\ref{solconstB42}),
  the use has been made of the symmetry for the commutator $[C^p,C^r]=[C^r,C^p]$,  Jacobi identities both for $(B^{pp_1},\, C^{p_2},C^{p_3})$ and for  $(C^{p_1},\, C^{p_2}, C^{p_3})$, which establish the absence of the $4$-th power in the fields
$C^p$ in the transformation for ${B}^{r_1r_2r_3}$ (\ref{4brsthB}) completely repeating the equations (\ref{Jacobiid}) for $N=3$ case, but with replacement: $\big(B^{p_1p_2}, \widehat{B}$,  $C^{p},\overleftarrow{s}{}^{p}_{[3]}\big)$ on  $\big(B^{r_1r_2}, {B}^{r_1r_2r_3}$, $C^{r},\overleftarrow{s}{}^{r}_{[3]}\big)$.

 Fourth, because of,
\begin{equation}\label{consistB34}
B^{r_1r_2r_3}\big(\overleftarrow{s}{}^{r_4}_{[3]}\overleftarrow{s}{}^{r_5}_{[3]}
+\overleftarrow{s}{}^{r_5}_{[3]}\overleftarrow{s}{}^{r_4}_{[3]}\big) \ne 0,
\end{equation}
we should determine for a nontrivial action of $\overleftarrow{s}{}^{r}_4$ on ${B}$, (vanishing when acting on $\mathcal{A}_\mu, C^r, B^{r_1r_2}, B^{r_1r_2r_3}$: $\big(\mathcal{A}_\mu,C^r, B^{r_1r_2}, B^{r_1r_2r_3}\big)\overleftarrow{s}{}^{r_4}_{4} \equiv 0$)
 a general ansatz  without new  Grassmann-odd field variable due to $5$-th order  nilpotency for  $\overleftarrow{s}{}^r$
 ($\prod_{l=1}^5\overleftarrow{s}{}^{r_l}\equiv 0$)
up to the fifth order in $C^r$ with a preservation of Grassmann homogeneity
in each summand, as in the case of (\ref{4brstC}), (\ref{4brstB}) and (\ref{4brsthB}),
\begin{eqnarray}
B \overleftarrow{s}{}^r_4 & = & (\varsigma_{{B}1})_{s}^{r}  \big[B,\,C^s\big]+ (\varsigma_{{B}2})_{s_1s_2s_3s_4s}^{r}  \big[\big[B^{s_1s_2s_3},\,C^{s_4}\big],\,C^{s}\big]\big] \nonumber \\
&+& (\varsigma_{{B}3})^{r}_{s_1s_2s_3s_4s}  \big[\big[B^{s_1s_2},\,B^{s_3s_4}\big],C^s\big] +(\varsigma_{{B}4})^{r}_{s_1s_2s_3s_4s}  \big[C^{s_1},\,\big[C^{s_2},\,\big[C^{s_3},\,\big[C^{s_4},C^s\big]\big]\big]\big].\label{4brstBlast}
\end{eqnarray}
 The above unknown real numbers, $(\varsigma_{{B}1})^{r}_{s}$, $(\varsigma_{{B}i})^{r}_{s_1s_2s_3s_4s}$, $i=2,3,4$,  obey the obvious  properties of (anti)symmetry, e,g,  as for the coefficients $(\varsigma_{{B}2})_{s_1s_2s_3s_4s}^{r}$ = $-(\varsigma_{{B}2})_{s_2s_1s_3s_4s}^{r}$=$(\varsigma_{{B}2})_{s_1s_3s_2s_4s}^{r}$.
They should be determined from the $4\times 4\times 4$ equations:
\begin{eqnarray}\label{consistB4l}
&&  B^{r_1r_2r_3}\big(\overleftarrow{s}{}^{r_4}_{[4]}\overleftarrow{s}{}^{r_5}_{[4]}+\overleftarrow{s}{}^{r_5}_{[4]}\overleftarrow{s}{}^{r_4}_{[4]}\big) =0, \ \mathrm{where} \ B \overleftarrow{s}{}^{r}_{l}\equiv 0,\,  l=0,1,2,3, \end{eqnarray}
whose general  solution has the form
\begin{eqnarray}\label{solconstB4l}
  && (\sigma_{B1})^{r}_{s}  = \frac{1}{2}  \delta^{r}_{s}, \quad  (\sigma_{{B}2})^{r}_{s_1s_2s_3s_4s}   =  -\frac{1}{4!}\delta^{r}_{s} \varepsilon_{s_1s_2s_3s_4}  ,\quad  (\sigma_{{B}3})^{r_1r_2r_3r}_{s_1s_2s_3s} = (\sigma_{{B}4})^{r_1r_2r_3r}_{s_1s_2s_3s} =0 \label{solconstB4l2},
\end{eqnarray}
providing the validity of the last  row in the table (\ref{minN4}).
In deriving (\ref{solconstB4l}),   we have used  the above mentioned properties found when establishing (\ref{solconstB41})--(\ref{solconstB42})  as well as the Jacobi identity for the fields  $\big({B}^{r_1r_2r_3}$, $C^{r_4},$  $C^{r_5})$ with the following representations for "$4$-cocycles", i.e. for $5$-th rank  tensors  being antisymmetric in $4$ indices:
\begin{eqnarray}
&\hspace{-1.2em}& \hspace{-1.2em}\frac{1}{3!}\sum_P(-1)^{P(rr_1r_2r_3)}\big[\big[B^{rr_1r_2},\, C^{r_3}\big],\,C^{r_4}\big]  \hspace{-0.1em}  =    \hspace{-0.1em} \varepsilon^{rr_1r_2r_3} P^{r_4}_4   \ \mathrm{for }\   P^{r_4}_4  = \frac{1}{3!} \big[\big[B^{rr_1r_2},C^{r_3}\big],C^{r_4}\big] \varepsilon_{rr_1r_2r_3}\hspace{-0.1em},  \label{ident14} \\
  &\hspace{-1.2em}&\hspace{-1.2em}   \frac{1}{2!}\sum_P(-1)^{P(rr_1r_2r_3)}\big[\big[B^{rr_1r_4},\, C^{r_2}\big],\,C^{r_3}\big]  \hspace{-0.1em}  =    \hspace{-0.1em} \varepsilon^{rr_1r_2r_3} Q^{r_4}_4  \ \mathrm{for}\ Q^{r_4}_4  \hspace{-0.1cm} =  \hspace{-0.1cm} \frac{1}{2}\big[\big[B^{rr_1r_4},\, C^{r_2}\big],\,C^{r_3}\big] \varepsilon_{rr_1r_2r_3}\hspace{-0.1em}, \label{ident24}
\end{eqnarray}
so that the latter quantities,  $Q^{r_4}_4$, (\ref{ident24}) do not presented in  the transformations for $B$ in (\ref{minN4}) as compared for the $N=3$ quantities, $Q^{p} \equiv Q^{p}_3$ (\ref{ident2}), (\ref{aident3}),  which are non-vanishing when enter into the transformations  for $\widehat{B}$ (\ref{minN3}).
One can  immediately check  that the  equations (\ref{consistB4l}) considered
for ${B}$, instead of $B^{r_1r_2r_3}$, are fulfilled as well:
\begin{equation}\label{consisttot4}
  {B}\big(\overleftarrow{s}{}^{r_1}_{[4]}\overleftarrow{s}{}^{r_2}_{[4]}+\overleftarrow{s}{}^{r_2}_{[4]}\overleftarrow{s}{}^{r_1}_{[4]}\big) =0 \Leftrightarrow \overleftarrow{s}{}^{\{r_1}_{[4]}\overleftarrow{s}{}^{r_2\}}_{[4]} =0.
\end{equation}
Therefore, $\overleftarrow{s}{}^{r} = \overleftarrow{s}{}^{r}_{[4]}$ are the generators of the irreducible representation of  $\mathcal{G}(4)$ superalgebra
of $N=4$-parametric SUSY transformations
in the field superspace, $\mathcal{M}^{(4)}_{tot}$, parameterized by the fields, $\Phi^{A_4}_{(4)}$.  That fact completes the proof of the Lemma~4.

Note, first, that the  transformations on the fields $B^{r_1r_2}, B^{r_1r_2r_3}, B$ do not contain the terms more than cubic in the fictitious fields, whereas they  depend linearly on the fields $B$'s in the  cubic terms. Second, the quantities, $Q^{r}_4$ do not enter into the transformations for Grassmann-even field $B$ as compared  to its  $N=3$ analogs, $Q^{p}$, which are essentially presented  in the transformations for Grassmann-odd  $\widehat{B}$.

Now, we have all necessary to  construct  $\mathcal{G}(4)$-invariant  quantum action for the Yang--Mills theory.

\section{N=4  BRST invariant gauge-fixing procedure and local path integral }\label{N4gf}
 \setcounter{equation}{0}

Let us determine according to the prescription (\ref{PintfpLlock}), (\ref{qe})  the local   path integral, ${Z}_{4}$,  generating functionals of Green functions in any admissible gauge, turning to the non-degenerate Faddeev-Popov matrix, for Yang-Mills theory underlying above constructed explicit  $N=4$ SUSY invariance (\ref{minN4})  in the total configuration space $\mathcal{M}^{(4)}_{tot}$, $\mathcal{M}^{(4)}_{tot}=\mathcal{M}^{(3)}_{tot}$, with quartet of anticommuting parameters $\lambda_r$  and the local  quantum action $S_{Y_{(4)}}(\Phi_{(4)})$  as follows:
\begin{eqnarray}
\label{Pintlocn4}
  &\hspace{-1em}& {Z}_{4|Y}(0) =  \int  d \Phi_{(4)}   \exp \Big\{\frac{\imath}{\hbar}S_{Y_{(4)}}\big(\Phi_{(4)}\big)\Big\}, \ \mathrm{with}  \ S_{Y_{(4)}}  = S_0(\mathcal{A}) -\frac{1}{4!}%
Y_{(4)}\overleftarrow{s}{}^{r_1}\overleftarrow{s}{}^{r_2}\overleftarrow{s}{}^{r_3}\overleftarrow{s}{}^{r_4} \varepsilon_{[r]_4},\\
\label{GFlocn4}
  &\hspace{-1em}& {Z}_{4|Y}({J}_{(4)}) =  \int  d \Phi_{(4)}  \exp \Big\{\frac{\imath}{\hbar}S_{Y_{(4)}}\big(\Phi_{(4)}\big) + J_{(4)}\Phi_{(4)}\Big\} \ =\ \exp \Big\{\frac{\imath}{\hbar}W_{4|Y}({J}_{(4)})\Big\}.  \end{eqnarray}
with  use of the compact notation  for, $\varepsilon_{r_1r_2r_3r_4} \equiv \varepsilon_{[r]_4}$. Here, $W_{4|Y}({J}_{(4)})$ is  the generating functional of connected correlated Green functions  and  gauge boson functional, $F_{(4)} = Y_{(4)}=Y_{(4)}\big({\Phi}_{(4)}\big)$, depends on the fields ${\Phi}_{(4)}$ as follows (confer with  $Y_{\xi }$ (\ref{Y(A,C)}) for $N=2$ BRST symmetry):
\begin{equation}\label{N4gaugeb}
  Y_{(4)}({\Phi}_{(4)}) =    Y^0_{(4)}({\Phi}_{(4)})  +  \widehat{Y}_{(4)}({\Phi}_{(4)}), \   \ \mathrm{for} \ deg_{{\Phi}}\widehat{Y}_{(4)}>2, \ \ deg_{\Phi}{} Y^0_{(4)}({\Phi}_{(4)})  = 2,
\end{equation}
and ${J}_{A^t_4}$ are the  external sources  (coinciding with ones for $N=3$ case, $\widetilde{J}_{A^t_3}$)  to the  Green functions related to  $\Phi^{A^t_4}_{(4)}$ with the same Grassmann parities: $\epsilon(J_{A^t_4}) = \epsilon(\Phi^{A^t_4}_{(4)})$.

 It is not difficult to check that both the' functional measure, $d \Phi_{(4)} $, as well as the quantum action,
 $S_{Y_{(4)}}$, are invariant with respect to the change of variables, ${\Phi}{}^{A^t}_{(4)}\to  {\Phi}{}^{\prime  A^t}_{(4)}$ generated by  $N=4$ SUSY transformations (\ref{minN4}) with accuracy   up to the first order in constant $\lambda_p$  (equally with infinitesimal  $\lambda_p$):
\begin{equation}\label{N4inv}
{\Phi}{}^{\prime  A^t}_{(4)}={\Phi}{}^{A^t}_{(4)}(1+\overleftarrow{s}{}^r\lambda_r): \ \delta_{\lambda}{\Phi}{}^{  A^t}_{(4)}={\Phi}{}^{  A^t}_{(4)}\overleftarrow{s}{}^r\lambda_r \Longrightarrow    \delta_{\lambda}S_{Y_{(4)}} =o(\lambda), \ \mathrm{sdet}\left\| \delta{\Phi}{}^{\prime}_{(4)}/  \delta{\Phi}_{(4)}  \right\| =1+o(\lambda).
\end{equation}
 These properties justify the definition of  the transformations:
 \begin{equation}\label{n4brst}
   \delta_{\lambda}{\Phi}{}^{  A^t}_{(4)}\ = \ {\Phi}{}^{\prime  A^t}_{(4)}-{\Phi}{}^{  A^t}_{(4)}\ = \ {\Phi}{}^{  A^t}_{(4)}\overleftarrow{s}{}^r\lambda_r,\
 \end{equation}
 with the explicit action of the generators $\overleftarrow{s}{}^r$ (\ref{minN4})  on  the component fields as $N=4$-\emph{parametric}  \emph{BRST transformations} for the functionals ${Z}_{4|Y}(0)$, ${Z}_{4|Y}({J}_{(4)})$.

 The particular representations for the path integrals (\ref{Pintlocn4}), (\ref{GFlocn4}) in  the Landau and Feynman gauges may be  obtained within the same $R_\xi$-family of the gauges as for the $N=1,2,3$ BRST invariant cases  (\ref{PintfpLlocn}), (\ref{Y(A,C)}), (\ref{Pintlocn3}).
 To do so  we determine the quadratic gauge boson functional, $Y^0_{(4)\xi}({\Phi}_{(4)}) $,  which should generate $R_\xi$-like gauges as follows:
\begin{eqnarray}
 \label{gBn4xi} Y^0_{(4)\xi}({\Phi}_{(4)}) = Y^0_{(4)}(\mathcal{A})+Y^B_{(4)\xi}(B^{rs}) = \int
d^{d}x\ tr \Big(\frac{1}{2}\mathcal{ A}_{\mu }\mathcal{A}^{\mu }-\frac{\xi g^2 }{4!}B^{q_1q_2}B^{q_3q_4}\varepsilon_{[q]_4}\Big)\footnotemark . \end{eqnarray}\footnotetext{Instead of the functional $Y^B_{(4)\xi}(B^{rs})$ which generates the $\xi$-dependent term it is possible to consider the functional $\tilde{Y}^B_{(4)\xi}(C,B^{rsq}) = \frac{\xi g^2 }{4!}\int
d^{d}x\ tr  C^{q_1}B^{q_2q_3q_4}\varepsilon_{[q]_4}$ still leading to the same quadratic term: ${\xi g^2} B^2$ in $S_{\mathrm{gf}(4)}$,  but with another non-quadratic in the fictitious fields summands in $S_{\mathrm{add}(4)}$.}
 The quantum action, $S_{Y_{(4)\xi}}$, has the representation:
\begin{align}\label{qexi4}
  S_{Y_{(4)\xi}}\big({\Phi}_{(4)}\big) & = S_0-\frac{1}{4!}%
Y_{(4)\xi}\overleftarrow{s}{}^{r_1}\overleftarrow{s}{}^{r_2}\overleftarrow{s}{}^{r_3}\overleftarrow{s}{}^{r_4} \varepsilon_{[r]_4}= S_0+S_{\mathrm{gf}(4)} +  S_{\mathrm{gh}(4)}+  S_{\mathrm{add}(4)} ,\\
 S_{\mathrm{gf}(4)} &  =  \int d^{d}x\ tr \Big[    \partial^{\mu}A_{\mu
}  + {\xi g^2} B\Big]  B,
\label{gfxi4}\\
   S_{\mathrm{gh}(4)}&  = \int d^{d}x\ tr \Big\{ %
  \frac{1}{3!} B^{r_1r_2r_3}M(\mathcal{A})  C^{r_4} + \frac{1}{8} B^{r_1r_2}M(\mathcal{A})  B^{r_3r_4}  \Big\}\varepsilon_{[r]_4}, \label{ghxi4}
 \end{align}
 \vspace{-1ex}
 \begin{align}
 S_{\mathrm{add}(4)} &=   \int d^{d}x\ tr   \frac{1}{4!}\Bigg\{
 (\partial^\mu\mathcal{ A}_{\mu })  \Big(2\big[B^{r_1r_2r_3}, C^{r_4}\big]  - \big[\big[B^{r_1r_2}, C^{r_3}\big], C^{r_4}\big]\Big) -  B^{r_1r_2}\Big(\big[ C^{r_3},  M(\mathcal{A})C^{r_4}\big]\nonumber \\
\phantom{S_{Y_{(4)\xi}}}&    +4\big[ \partial_\mu C^{r_3},  D^\mu C^{r_4}\big]\Big)+
 C^{r_1}\partial_\mu \Big( \big[D^\mu C^{r_2},\,    B^{r_3r_4}\big] -\big[C^{r_2},\, D^\mu B^{r_3r_4}   \big] + \big[\big[D^\mu C^{r_2},\, C^{r_3}\big],  C^{r_4}\big]\Big)\nonumber \\
\phantom{S_{Y_{(4)\xi}}\big({\Phi}_{(4)}\big)} &
+ \frac{{\xi g^2}}{4}\Big(\frac{1}{4}\big[B^{q_1q_2}, B^{r_1r_2}\big]
\big[B^{q_3q_4}, B^{r_3r_4}\big]   + \frac{1}{(3!)^2}\big[C^{q_1},\big[\big[C^{q_2},\,C^{r_2}\big] , C^{r_1}\big]\big]\times \nonumber\\
 \phantom{S_{Y_{(4)\xi}}\big({\Phi}_{(4)}\big)} &    \times \big[C^{q_3},\big[\big[C^{q_4},\,C^{r_4}\big] , C^{r_3}\big]\big]\Big)\varepsilon_{[q]_4} \Bigg\}\varepsilon_{[r]_4}+ \widetilde{S}_\xi,  \label{addxi4}
\end{align}
with some Grassmann-even functional  $\widetilde{S}_\xi$ vanishing in the Landau gauge ($\xi=0$).  To derive (\ref{qexi4})--(\ref{addxi4})  we have used the relations (\ref{auxsaK})--(\ref{s3(AB)}), (\ref{sABcom}) being adapted for $N=4$ BRST symmetry, as well as the following from (\ref{s3(AB)})
 Leibnitz-like property of the generators, $\overleftarrow{s}{}^{r}$ acting on the product of
any functions $A$, $B$ with definite Grassmann grading:
\begin{eqnarray}\label{s4(AB)}
&& \left(  AB\right) \overleftarrow{s}{}^{r_1}\overleftarrow{s}{}^{r_2}\overleftarrow{s}{}^{r_3}\overleftarrow{s}{}^{r_4}\varepsilon_{[r]_4}
=  \Big[ A \overleftarrow{s}{}^{r_1}\overleftarrow{s}{}^{r_2}\overleftarrow{s}{}^{r_3}\overleftarrow{s}{}^{r_4} B+4A\overleftarrow{s}{}^{r_1}
\left(  B\overleftarrow{s}{}^{r_2}\overleftarrow{s}{}^{r_3}\overleftarrow{s}{}^{r_4}\right)  \left(  -1\right) ^{\epsilon(B)} \\
&& \quad +6A\overleftarrow{s}{}^{r}_1\overleftarrow{s}{}^{r_2}
\left(  B\overleftarrow{s}{}^{r_3}\overleftarrow{s}{}^{r_4}\right)+ 4A\overleftarrow{s}{}^{r_1}\overleftarrow{s}{}^{r_2}\overleftarrow{s}{}^{r_3}
\left(  B\overleftarrow{s}{}^{r_4}\right)  \left(  -1\right) ^{\epsilon(B)} +A\left(
B\overleftarrow{s}{}^{r_1}\overleftarrow{s}{}^{r_2}\overleftarrow{s}{}^{r_3}\overleftarrow{s}{}^{r_4}\right)\Big]\varepsilon_{[r]_4} .  \nonumber %
\end{eqnarray}
The detailed derivation for the quantum action, structure of the additional $\xi$-dependent term,  $\widetilde{S}_\xi$, are considered in the   Appendix~\ref{AppB}.
Note, the each term in $S_{\mathrm{add}(4)}$ contains space-time derivative and, in particular, the second-order differential operator (Faddeev-Popov operator)  for any gauge from $R_\xi$-gauges, as for the   $S_{\mathrm{add}(3)}$ (\ref{eaddxi3}) for $N=3$ BRST symmetry. For the Landau gauge, the summands in $S_{\mathrm{add}(4)}$ proportional to the Lorentz condition: $(\partial^\mu\mathcal{ A}_{\mu })=0$, may be omitted therein due to the presence of $\delta((\partial^\mu\mathcal{ A}_{\mu }))$ in the functional integral (\ref{Pintlocn4}) after integrating over the fields $B$.

The equivalence of $N=4$ and $N=1$
 BRST invariant path integrals
 ${Z}_{4|Y}(0)$  (\ref{Pintlocn4}),  $
Z_\Psi $  (\ref{PintfpLlocn}),. e.g in the Landau gauge determined by the gauge functional $Y^0_{(4)}(\mathcal{A})$  (\ref{gBn4xi}) follows  analogously  to the derivation (\ref{N3N1eqv}) for $N=3$ case
from  the structure of the quantum action $S_{Y_{(4)\xi}}$,   (\ref{qexi4})--(\ref{addxi4}).  Indeed, using the representation for $S_{Y_{(4)\xi}}$  (\ref{qexi4appgen})  in terms of  dual  $\mathcal{G}(4)$-tensor fields  $B_{r_1r_2}$, $C_{r}$  (\ref{dual42}), (\ref{dual43})
let us divide  the quartets of ghost Grassman-odd  fields  $C_{r}, C^{r}$  as $\mathcal{G}(3)$-triplets and singlets which  permits to present the respective term in the ghost part of the action as:
 \begin{equation}\label{N4N31}
   \big(C_{r}; C^{r}\big) = \big((\overline{C},{C}_{p}); (C, C^{p})\big) \ \Rightarrow\  C_{r}M(\mathcal{A})  C^{r} = \overline{C}M(\mathcal{A})  C+ C_{p}M(\mathcal{A})  C^{p},
 \end{equation}
 for $r=(1,p), p=2,3,4$ and $\overline{C} \equiv -B^{234}$.  Because of the remark above we may omit the terms with $(\partial^\mu\mathcal{ A}_{\mu })$ with except for Nakanishi-Lautrup field $B$ and therefore integrate  by the fields $C_{p}$, second,   with respect to $C^{p}$, and then trivially  with respect  to $B^{r_1r_2}$  and $B_{r_1r_2}$ for $1\leq r_1<r_2 \leq 3$
 as follows:
\begin{eqnarray}
  {Z}_{4|Y(0)}(0) &=& \hspace{-0.15em}\int \hspace{-0.15em}  d {\Phi}d {C^p} dB^{r_1r_2}  dB_{r_1r_2} {\det}^{3} M(\mathcal{A})  \delta(C^p) \exp \hspace{-0.15em}\Big\{\frac{\imath}{\hbar}\Big(S_{Y_{(4)0}}\big({\Phi}_{(4)} \big) - \int d^{d}x\ tr\, {C}{}_{p}M(\mathcal{A}) {C}{}^{p}  \Big)\Big\} \nonumber \\
   & =&   \hspace{-0.15em}\int \hspace{-0.15em}  d {\Phi} dB_{r_1r_2}  {\det}^{3} M(\mathcal{A}) {\det}^{-3} M(\mathcal{A})  \delta(B^{r_1r_2})  \exp \hspace{-0.15em}\Big\{\frac{\imath}{\hbar}S_{\Psi{\xi}}\big({\Phi} \big )\vert_{\xi=0} \Big\} = {Z}_{\Psi}.\label{N4N1eqv}
\end{eqnarray}
The functional ${Z}_{\Psi}$ exactly coincides with one given in (\ref{PintfpLlocn}) in the Landau gauge.

Again, the  $N=4$  BRST invariance, for the corresponding generating functionals
of Green's functions, ${Z}_{4|Y}({J}_{(4)})$ , ${W}_{4|Y}({J}_{(4)})$ and effective action, $\Gamma_{4|Y}(\langle{\Phi}_{(4)}\rangle)$ determined by the same rule as for its $N=3$ analog (\ref{GFGF3})
with a given gauge condition $Y_{(4)}({\Phi}_{(4)})$,
leads to the presence of an $\mathcal{G}(4)$-quartet of Ward identities:
\begin{equation}\label{GFGFWIN4}
{J}_{A^t_4}\langle {\Phi}{}^{A^t_4}_{(4)}\overleftarrow{s}{}^r\rangle_{Y_{(4)},{J}}=0, \ \ {J}_{A^t_4}\langle\langle {\Phi}^{A^t_4}_{(4)} \overleftarrow{s}{}^r\rangle\rangle_{Y_{(4)},{J}}=0, \ \ \frac{\delta \Gamma_{4|Y}}{\delta \langle {\Phi}^{A^t_4}_{(4)}\rangle}\langle\langle {\Phi}{}^{A^t_4}_{(4)} \overleftarrow{s}{}^r\rangle\rangle_{Y_{(4)} \langle {\Phi}\rangle} =0,
 \end{equation}
with corresponding normalized  average expectation values (as in  (\ref{GFGFWIN3}))  in the presence of the external sources ${J}_{A^t_4}$ and mean fields
$\langle {\Phi}{}^{A^t_4}_{(4)}\rangle$. The gauge independence of the path integral $Z_{4|Y}(0)\equiv Z_{4|Y_{(4)}}(0)$
under an infinitesimal variation of the gauge condition, $Y_{(4)} \to Y_{(4)}+\delta Y_{(4)}$:
\begin{equation}\label{N4GIinf}
  Z_{4|Y_{(4)}+\delta Y_{(4)}}(0) =     Z_{4|Y_{(4)}}(0)
\end{equation}
is established using the \emph{infinitesimal FD} $N=4$ BRST transformations
 with the functional parameters,
  \begin{equation}\label{N4FDP}
 \lambda_{r_1}({\Phi}_{(4)})= \frac{1}{4!}\big(\imath / \hbar\big) \delta Y_{(4)}({\Phi}_{(4)}) \prod_{k=2}^4\overleftarrow{s}{}^{r_k}\varepsilon_{[r]_4},
.\end{equation}
 which will be carefully  elaborated    in the next Section~\ref{N=kgauge} as well as
  some important consequences of the suggested $N=3$ and $N=4$   BRST transformations,
 respective quantum actions and gauge-fixing procedures.

\section{$N=k$,  $k=3,4$ infinitesimal and finite BRST transformations and their Jacobians} \label{N=kgauge}

 \setcounter{equation}{0}

Here, we consider  the algorithm of construction of finite  $N=k$  BRST transformations starting from its algebraic (infinitesimal) proposals respectively for $k=3, 4$ cases and calculate theirs Jacobians together with some  physical corollaries.

\subsection{$N=3$ BRST transformations} \label{N=3gauge}

The finite $N=3$ BRST transformations acting on the fields $\widetilde{\Phi}{}^{A^t_3}_{(3)}$, parameterizing  configuration space $\mathcal{M}^{(3)}_{tot}$, are restored
from the algebraic (equivalently, infinitesimal for small $\lambda_p$) $N=3$ BRST transformations, generalizing the recipe \cite{re1} for
$N=2$ BRST symmetry and following to  \cite{UMR}, \cite{BLTf} 
 in two equivalent ways. First,  the derivation   is based on the condition which follows  for any
$\overleftarrow{s}^{p }$-closed regular functional $K\big(\widetilde{\Phi}_{(3)} \big)$
to be invariant with respect to right-hand supergroup transformations  and, second,  from
the \emph{Lie equations}:
\begin{eqnarray}
&&1) \quad \left\{ K\left( g(\lambda_p)\widetilde{\Phi}_{(3)} \right) =K\left( \widetilde{\Phi}_{(3)}
\right) \ \mathrm{and}\ K\overleftarrow{s}{}^{p}=0\right\} \hspace{-0.1em}%
\Rightarrow g\left( \hspace{-0.1em}\lambda_{p}\hspace{-0.1em}\right) = \exp \left\{ \hspace{-0.1em}\overleftarrow{s%
}{}^{p}\lambda_{p}\hspace{-0.1em}\right\} ,  \label{bab3}
\\
&&2) \quad   \widetilde{\Phi}{}^{A^t_3}_{(3)} \big(\widetilde{\Phi}_{(3)}|\lambda \big)\overleftarrow{\partial }^{p} =\widetilde{\Phi}{}^{A^t_3}_{(3)}\big(\widetilde{\Phi}_{(3)}|\lambda\big)%
\overleftarrow{s}^{p}\ \left( \mathrm{for}\ \overleftarrow{\partial }%
^{p}\equiv \frac{\overleftarrow{\partial }}{\partial \lambda_{p}}\right)\footnotemark .
\label{Lieeq}
\end{eqnarray}\footnotetext{For a $t$-rescaled argument $\lambda_{p}\rightarrow t\lambda_{p}$ of $\widetilde{\Phi}{}^{A^t_3}_{(3)}\big(\widetilde{\Phi}_{(3)}|t\lambda\big)$, the form of Lie equations: $
\frac{d}{dt}\widetilde{\Phi}{}^{A^t_3}_{(3)}\big(\widetilde{\Phi}_{(3)}|t\lambda\big)=\widetilde{\Phi}{}^{A^t_3}_{(3)}\big(\widetilde{\Phi}_{(3)}|t\lambda\big)
\overleftarrow{s}{}^p \lambda_p $,  is
equivalent to (\ref{Lieeq}) with a formal solution for constant $\lambda^p$:  $\widetilde{\Phi}{}^{A^t_3}_{(3)}\big(\widetilde{\Phi}_{(3)}|t\lambda\big)= \widetilde{\Phi}{}^{A^t_3}_{(3)}\exp \big\{t%
\overleftarrow{s}{}^p \lambda_p\big\}$}
whose set forms an Abelian  $3$-parametric  supergroup,
\begin{equation}\label{N3finBRST}
G(3) = \left\{g(\lambda_p)
: g(\lambda_p)= 1 + \sum_{e=1}^3 \frac{1}{e!}\prod_{l=1}^e\overleftarrow{s}{}^{p_l}\lambda_{p_l}   = \exp \left( \hspace{-0.1em}\overleftarrow{s%
}{}^{p}\lambda_{p}\hspace{-0.1em}\right)\right\},
\end{equation}
where  $\overleftarrow{s}{}^{p}$, $%
\overleftarrow{s}{}^{p_1}\overleftarrow{s}{}^{p_2} \varepsilon_{[p]_3}$ and  $\overleftarrow{s}{}^{p_1}\overleftarrow{s}{}^{p_2}\overleftarrow{s}{}^{p_3} \varepsilon_{[p]_3}$
are respectively  the generators of $N=3$  BRST,   quadratic  mixed and cubic mixed  $N=3$ BRST  transformations
in the space of fields  $\widetilde{\Phi}{}^{A^t_3}_{(3)}$.

For the \emph{field-dependent}  $\mathcal{G}(3)$ triplet of odd-valued functionals $\lambda_p(\widetilde{\Phi}_{(3)})$,
which is not closed under $N=3$ BRST transformations, $\lambda_p \overleftarrow{s}{}^p \ne 0$, but
for, $\partial/\partial x^\mu  \lambda_p = 0$,  the  finite element $g\big(\lambda_p(\widetilde{\Phi}_{(3)})\big)$ cannot be presented as group element (using an exp-like relation) in (\ref{N3finBRST}). In this case, the set of algebraic elements
$\widetilde{\mathcal{G}}(3) =\big\{\tilde{g}_{lin}(\lambda (\widetilde{\Phi}_{(3)})):=1+\overleftarrow{s}^p\lambda_p(\widetilde{\Phi}_{(3)})\big\}$
forms a non-linear superalgebra which corresponds to a set of formal group-like finite elements:
\begin{equation}
\tilde{G}(3)=\left\{ \tilde{g}\big(\lambda_p(\widetilde{\Phi}_{(3)})\big):\tilde{g}=1+ \overleftarrow{s}{}^p\lambda_p + \frac{1}{2}\overleftarrow{s}{}^p\overleftarrow{s}{}^q \lambda_q\lambda_p+ \frac{1}{3!}\overleftarrow{s}{}^p\overleftarrow{s}{}^q \overleftarrow{s}{}^r \lambda_r\lambda_q\lambda_p\right\} ,
\label{tildeG3}
\end{equation}
with loss of the commutativity property: $\big[\tilde{g}\big(\lambda^{(1)}_p(\widetilde{\Phi}_{(3)})\big),\,\tilde{g}\big(\lambda^{(2)}_p(\widetilde{\Phi}_{(3)})\big)\big] \ne 0$.
The Jacobian of a change of variables:
$\widetilde{\Phi}{}^{A^t_3}_{(3)}\to \widetilde{\Phi}{}^{\prime A^t_3}_{(3)}=
\widetilde{\Phi}{}^{A^t_3}_{(3)}\tilde{g}\big(\lambda_p(\widetilde{\Phi}_{(3)})\big)$, in $\mathcal{M}^{(3)}_{tot}$, in the path integral ${Z}_{3|\Psi}(0)$ (\ref{Pintlocn3}) 
generated by finite FD  $N=3$ BRST transformations may be calculated
explicitly, following a generalization of the recipe proposed in \cite{re1}
for an irreducible gauge theory with a closed algebra (including the Yang--Mills theory, see as well \cite{MR5})
in the $N=2$ case, or following the recipe of \cite{UMR} for $N=m$ finite
FD SUSY transformations. The results are as follows:
\begin{equation}
\mathrm{sdet}\left\|\widetilde{\Phi}{}^{A^t_3}_{(3)}\tilde{g}\big(\lambda_p(\widetilde{\Phi}_{(3)})\big)\frac{\overleftarrow{\delta}}{\delta \widetilde{\Phi}{}^{B^t_3}_{(3)}} \right\| = \exp \Big\{-\mathrm{tr}_{G(3)}\ln \left( [e+m]^p_q\right) \Big\}, \ \mathrm{for} \ (e^p_q, m^p_q)\equiv \big(\delta^p_q, \lambda_q \overleftarrow{s}^p\big),
\label{jacobianresN3}
\end{equation}%
where $\mathrm{tr}_{G(3)}$ denotes  trace over matrix $G(3)$-indices.
Representation (\ref{jacobianresN3}) is based on the explicit calculation which generalize the algorithm for the Jacobian of the change of variables generated by  $N=2$ BRST transformations for Yang-Mills theory \cite{MR5}, \cite{reshmosh} as follows
\begin{eqnarray}
&\hspace{-0.5em}& \hspace{-0.5em}\mathrm{sdet}\left\|\widetilde{\Phi}{}^{A^t_3}_{(3)}\tilde{g}\big(\lambda_p(\widetilde{\Phi}_{(3)})\big)\frac{\overleftarrow{\delta}}{\delta \widetilde{\Phi}{}^{B^t_3}_{(3)}} \right\| =\exp \hspace{-0.1em%
}\left\{ \mathrm{Str}\,\mathrm{ln}\left( \delta
_{B_3}^{A_3}+M_{B_3}^{A_3}\right) \right\} ,\ \ \mathrm{for}\ \
M_{B_3}^{A_3}=P_{B_3}^{A_3}+\sum_{i=1}^3 (Q_i)_{B_3}^{A_3}  \label{jacobianN3YM} \\
&& \left\{\begin{array}{l}
\ P_{B_3}^{A_3} =\widetilde{\Phi}{}^{A_3}_{(3)}\overleftarrow{s}{}^{p}\big(\lambda _{p}\overleftarrow{\partial }%
_{B_3}\big) , \\
 (Q_1)_{B_3}^{A_3}=\lambda _{p}\Big\{\big(\widetilde{\Phi}{}^{A_3}_{(3)}\overleftarrow{s}{}^{p}\big)\overleftarrow{\partial }%
_{B_3}- \big(\widetilde{\Phi}{}^{A_3}_{(3)}\overleftarrow{s}{}^{q}\overleftarrow{s}{}^{p}\big)\big(\lambda_{q}%
\overleftarrow{\partial }_{B_{3}}\big)  \Big\}(-1)^{\epsilon _{A_3}+1},\\
\ (Q_2)_{B_3}^{A_3}=  \frac{1}{2}\lambda_p \lambda_q  \Big\{\big(\widetilde{\Phi}{}^{A_3}_{(3)}\overleftarrow{s}{}^{p}\overleftarrow{s}{}^{q}\big)%
\overleftarrow{\partial }_{B_{3}}- \frac{1}{ 3!}\varepsilon^{pqr} \big(\widetilde{\Phi}{}^{A_3}_{(3)}(\overleftarrow{s}){}^{3}\big)\big(\lambda_{r}%
\overleftarrow{\partial }_{B_3}\big)\Big\},\\
\  (Q_3)_{B_3}^{A_3}= \frac{1}{%
(3!)^2}(\lambda)^{3}\big(\widetilde{\Phi}{}^{A_3}_{(3)}(\overleftarrow{s}){}^{3}\overleftarrow{\partial }_{B_3}\big)(-1)^{\epsilon _{A_3}+1},
\end{array}\right. \label{PQiexp}\\
&&\ \Longrightarrow \mathrm{Str}\big(P+\sum_{i=1}^3 Q_i\big)^{n}=\mathrm{Str}\big(P+\sum_{i=1}^2 Q_i\big)^{n}+n\,\mathrm{Str}%
P^{n-1}Q_3,\
\label{jacobianN3YMrules} \\
&&\ \mathrm{Str}\big(P+\sum_{i=1}^2 Q_i\big)^{n}=
\mathrm{Str}P^{n}+ \mathrm{Str}F_n\big(P,Q_1,Q_2\big)
 \ \mathrm{with }\    F_n\big(P,Q_i\big)\big|_{(Q_i=0)}=0, \label{PQi}
\end{eqnarray}
(where we imply: $A^t_3\equiv A_3$;  $(\lambda)^{3} \equiv \lambda_{q_1}\lambda_{q_2}\lambda_{q_3}\varepsilon
^{[q]_3}$ and  $(\overleftarrow{s}){}^{3}$ given by (\ref{s3pres})), so that the only supermatrix $P$ gives the non-vanishing contribution into the Jacobian (\ref{jacobianresN3}):
\begin{eqnarray}
&&\mathrm{sdet}\left\|\widetilde{\Phi}{}^{A^t_3}_{(3)}\tilde{g}\big(\lambda_p(\widetilde{\Phi}_{(3)})\big)\frac{\overleftarrow{\delta}}{\delta \widetilde{\Phi}{}^{B^t_3}_{(3)}} \right\| =\exp \hspace{%
-0.1em}\Big\{\hspace{-0.1em}-\sum_{n=1}\frac{(-1)^{n}}{n}\mathrm{Str}%
(P_{B_3}^{A_3})^{n}\Big\},  \label{jacobianN3res}
\end{eqnarray}%
as compared  with  the nilpotent  supermatrices  $(Q_i)_{B_{3}}^{A_{3}}$  (entering in $F_n$ (\ref{PQi})) which do not contribute  to the Jacobian (\ref{jacobianresN3}), due to: $\prod_{k=1}^m \lambda_{p_k} \equiv 0$ for $m>3$.

For functionally-independent FD $\lambda_p\big(\widetilde{\Phi}_{(3)}\big)$, the Jacobian
(\ref{jacobianresN3}) is not $\overleftarrow{s}^p$-closed in general. For $\overleftarrow{s}{}^p$-potential (thereby, functionally-dependent)  parameters
\begin{equation}
\hat{\lambda}_{p_1}\big(\widetilde{\Phi}_{(3)}\big)=\frac{1}{2!}\Lambda \big(\widetilde{\Phi}_{(3)}\big)\varepsilon
_{[p]_3}\overleftarrow{s}{}^{p_2}
\overleftarrow{s}{}^{p_3},\
\label{fdepparam}
\end{equation}%
with an arbitrary potential being by  Grassmann-odd-valued functional $\Lambda\big(\widetilde{\Phi}_{(3)}\big)$
the Jacobian (\ref{jacobianresN3}) simplifies to $N=3$ BRST exact functional determinant:
\begin{equation}
\mathbf{J}_{(3)}\big(\widetilde{\Phi}_{(3)}\big) = \mathrm{sdet}\left\|\widetilde{\Phi}{}^{A^t_3}_{(3)}\tilde{g}\big(\hat{\lambda}_p(\widetilde{\Phi}_{(3)})\big)\frac{\overleftarrow{\delta}}{\delta \widetilde{\Phi}{}^{B^t_3}_{(3)}} \right\| =  \Big\{1 + \frac{1}{3!}\Lambda (\widetilde{\Phi}_{(3)})\varepsilon
_{[p]_3}\prod_{k=1}^3\overleftarrow{s}{}^{p_k}\Big\}^{-3}, \ \mathbf{J}_{(3)}(\widetilde{\Phi}_{(3)})\overleftarrow{s}{}^p = 0,
\label{jacobianresN3fd}
\end{equation}%
by virtue of the fact that the tensor quantity $\big(\overleftarrow{s}{}^{p_1}\overleftarrow{s}{}^{p_2}
\overleftarrow{s}{}^{p_3}\big)$ is  completely antisymmetric in  $(p_1,p_2,p_3)$  indices and can be presented as:
\begin{equation}\label{s3pres}
  \overleftarrow{s}{}^{p_1}\overleftarrow{s}{}^{p_2}
\overleftarrow{s}{}^{p_3} = \frac{1}{3!}\varepsilon
^{[p]_3} \left(\overleftarrow{s}\right)^3 \ \ \mathrm{for} \ \  \left(\overleftarrow{s}\right)^3 \equiv     \overleftarrow{s}{}^{q_1}\overleftarrow{s}{}^{q_2}
\overleftarrow{s}{}^{q_3}\varepsilon
_{[q]_3}
\end{equation}
 which permits, because of: $\prod_{k=1}^4\overleftarrow{s}{}^{q_k} \equiv 0$, to have the  representation
\begin{eqnarray}\label{repreN3}
  &&  \delta^p_q+\hat{\lambda}_{q}\big(\widetilde{\Phi}_{(3)}\big) \overleftarrow{s}{}^p = \delta^p_q+  \frac{1}{2!}\Lambda \big(\widetilde{\Phi}_{(3)}\big)\varepsilon
_{qp_2p_3}\overleftarrow{s}{}^{p_2}
\overleftarrow{s}{}^{p_3}\overleftarrow{s}{}^p  \\
&& \quad =  \Big(\delta^p_q+ \frac{1}{2 \cdot 3! }\Lambda \varepsilon
_{qp_2p_3}\varepsilon
^{p_2p_3p}\left(\overleftarrow{s}\right)^3\Big) = \delta^p_q\Big(1+ \frac{1}{3!}\Lambda \left(\overleftarrow{s}\right)^3\Big), \nonumber \\
&&  \Longrightarrow  \  \mathrm{tr}_{G(3)}\ln \left( [\delta^p_q+ \hat{\lambda}_q \overleftarrow{s}^p]\right) = \mathrm{tr}_{G(3)}\ln \left( \delta^p_q\Big[1+ \frac{1}{3!}\Lambda \left(\overleftarrow{s}\right)^3\Big]\right) = \delta^q_q\ln \left( 1+ \frac{1}{3!}\Lambda \left(\overleftarrow{s}\right)^3\right) \label{repreN3f}
\end{eqnarray}
that proves (\ref{jacobianresN3fd}).

In the case of $\overleftarrow{s}^p$-closed parameters ${\lambda}_p$,
${\lambda}_p\overleftarrow{s}^q = 0$, including constant ${\lambda}_p$,
i.e., for $G(3)$ group elements, the Jacobian becomes trivial: $\mathbf{J}_{(3)}=1$.
In turn, for the infinitesimal FD triplet $\hat{\lambda}_p\big(\widetilde{\Phi}_{(3)}\big)$ (\ref{fdepparam})
the Jacobian (\ref{jacobianresN3fd}) reduces to:
\begin{equation}
\mathbf{J}_{(3)}(\widetilde{\Phi}_{(3)}) = 1 - \frac{1}{2}\Lambda (\widetilde{\Phi}_{(3)})\big(\overleftarrow{s}\big){}^{3} +o(\Lambda) = \exp\Big\{- \frac{1}{2}\Lambda (\widetilde{\Phi}_{(3)})\big(\overleftarrow{s}\big){}^{3}\Big\}+o(\Lambda),
\label{jN3fdinf}
\end{equation}%
which permits to justify  the gauge independence for the path integral $Z_{\Psi_{(3)}}$ (and therefore for the conventional S-matrix) under small variation of the gauge condition: $\Psi_{(3)} \to \Psi_{(3)}+\delta \Psi_{(3)}$,  announced in  (\ref{N3GIinf}) because of
\begin{equation}\label{N3GI}
  Z_{3|\Psi_{(3)}+\delta \Psi_{(3)}}(0) =  \int  d \widetilde{\Phi}_{(3)} \  \mathrm{sdet}\|\widetilde{\Phi}{}^{\prime A_3} \overleftarrow{\partial}_{B_3} \| \exp \Big\{\frac{\imath}{\hbar}S_{\Psi_{(3)}+\delta \Psi_{(3)}}(\widetilde{\Phi})\Big\}=   Z_{3|\Psi_{(3)}}(0).
\end{equation}
in accordance with the choice (\ref{N3FDP})   for $\delta \Psi_{(3)}$ in terms of $\Lambda (\widetilde{\Phi}_{(3)})$ and therefore of  $\hat{\lambda}_p = \hat{\lambda}_p(\Lambda)$
\begin{equation}\label{N3FDPder}
 \Lambda \big(\widetilde{\Phi}_{(3)}|\delta \Psi_{(3)}\big) =  \frac{1}{3}\big(\imath / \hbar\big) \delta \Psi_{(3)}\big(\widetilde{\Phi}_{(3)}\big)   \Longrightarrow \hat{\lambda}_p(\Lambda) =  \frac{1}{3!}\big(\imath / \hbar\big) \delta \Psi_{(3)}  \overleftarrow{s}{}^q\overleftarrow{s}{}^r \varepsilon_{pqr}.
.\end{equation}

Another properties for the generating functionals of Green functions  related to the finite FD $N=3$ BRST transformations we will  consider in the Section ~\ref{NkWI}.

\subsection{$N=4$ BRST transformations} \label{N=4gauge}

The results of the above subsection  are easily adapted for $N=4$ BRST transformations with some specificity.
Thus,
the finite $N=4$ BRST transformations acting on the fields ${\Phi}{}^{A^t_4}_{(4)}$, parameterizing  configuration space $\mathcal{M}^{(4)}_{tot}$ coinciding with $\mathcal{M}^{(3)}_{tot}$ by dimension,  are restored
from the algebraic  $N=4$ BRST transformations
 by two equivalent ways:  or from the condition which follows  for any
$\overleftarrow{s}^{r }$-closed regular functional $K\big({\Phi}_{(4)} \big)$
to be invariant with respect to right-hand supergroup transformations $\{g(\lambda_r)\}$, $r=1,2,3,4$,  or  from
the {Lie equations}:
\begin{eqnarray}
&&1) \quad \left\{ K\left( g(\lambda_r){\Phi}_{(4)} \right) =K\left( {\Phi}_{(4)}
\right) \ \mathrm{and}\ K\overleftarrow{s}{}^{r}=0\right\} \hspace{-0.1em}%
\Rightarrow g\left( \hspace{-0.1em}\lambda_{r}\hspace{-0.1em}\right) = \exp \left\{ \hspace{-0.1em}\overleftarrow{s%
}{}^{r}\lambda_{r}\hspace{-0.1em}\right\} ,  \label{bab4}
\\
&&2) \quad   {\Phi}{}^{A^t_4}_{(4)} \big({\Phi}_{(4)}|\lambda \big)\overleftarrow{\partial }^{r} ={\Phi}{}^{A^t_4}_{(4)}\big({\Phi}_{(4)}|\lambda\big)%
\overleftarrow{s}^{r}\ \left( \mathrm{for}\ \overleftarrow{\partial }%
^{r}\equiv \frac{\overleftarrow{\partial }}{\partial \lambda_{r}}\right).
\label{Lieeq4}
\end{eqnarray}
The set of such  $\{g(\lambda_r)\}$ forms an Abelian  $4$-parametric  supergroup,
\begin{equation}\label{N4finBRST}
G(4) = \left\{g(\lambda_r)
: g(\lambda_r)= 1 + \sum_{e=1}^4 \frac{1}{e!}\prod_{l=1}^e\overleftarrow{s}{}^{r_l}\lambda_{r_l}   = \exp \left( \hspace{-0.1em}\overleftarrow{s%
}{}^{r}\lambda_{r}\hspace{-0.1em}\right)\right\},
\end{equation}
For the \emph{field-dependent}  $\mathcal{G}(4)$ quartet of odd-valued functionals $\lambda_r({\Phi}_{(4)})$,
which is not closed under $N=4$ BRST transformations, $\lambda_r \overleftarrow{s}{}^r \ne 0$,  the  finite element $g\big(\lambda_r({\Phi}_{(4)})\big)$ cannot be presented as group element  in (\ref{N4finBRST}). The set of algebraic elements
$\widetilde{\mathcal{G}}(4) =\big\{\tilde{g}_{lin}(\lambda ({\Phi}_{(4)})):=1+\overleftarrow{s}^r\lambda_r({\Phi}_{(4)})\big\}$
forms a non-linear superalgebra which again corresponds to a set of formal group-like finite elements:
\begin{equation}
\tilde{G}(4)=\left\{ \tilde{g}\big(\lambda_r({\Phi}_{(4)})\big):\tilde{g}=1+  \sum_{e=1}^4\frac{1}{e!}\prod_{k=1}^e  \overleftarrow{s}{}^{r_k} \prod_{k=1}^e \lambda_{r_{e+1-k}}({\Phi}_{(4)}) \right\} ,
\label{tildeG4}
\end{equation}
The Jacobian of a change of variables:
${\Phi}{}^{A^t_4}_{(4)}\to {\Phi}{}^{\prime A^t_4}_{(4)}=
{\Phi}{}^{A^t_4}_{(4)}\tilde{g}\big(\lambda_r({\Phi}_{(4)})\big)$, in $\mathcal{M}^{(4)}_{tot}$, in the path integral ${Z}_{4|Y}(0)$ (\ref {Pintlocn4}) and in ${Z}_{4|Y}({J}_{(4)})$ (\ref{GFlocn4})
 generated by finite FD  $N=4$ BRST transformations may be calculated
explicitly  following to the same way as for the Jacobian (\ref{jacobianresN3}) in $N=3$ case:
\begin{equation}
\mathrm{sdet}\left\|{\Phi}{}^{A^t_4}_{(4)}\tilde{g}\big(\lambda_r({\Phi}_{(4)})\big)\frac{\overleftarrow{\delta}}{\delta {\Phi}{}^{B^t_4}_{(4)}} \right\| = \exp \Big\{-\mathrm{tr}_{G(4)}\ln \left( [e+m]^{r_1}_{r_2}\right) \Big\}, \ \mathrm{for} \ (e^{r_1}_{r_2}, m^{r_1}_{r_2})\equiv \big(\delta^{r_1}_{r_2}, \lambda_{r_2} \overleftarrow{s}{}^{r_1}\big),
\label{jacobianresN4}
\end{equation}%
where $\mathrm{tr}_{G(4)}$ denotes  trace over matrix $G(4)$-indices. The justification of the representation (\ref{jacobianresN4}) is based on the same points (\ref{jacobianN3YM})--(\ref{jacobianN3res})  as for its $N=3$ analog (\ref{jacobianresN3}), whose detailed calculation we leave out of the paper scope.

 For $\overleftarrow{s}{}^r$-potential, therefore functionally-dependent  parameters
\begin{equation}
\hat{\lambda}_{r_1}\big({\Phi}_{(4)}\big)=- \frac{1}{3!}\Lambda \big({\Phi}_{(4)}\big)\varepsilon
_{[r]_4}\overleftarrow{s}{}^{r_2}
\overleftarrow{s}{}^{r_3}\overleftarrow{s}{}^{r_4},\
\label{fdepparam4}
\end{equation}%
with an arbitrary potential being by  Grassmann-even-valued functional $\Lambda_{(4)} = \Lambda_{(4)}\big({\Phi}_{(4)}\big)$
the Jacobian (\ref{jacobianresN4}) reduces  to $N=4$ BRST exact functional determinant:
\begin{equation}
\mathbf{J}_{(4)}\big({\Phi}_{(4)}\big) = \mathrm{sdet}\left\|{\Phi}{}^{A^t_4}_{(4)}\tilde{g}\big(\hat{\lambda}_r({\Phi}_{(4)})\big)\frac{\overleftarrow{\delta}}{\delta {\Phi}{}^{B^t_4}_{(4)}} \right\| =  \Big\{1 + \frac{1}{4!}\Lambda_{(4)} ({\Phi}_{(4)}) \big(\overleftarrow{s}\big){}^{4}\Big\}^{-4}, \ \mathbf{J}_{(4)}({\Phi}_{(4)})\overleftarrow{s}{}^r = 0,
\label{jacobianresN4fd}
\end{equation}%
where we have used the property for tensor quantity $\prod_{k=1}^4\overleftarrow{s}{}^{r_k}$  to be  completely antisymmetric in  $(r_1,r_2,r_3, r_4)$  indices  that makes natural the definition:
\begin{equation}\label{s4pres}
\prod_{k=1}^4\overleftarrow{s}{}^{r_k}  = \frac{1}{4!}\varepsilon
^{[r]_4} \left(\overleftarrow{s}\right)^4 \ \ \mathrm{for} \ \  \left(\overleftarrow{s}\right)^4 \equiv     \prod_{k=1}^4\overleftarrow{s}{}^{r_k} \varepsilon
_{[r]_4}.
\end{equation}
Again,  for the  case of $\overleftarrow{s}^r$-closed parameters ${\lambda}_r$,
${\lambda}_{r_1}\overleftarrow{s}{}^{r_2} = 0$, including constant ${\lambda}_r$,
i.e., for $G(4)$ group elements, the Jacobian becomes trivial: $\mathbf{J}_{(4)}=1$, whereas for the infinitesimal FD quartet $\hat{\lambda}_r\big({\Phi}_{(4)}\big)$ (\ref{fdepparam4})
the Jacobian (\ref{jacobianresN4fd}) reduces to:
\begin{equation}
\mathbf{J}_{(4)}({\Phi}_{(4)}) = 1 - \frac{1}{3!}\Lambda_{(4)} ({\Phi}_{(4)})\big(\overleftarrow{s}\big){}^{4} +o(\Lambda_{(4)}) = \exp\Big\{- \frac{1}{3!}\Lambda_{(4)} ({\Phi}_{(4)})\big(\overleftarrow{s}\big){}^{4}\Big\}+o(\Lambda_{(4)}),
\label{jN4fdinf}
\end{equation}%
which immediately leads to  the gauge independence of the path integral ${Z}_{4|Y_{(4)}}(0)$ (and therefore for the conventional S-matrix) under small variation of the gauge condition: $Y_{(4)} \to Y_{(4)}+\delta Y_{(4)}$,  announced in  (\ref{N4GIinf}) because of
\begin{equation}\label{N4GI}
  {Z}_{4|Y_{(4)}+\delta Y_{(4)}}(0) =  \int  d {\Phi}_{(4)} \  \mathrm{sdet}\|{\Phi}{}^{\prime A^t_4} \overleftarrow{\partial}_{B^t_4} \| \exp \Big\{\frac{\imath}{\hbar}S_{Y_{(4)}+\delta Y_{(4)}}({\Phi}_{(4)})\Big\}=   Z_{4|Y_{(4)}}(0).
\end{equation}
according to the choice (\ref{N4FDP})   for $\delta Y_{(4)}$ in terms of $\Lambda_{(4)} ({\Phi}_{(4)})$ and therefore of  $\hat{\lambda}_r = \hat{\lambda}_r(\Lambda_{(4)})$
\begin{equation}\label{N4FDPder}
 \Lambda_{(4)} \big({\Phi}_{(4)}|\delta Y_{(4)}\big) =  -\frac{1}{4}\big(\imath / \hbar\big) \delta Y_{(4)}\big({\Phi}_{(4)}\big)   \Longrightarrow \hat{\lambda}_{r_1}(\Lambda_{(4)}) =   \frac{1}{4!}\big(\imath / \hbar\big) \delta Y_{(4)}  \prod_{k=2}^4\overleftarrow{s}{}^{r_k} \varepsilon_{[r]_4}.
\end{equation}

\section{Correspondence between the gauges, Ward identities, gauge dependence,
gauge-invariant Gribov--Zwanziger model.}
\label{NkWI}
 \setcounter{equation}{0}

Here we consider the physical properties of the respective $N=3$, $N=4$ finite BRST transformations, including  extended by sources (antifields) to the $N=3$ or $N=4 $ BRST transformations effective actions
  in the Subsection~\ref{NkWI1} and its applications in the Subsection~\ref{NkWIGZ} to the  Gribov--Zwanziger model \cite{Zwanziger} with gauge-invariant  horizon functional suggested in \cite{Pereira} with preservation of the local $N=1,2$ BRST invariance, shown in \cite{mr}, \cite{reshmosh}.

\subsection{FD Finite $N=3,4$ BRST Symmetry  for  Ward identities and Gauge Dependence Problem.}

\label{NkWI1}

First, let us study a relation that exists among the path integrals underlying $N=3$ BRST symmetry,
$Z_{3|\Psi_{(3)0}}(0)$ and $Z_{3|\Psi_{(3)0}+\Psi'_{(3)}}(0)$
in different admissible gauges,  one of which being described
by a Grassmann-odd  gauge functional $\Psi_{(3)0}$  corresponding to
the Landau gauge (\ref{N3gaugef}) for $\xi=0$. The other one $(\Psi_{(3)0}+\Psi'_{(3)})$
corresponds to any family from the gauges within the $\Psi_{(3)}(\widetilde{\Phi}_{(3)})$, including $R_\xi$-gauges
 for $\widehat{\Psi}_{(3)}=0$ in (\ref{N3gaugef}) and for $\chi(\mathcal{A},B)=(\partial^\mu A_{\mu}+\xi g^2 B=0) $ within the functional $\Psi_{(3)\xi}(\widetilde{\Phi}_{(3)})$.
To this end, we use a finite FD $N=3$ BRST transformation with
functionally-dependent parameters $\hat{\lambda}_{p_1}\big(\Lambda|\widetilde{\Phi}_{(3)}\big)$ (\ref{fdepparam}),  the $N=3$ BRST invariance of the quantum
action,  $
  S_{\Psi_{(3)\xi}}\big(\widetilde{\Phi}_{(3)}\big)$ (\ref{qexi3})  for $\xi=0$, and the form of the Jacobian, $\mathbf{J}_{(3)}\big(\widetilde{\Phi}_{(3)}\big)$,
(\ref{jacobianresN3fd}) of a corresponding change of variables,
$\widetilde{\Phi}_{(3)} \to \widetilde{\Phi}_{(3)}\tilde{g}(\hat{\lambda})$, given as follows
\begin{eqnarray}
 Z_{3|\Psi_{(3)0}}(0) & \hspace{-1em}\stackrel{\widetilde{\Phi}_{3)} \rightarrow \widetilde{\Phi}_{(3)}\tilde{g}(\hat{\lambda})}{=}& \hspace{-1em}\int  d \widetilde{\Phi}_{(3)}\;\exp\left\{  \frac{i}{\hbar}  \left[
 S_{\Psi_{(3)0}} +3 i\hbar\,\mathrm{\ln}\left(
1+ \frac{1}{3!}\Lambda (\widetilde{\Phi}_{(3)})(\overleftarrow{s})^{3}\right)  \right]  \right\} \nonumber \\
   &\hspace{-1em}=& \hspace{-1.8em} \int \hspace{-0.15em}d\widetilde{\Phi}_{(3)}\hspace{-0.15em}\exp\left\{ \hspace{-0.2em} \frac{i}{\hbar}  \left[\hspace{-0.1em}
 S_{\Psi_{(3)0}+\Psi'_{(3)}} \hspace{-0.2em}+ 3i\hbar\,\mathrm{\ln}\left(\hspace{-0.1em}
1+ \frac{1}{3!}\Lambda (\widetilde{\Phi}_{(3)})(\overleftarrow{s})^{3}\hspace{-0.15em}\right)
-\frac{1}{3!} \Psi'_{(3)}  (\overleftarrow{s})^{3} \hspace{-0.2em}\right]  \hspace{-0.2em}\right\}\hspace{-0.1em}.
\label{zfDtrans3}
\end{eqnarray}
The coincidence of the vacuum functionals $Z_{3|\Psi_{(3)0}}(0)$ and
$Z_{3|\Psi_{(3)0}+\Psi'_{(3)}}(0)$, evaluated with the respective fermionic
functionals $\Psi_{(3)0}$ and $\Psi_{(3)0}+\Psi'_{(3)}$, takes
place in case there holds a \emph{compensation equation}
for an unknown Fermionic functional $\Lambda=\Lambda(\widetilde{\Phi}_{(3)})$:
\begin{equation}
3\imath\hbar\,\mathrm{\ln}\left(
1+\frac{1}{3!}\Lambda (\overleftarrow{s})^{3}\right)
=\frac{1}{3!} \Psi'_{(3)} (\overleftarrow{s})^{3}  \Longleftrightarrow  \frac{1}{3!}\Lambda(\overleftarrow{s})^{3}=\exp\left(  -{\frac{\imath}{3\cdot 3!\hbar}%
} \Psi'_{(3)}(\overleftarrow{s})^{3}\right)-1 .
\label{eqexplN3}%
\end{equation}
The solution of equation (\ref{eqexplN3}) for an unknown  $\Lambda\big(
\widetilde{\Phi}_{(3)}\big)  $, which determines
$\hat{\lambda}_{p}\big(  \widetilde{\Phi}_{(3)}\big)  $,
according to (\ref{fdepparam}), with accuracy up to $N=3$ BRST exact terms, is given by%
\begin{equation}
\Lambda\big(\widetilde{\Phi}_{(3)}|\Psi'_{(3)}\big)=-\frac{\imath}{3\hbar}g(y)\Psi'_{(3)}\ ,\ \ \mathrm{for}%
\ \ g(y)=\left[  \exp(y)-1\right]  /y\ \ \mathrm{and}\ \ y\equiv-\frac
{i}{3\cdot3!\hbar}\Psi'_{(3)}\big(\overleftarrow{s}\big){}^{3}\ , \label{solcompeq3}%
\end{equation}
and therefore the corresponding triplet of field-dependent parameters have the form%
\begin{equation}
\hat{\lambda}_{p}\left( \widetilde{\Phi}_{(3)}|\Psi'_{(3)}\right)  = - \frac{i}{3!\hbar}g(y)
\Psi'_{(3)}\overleftarrow{s}^q\overleftarrow{s}^r \varepsilon_{pqr}  \, , \label{funcdeplafinN3}%
\end{equation}
whose approximation linear in $\Psi'_{(3)}$ is given    by
\begin{equation}
\hat{\lambda}_{p}\left(\widetilde{\Phi}_{(3)} |\Psi'_{(3)} \right)  =- \frac{i}{3!\hbar}\left(
\Psi'_{(3)}\overleftarrow{s}^q\overleftarrow{s}^r \varepsilon_{pqr} \right)  + {o}\big(\Psi'_{(3)}\big) ,
\label{funcdeplainfN3}%
\end{equation}
with opposite sign than in (\ref{N3FDP}) because of we  started here  from the gauge determined by $\Psi_{(3)0}$ instead of $\Psi_{(3)0}+\Psi'_{(3)}$  in (\ref{N3GIinf}).
Therefore, for any change $\Psi'_{(3)}$ of the gauge condition $\Psi_{(3)0}\to \Psi_{(3)0}+  \Psi'_{(3)}$,
we can construct a unique FD $N=3$ BRST transformation
with functionally-dependent parameters (\ref{funcdeplafinN3}) that allows one to
preserve the form of the path integral (\ref{zfDtrans3}) for the same Yang--Mills
theory.    On the another hand, if we consider the inverse form of compensation equation (\ref{eqexplN3})
for an unknown gauge variation $ \Psi'_{(3)}$ with a given $\Lambda\big( \widetilde{\Phi}_{(3)}\big) $,
we can present it in the form
\begin{equation}
3\cdot3!\imath\hbar\ln\left( 1+\frac{1}{3!}\Lambda(\overleftarrow{s})^3\right)
= \Psi'_{(3)}(\overleftarrow{s})^3\ \Longleftrightarrow\ 3\cdot3! \imath\hbar\left[
\sum_{n=1}\frac{-(-1)^{n}}{(3!)^{n} n}\left( \Lambda\big(\overleftarrow{s}\big)^3\right)
{}^{n-1} \Lambda\right] \big(\overleftarrow{s}\big)^3 =  \Psi'_{(3)}\big(\overleftarrow{s}\big)^3\,, \label{eqexpN3}%
\end{equation}
whose solution, with accuracy up to an $\overleftarrow{s}{}^{p}$-exact term,
is given by
\begin{equation}
\label{oppN3sol} \Psi'_{(3)}\big(\widetilde{\Phi}_{(3)}| \Lambda\big) = 3\cdot3!\imath\hbar\left[ \sum
_{n=1}\frac{-(-1)^{n}}{(3!)^{n} n}\left( \Lambda\big(\overleftarrow{s}\big)^{3}\right)
{}^{n-1} \Lambda\right]  = 3i\hbar\left[\hspace{-0.1em} \sum_{n=1}\frac{-(-1)^{n}%
}{3^{n-1} n}\left( \hat{\lambda}_{p}\overleftarrow{s}^{p}\right) {}^{n-1}
\Lambda\big(\widetilde{\Phi}_{(3)}\big)\right]\hspace{-0.1em} .
\end{equation}
Thereby, for any change of variables in the path integral $Z_{\Psi_{(3)0}}$
given by finite FD $N=3$ BRST transformations with the parameters
$\hat{\lambda}_{p} = \frac{1}{2}\Lambda \overleftarrow{s}^q\overleftarrow{s}^r \varepsilon_{pqr}$,
we obtain 
the same path integral $Z_{\Psi_{(3)0}+ \Psi'_{(3)}}$, evaluated, however,
in a gauge determined by the Fermionic functional $\Psi_{(3)0}+ \Psi'_{(3)}$,
in complete agreement with (\ref{oppN3sol}).

This latter, in particular, implies that we are able to reach any gauge condition
for the partition function within the $R_\xi$-like family of gauges, starting, e.g.,
from the Landau gauge and choosing:  $\Psi'_{(3)} =  \xi g^2 \int d^dx tr\, \big(\overline{C} B\big) $  (for $\xi=1$ in the Feynman gauge).

Making in $Z_{\Psi_{(3)}}(\widetilde{J}_{(3)})$ an FD $N=3$ BRST  transformation,
$\widetilde{\Phi}_{(3)} \to \widetilde{\Phi}_{(3)}\tilde{g}(\hat{\lambda})$
and using the relations (\ref{jacobianresN3fd}) and (\ref{solcompeq3}),
we obtain a \emph{modified Ward} (\emph{Slavnov--Taylor}%
) \emph{identity:}%
\begin{eqnarray}
&& \left\langle \exp\left\{ \frac{i}{\hbar}\widetilde{J}_{C^t_3}\widetilde{\Phi}{}^{C^t_3}_{(3)}\left[ \tilde{g}\big(\hat{\lambda}_p\big(\widetilde{\Phi}_{(3)}|\Lambda\big)\big)-1\right]\right\}  \left(  1+\frac{1}{3!}\Lambda(\overleftarrow
{s})^{3}\right)  {}^{-3}\right\rangle _{\Psi_{(3)},\widetilde{J}_{(3)}} =1, \label{mWIN3}%
\end{eqnarray}
where the  source-dependent average expectation value corresponding to a gauge-fixing
$\Psi_{(3)}\big(\widetilde{\Phi}_{(3)}\big)$, as in (\ref{GFGFWIN3}), explicitly for regular functional $L=L\big( \widetilde{\Phi}_{(3)}\big)$:
\begin{equation}
\left\langle L\right\rangle _{\Psi_{(3)},\widetilde{J}_{(3)}}=Z_{3|\Psi_{(3)0}}^{-1}\big(\widetilde{J}_{(3)}\big)\int d \widetilde{\Phi}_{(3)}
\ L  \exp\left\{  \frac{\imath}{\hbar}\left[
{{S}}_{\Psi_{(3)}}  +\widetilde{J}_{C_3^t} \widetilde{\Phi}{}^{C_3^t}_{(3)}\right]  \right\}
\ ,\ \ \mathrm{with\ \ }\left\langle 1\right\rangle _{\Psi_{(3)},J_{(3)}}=1\ . \label{aexvN3}%
\end{equation}
Due to the presence of $\Lambda\big(\widetilde{\Phi}_{(3)}\big)$, which implies functionally dependent
$\hat{\lambda}_{p}(\Lambda)$, the modified Ward identity depends on a choice
of the gauge Fermion $\Psi_{(3)}\big(\widetilde{\Phi}_{(3)}\big)$ for non-vanishing
$\widetilde{J}_{(3)}$, according to (\ref{solcompeq3}), (\ref{funcdeplafinN3}),
and therefore the corresponding Ward identities for Green's functions, obtained
by differentiating (\ref{mWIN3}) with respect to the sources, contain the functionals
$\hat{\lambda}_{p}(\Lambda)$ and their derivatives as weight functionals.
Due to (\ref{mWIN3}) for constant ${\lambda}_{p}$, the usual $G(3)$-triplet
of the Ward identities (\ref{GFGFWIN3}) for $Z_{3|\Psi_{(3)}}(\widetilde{J}_{(3)})$ follow
from the first order in $\lambda_{p}$.

Then, taking account of (\ref{funcdeplafinN3}),
we find that (\ref{mWIN3}) implies a relation which describes
the gauge dependence of $Z_{3|\Psi_{(3)}}(\widetilde{J}_{(3)})$ for a finite change
of the gauge, $\Psi_{(3)}\rightarrow \Psi_{(3)}+ \Psi'_{(3)}$:%
\begin{align}
Z_{3|\Psi_{(3)}+ \Psi'_{(3)}}\big(\widetilde{J}_{(3)}\big)  &  =Z_{3|\Psi_{(3)}}\big(\widetilde{J}_{(3)}\big)\left\langle\exp\left\{ \frac{i}{\hbar}\widetilde{J}_{C^t_3}\widetilde{\Phi}^{C^t_3}_{(3)}\left[ \tilde{g}\big(\hat{\lambda}_p\big(\widetilde{\Phi}_{(3)}|-\Psi'_{(3)}\big)\big)-1\right]\right\}
\right\rangle _{\Psi_{(3)},\widetilde{J}_{(3)}} , \label{GDInewN3}%
\end{align}
so that on the mass-shell for $Z_{3|\Psi_{(3)}}\big(\widetilde{J}_{(3)}\big)$:  $\widetilde{J}_{(3)}=0$, the path integral (and therefore
the conventional physical S-matrix) does not depend on the choice
of $\Psi'_{(3)}\big(\widetilde{\Phi}_{(3)}\big)$.

Let us introduce extended generating functionals of Green's functions
by means of sources $K_{C^t_3|p}$,  $K_{C^t_3|pq}=-K_{C^t_3|qp}$, $\overline{K}_{C^t_3}$,
($\epsilon(K_{C^t_3|p})=\epsilon(K_{C^t_3|pq})+1$ = $\epsilon(\overline{K}_{C^t_3})=\epsilon(\widetilde{\Phi}{}^{C^t_3})+1$),
introduced respectively to $N=3$ BRST variations $\widetilde{\Phi}^{C^t_3}_{(3)} \overleftarrow{s}{}^p$,
$\widetilde{\Phi}{}_{(3)}^{C^t_3} \overleftarrow{s}{}^p\overleftarrow{s}{}^q$, and $\widetilde{\Phi}{}_{(3)}^{C^t_3} (\overleftarrow{s})^3$:
\begin{eqnarray}
&& Z_{3|\Psi_{(3)}}\big(\widetilde{J}_{(3)}, K_{p},  K_{pq}, \overline{K}\big)=\int d\widetilde{\Phi}_{(3)}\ \exp\left\{
\frac{i}{\hbar}\left[  {S}_{\Psi_{(3)}}\big(\widetilde{\Phi}_{(3)}\big) +  K_{C^t_3|p} \widetilde{\Phi}_{(3)}^{C^t_3} \overleftarrow{s}^p + K_{C^t_3|pq} \widetilde{\Phi}_{(3)}^{C^t_3} \overleftarrow{s}^p\overleftarrow{s}^q \right.\right. \nonumber \\
&& \ \  \left.\left.+ \overline{K}_{C^t_3}  \widetilde{\Phi}_{(3)}^{C^t_3} \big(\overleftarrow{s}\big)^3%
+\widetilde{J}_{(3)}\widetilde{\Phi}_{(3)}\right]  \right\}    \ \ \mathrm{for}\ \ Z_{3|\Psi_{(3)}}\big(\widetilde{J}_{(3)},0,0,0\big)=Z_{3|\Psi_{(3)}}\big(\widetilde{J}_{(3)}\big).\label{ZPIextN3}%
\end{eqnarray}
If we make in (\ref{ZPIextN3}) a change of variables in the
extended space of $\big(\widetilde{\Phi}_{(3)}^{C_3^t}, K_{C_3^t|p},  K_{C_3^t|pq}, \overline{K}_{C_3^t}\big)$%
\begin{align}
& \widetilde{\Phi}_{(3)}^{C_3^t}\rightarrow\widetilde{\Phi}_{(3)}^{C_3^t} g(\lambda),&&  K_{{C_3^t}|p}\rightarrow K_{{C_3^t}|p},\label{extBRSTN3}\\
& K_{{C_3^t}|pq}\rightarrow K_{{C_3^t}|pq} + \frac{1}{2}\lambda_{[q}K_{{C_3^t}|p]}  , &&  \overline{K}_{C_3^t} \rightarrow \overline{K}_{C_3^t}  + \frac{1}{3!} \varepsilon^{pqr}\lambda_{r} \big(K_{{C_3^t}|pq} + \frac{1}{4}\lambda_{[q}K_{{C_3^t}|p]}\big)\nonumber
\end{align}
for $\widetilde{J}_{(3)}^{C_3^t}=0$, with finite constant parameters $\lambda_{p}$, we find that the
integrand in (\ref{ZPIextN3}) is unchanged, due to $(\overleftarrow{s})^4%
\equiv0$ , which means that the transformations
(\ref{extBRSTN3}) are \emph{extended} $N=3$ \emph{finite BRST transformations} for
the functional $Z_{3|\Psi_{(3)}}\big(\widetilde{J}_{(3)},K_{p},  K_{pq}, \overline{K}\big)$. For the linearized  in the parameters  $\lambda_p$  transformations (\ref{extBRSTN3}) the
integrand in (\ref{ZPIextN3}) is invariant with accuracy up to $o(\lambda)$ justifying to call  them as the algebraic \emph{extended} $N=3$ \emph{BRST transformations}.

Making in (\ref{ZPIextN3}) a change of variables, which corresponds only to
$N=3$ BRST transformations $\widetilde{\Phi}_{(3)}^{C^t_3}\rightarrow\widetilde{\Phi}_{(3)}^{C^t_3} \tilde{g}(\hat{\lambda})$
with an arbitrary functional $\hat{\lambda}_{p}(\widetilde{\Phi}_{(3)}) $ from (\ref{funcdeplafinN3}), we obtain a \emph{modified
Ward identity} for $Z_{3|\Psi_{(3)}}\big(\widetilde{J}_{(3)},K_{p},  K_{pq}, \overline{K}\big)$:
\begin{align}
& \left\langle \exp\left\{ \frac{\imath}{\hbar}\Big[\widetilde{J}_{C^t_3}\widetilde{\Phi}^{C^t_3}_{(3)}\left[ \tilde{g}\big(\hat{\lambda}(\widetilde{\Phi}_{(3)}|\Lambda)\big)-1\right]+  K_{C^t_3|p} (\widetilde{\Phi}_{(3)}^{C^t_3}) \overleftarrow{s}{}^p\left[ \tilde{g}\big(\hat{\lambda}(\widetilde{\Phi}_{(3)}|\Lambda)\big)-1\right]  \right.\right.\nonumber \\
& \ \left.\left.+ \frac{1}{3!}\varepsilon^{pqr}K_{C^t_3|pq} (\widetilde{\Phi}_{(3)}^{C^t_3}) \big(\overleftarrow{s}\big)^3\hat{\lambda}_r\Big]\right\}  \left(  1+\frac{1}{3!}\Lambda\big(\overleftarrow
{s}\big)^{3}\right)  {}^{-3}\right\rangle _{\Psi_{(3)},\widetilde{J}_{(3)},K_p, K_{pq}, \overline{K}} =1\ ,\label{mWIextN3}%
\end{align}
where the symbol \textquotedblleft$\langle{L}\rangle_{\Psi_{(3)},\widetilde{J}_{(3)},K_p, K_{pq}, \overline{K}}$\textquotedblright\ for any ${L}%
=L\big(\widetilde{\Phi}_{(3)}, K_p, K_{pq}, \overline{K}\big)$ stands for a source-dependent average expectation value
for a gauge $\Psi_{(3)}$ in the presence of sources (extended Zinn--Justin fields)
$K_{C_3^t|p},  K_{C_3^t|pq}, \overline{K}_{C^t_3}$:
\begin{align}
& \left\langle L\right\rangle _{\Psi_{(3)},\widetilde{J}_{(3)},\mathbf{K}_{(3)}}  \  =\ Z_{3|\Psi_{(3)}}^{-1}\big(\widetilde{J}_{(3)},\mathbf{K}_{(3)}\big)
\int d\widetilde{\Phi}_{(3)}\ L\exp\left\{  \frac{\imath}{\hbar}\left[ {S}_{\Psi_{(3)}}(\widetilde{\Phi}_{(3)},\mathbf{K}_{(3)}\big) %
+\widetilde{J}_{(3)}\widetilde{\Phi}_{(3)}\right]
\right\}  ,\label{aexvextN3}\\
& \mathrm{with}\ \ {S}_{\Psi_{(3)}}\big(\widetilde{\Phi}_{(3)},\mathbf{K}_{(3)}\big) ={S}_{\Psi_{(3)}}\big(\widetilde{\Phi}_{(3)}\big)+  K_{C^t_3|p} \widetilde{\Phi}_{(3)}^{C^t_3}\overleftarrow{s}{}^p + K_{C^t_3|pq} \widetilde{\Phi}_{(3)}^{C^t_3} \overleftarrow{s}{}^p\overleftarrow{s}{}^q  + \overline{K}_{C^t_3}  \widetilde{\Phi}_{(3)}^{C^t_3} \big(\overleftarrow{s}\big)^3 ,
\nonumber
\end{align}
for $\mathbf{K}_{(3)}\equiv \big(K_p, K_{pq}, \overline{K}\big)$. We can see that (\ref{mWIN3}) and (\ref{mWIextN3}) differ by definitions
(\ref{aexvN3}) and (\ref{aexvextN3}), as well as by the presence of terms
proportional to the sources $K_{C^t_3|p}, K_{C^t_3|pq}$, except for the Jacobian.

For constant parameters $\lambda_{p}$, we deduce from (\ref{mWIextN3}),
in the first order in $\lambda_{p}$
\begin{align}
&\left\langle \widetilde{J}_{C^t_3}\widetilde{\Phi}^{C^t_3}_{(3)} \overleftarrow{s}{}^p +  K_{C^t_3|q} \widetilde{\Phi}_{(3)}^{C^t_3} \overleftarrow{s}{}^q \overleftarrow{s}{}^p + \frac{1}{3!}\varepsilon^{pqr}K_{C^t_3|qr} \widetilde{\Phi}_{(3)}^{C^t_3} \big(\overleftarrow{s}\big)^3  \right\rangle _{\Psi_{(3)},\widetilde{ J}_{(3)},K_p, K_{pq}, \overline{K}} =0\ ,\label{WIN3ext}
\end{align}
Identities (\ref{WIN3ext}) can be presented as
\begin{align}
&   \Big[\widetilde{J}_{C^t_3}\frac{\overrightarrow{\delta}}
{\delta K_{C^t_3|p}} -
 K_{C^t_3|q}
\frac{\overrightarrow{\delta}}{\delta K_{C^t_3|pq}}+ \frac{1}{3!}
\varepsilon^{pqr} K_{C^t_3|qr}
\frac{\overrightarrow{\delta}}{ \delta \overline{K}_{C^t_3}} \Big]
\ln  Z_{3|\Psi_{(3)}}\big(\widetilde{J}_{(3)}, K_{p},  K_{pq}, \overline{K}\big)%
=0\ .\label{WIN3exteq}
\end{align}
Let us consider an extended generating functional of vertex Green's functions,
$\Gamma\big(\langle\widetilde{\Phi}_{(3)}\rangle, K_{p},  K_{pq}, \overline{K}\big)$, being a functional
Legendre transform of $\ln Z_{3|\Psi_{(3)}}\big(\widetilde{J}_{(3)}, K_{p},  K_{pq}, \overline{K}\big)$
with respect to the sources $\widetilde{J}_{(3)}$:%
\begin{align}
&  \Gamma\big(\langle\widetilde{\Phi}_{(3)}\rangle, K_{p},  K_{pq}, \overline{K}\big)=\frac{\hbar}{i}\ln
Z_{3|\Psi_{(3)}}\big(\widetilde{J}_{(3)}, K_{p},  K_{pq}, \overline{K}\big) -
\widetilde{J}_{C^t_3} \langle\widetilde{\Phi}^{C^t_3}_{(3)}\rangle ,\label{avfields1N3}\\
& \mathrm {where}\ \widetilde{J}_{C^t_3} = -\Gamma\big(\langle\widetilde{\Phi}_{(3)}\rangle, K_{p},  K_{pq}, \overline{K}\big)\frac{\overleftarrow{\delta}}{\delta\langle\widetilde{\Phi}_{(3)}^{C^t_3}\rangle} \ \mathrm{and } \ \langle\widetilde{\Phi}^{C^t_3}\rangle=\frac{\hbar}{i}\frac{\overrightarrow{\delta}}{\delta \widetilde{J}_{C^t_3}}\ln
Z_{3|\Psi_{(3)}}(\widetilde{J}_{(3)}, K_{p},  K_{pq}, \overline{K}).\label{avfields2N3}%
\end{align}
From (\ref{WIN3exteq})--(\ref{avfields2N3}), we deduce for
$\Gamma_{(3)}=\Gamma\big(\langle\widetilde{\Phi}_{(3)}\rangle, K_{p},  K_{pq}, \overline{K}\big)$
an $G(3)$-triplet of independent Ward identities:%
\begin{equation}
   \frac{\Gamma_{(3)} \overleftarrow{\delta}}{\delta\langle\widetilde{\Phi}_{(3)}^{C^t_3}\rangle} \frac{\overrightarrow{\delta}\Gamma_{(3)}}
{\delta K_{C^t_3|p}} +
 \Big\{ K_{C^t_3|q}
\frac{\overrightarrow{\delta}}{\delta K_{C^t_3|pq}}- \frac{1}{3!}
\varepsilon^{pqr} K_{C^t_3|qr}
\frac{\overrightarrow{\delta}}{ \delta \overline{K}_{C^t_3}}\Big\}\Gamma_{(3)}   \ = \  \frac{1}{2}\left(  \Gamma_{(3)}, \, \Gamma_{(3)}\right) _{(3)}^{p}+ V_{(3)}^{p}\Gamma_{(3)}=0,  \label{meq3}%
\end{equation}
for $p=1,2,3,$  in terms of $G(3)$-triplets of extended antibrackets, $\left(  \bullet , \, \bullet\right)_ {(3)}^{p}$,  and operators $V_{(3)}^{p}$, extending
the familiar $\mathrm{Sp}\left(2\right)$-covariant Lagrangian quantization for gauge
theories \cite{bltsp2, BLT2} (see also \cite{henneauxsp2, hullsp2, Barnich} as well as  \cite{MR2, MR3, MR4}) in the $N=2$ case, introduced for general gauge theories
\begin{equation}
\left(  F,G\right)_{(3)} ^{p}= F\left(  \frac{\overleftarrow{\delta}%
}{\delta\widetilde{\Phi}_{(3)}^{C^t_3}}\frac{\overrightarrow{\delta}}{\delta K_{C^t_3|p}}-\frac{\overleftarrow{\delta}}{\delta K_{C^t_3|p}}%
\frac{\overrightarrow{\delta}}{\delta\widetilde{\Phi}_{(3)}^{C^t_3}}\right)  G\ ,\ \ \ V_{(3)}^{p}%
=K_{{C^t_3}|q}\frac{\overrightarrow{\delta}%
}{\delta{K_{{C^t_3}|pq}}}- \frac{1}{3!} \varepsilon^{pqr} K_{{C^t_3}|qr} \frac{\overrightarrow{\delta}%
}{\delta \overline{K}_{C^t_3}}  \label{eN3br}%
\end{equation}
for any functionals $F,G$ with omitting the sign of averaging for the fields $\widetilde{\Phi}{}_{(3)}^{C^t_3}$ and with the usual convention: $\big({\overrightarrow{\delta}}/{\delta K_{C^t_3|pp_1}}\big)K_{D^t_3|qq_1} = (1/2) \delta^{[p}_q\delta^{p_1]}_{q_1}\delta^{C^t_3}_{D^t_3}$. Note that the algebra
of operators $V^p_{(3)}$ repeats, by the construction of the extended $N=3$ BRST transformations (\ref{extBRSTN3}),  the algebra of generators $\overleftarrow{s}{}^p$, i.e.,
$V_{(3)}^{\{p}V_{(3)}^{q\}}=0$.

The Ward identities (\ref{meq3}) are interesting as they remind of the behavior
of the extended quantum action ${S}_{\Psi_{(3)}}\big(\widetilde{\Phi}_{(3)},
K_p, K_{pq}, \overline{K}\big)$ (\ref{aexvextN3}) -- being the tree approximation for the extended
effective action $\Gamma_{(3)}$ within the loop expansion -- and serve as generating equations
for a corresponding $G(3)$-covariant method of Lagrangian quantization, covering the case
more general than a gauge group.

In turn, the case of  $N=4$ finite BRST transformations permits to get the same results with some peculiarities. We  restrict ourselves by only derivation of the  respective modified Ward identity and description of the gauge dependence problem, being based on the solution of the compensation equation from the change of variables in ${Z}_{4|Y}(0)$ (\ref{Pintlocn4}) generated by FD $N=4$  BRST transformations with quartet of the parameters $\hat{\lambda}{}_r(\Phi_{(4)})$ (\ref{fdepparam4}) with jacobian $\mathbf{J}_{(4)}({\Phi}_{(4)})$ (\ref{jacobianresN4fd})
\begin{equation}
4\imath\hbar\,\mathrm{\ln}\left(
1+\frac{1}{4!}\Lambda(\Phi_{(4)}) (\overleftarrow{s})^{4}\right)
= - \frac{1}{4!} Y'_{(4)} (\overleftarrow{s})^{4}  \Longleftrightarrow  \frac{1}{4!}\Lambda(\Phi_{(4)})(\overleftarrow{s})^{4}=\exp\left(  {\frac{\imath}{4\cdot 4!\hbar}%
} Y'_{(4)}(\overleftarrow{s})^{4}\right)-1 .
\label{eqexplN4}%
\end{equation}
to guarantee the coincidence of the path integrals, ${Z}_{4|Y}(0)$, (\ref{Pintlocn4}) and ${Z}_{4|Y+Y'}(0)$ evaluated in different admissible gauges corresponding to  the Bosonic gauge functionals $Y_{(4)}\big(\Phi_{(4)}\big)$ (e.g. for the Landau gauge $Y^0_{(4)0}$ (\ref{gBn4xi})) and, $Y_{(4)}+Y'_{(4)}$, (e.g. for the Feynman gauge $Y^0_{(4)\xi}$ (\ref{gBn4xi}) within $R_\xi$-like gauges for $\xi=1$) for finite $Y'_{(4)}\big(\Phi_{(4)}\big)$.
The solution of  (\ref{eqexplN4}) for an unknown  $\Lambda\big(
{\Phi}_{(4)}\big)  $ and hence of
$\hat{\lambda}_{r}\big(  {\Phi}_{(4)}\big)  $,
 with accuracy up to $N=4$ BRST exact terms, is given in terms of  the function $g(z)$ (\ref{solcompeq3})
\begin{eqnarray}
&& \Lambda\big({\Phi}_{(4)}|Y'_{(4)}\big)=\frac{\imath}{4\hbar}g(z)Y'_{(4)}\ ,\ \ \mathrm{for}%
\ z\equiv \frac
{i}{4\cdot 4!\hbar}Y'_{(4)}\big(\overleftarrow{s}\big){}^{4}\ , \label{solcompeq4}
\\
&&  \hat{\lambda}_{r_1}\left( {\Phi}_{(4)}|Y'_{(4)}\right)  = - \frac{i}{4!\hbar}g(z)
Y'_{(4)}\prod_{k=2}^4\overleftarrow{s}{}^{r_k} \varepsilon_{[r]_4}  \, , \label{funcdeplafinN4} %
\end{eqnarray}
whose approximation linear in $Y'_{(4)}$ coincide with (\ref{N4FDP}) with  opposite sign because of we  started here  from the gauge determined by $Y_{(4)}$  instead of  $Y_{(4)}+Y'_{(4)}$ in  (\ref{N4GIinf}).
From the inverse form of compensation equation (\ref{eqexplN4})
for an unknown gauge variation $ Y'_{(4)}$ with a given $\Lambda\big( {\Phi}_{(4)}\big) $:
\begin{equation}
4!4\imath\hbar\ln\left( 1+\frac{1}{4!}\Lambda(\overleftarrow{s})^4\right)
= -Y'_{(4)}(\overleftarrow{s})^4\ \Longleftrightarrow\ 4!4 \imath\hbar\left[
\sum_{n=1}\frac{-(-1)^{n}}{(4!)^{n} n}\left( \Lambda\big(\overleftarrow{s}\big)^4\right)
{}^{n-1} \Lambda\right] \big(\overleftarrow{s}\big)^4 =  -Y'_{(4)}\big(\overleftarrow{s}\big)^4, \label{eqexpN4}%
\end{equation}
we find  with accuracy up to an $\overleftarrow{s}{}^{r}$-exact term,
that
\begin{equation}
\label{oppN4sol} Y'_{(4)}\big({\Phi}_{(4)}| \Lambda\big) = 4\cdot4!\imath\hbar\left[ \sum
_{n=1}\frac{(-1)^{n}}{(4!)^{n} n}\left( \Lambda\big(\overleftarrow{s}\big)^{4}\right)
{}^{n-1} \Lambda\right]  = 4i\hbar\left[\hspace{-0.1em} \sum_{n=1}\frac{(-1)^{n}%
}{4^{n-1} n}\left( \hat{\lambda}_{r}\overleftarrow{s}^{r}\right) {}^{n-1}
\Lambda\big({\Phi}_{(4)}\big)\right]\hspace{-0.1em} .
\end{equation}
Thus, for any change of variables in the path integral $Z_{4|Y_{(4)}}$
given by finite FD $N=4$ BRST transformations with the parameters
$\hat{\lambda}_{r}$ (\ref{fdepparam4}),
we obtain 
the same path integral $Z_{4|Y_{(4)}+ Y'_{(4)}}$, evaluated, however,
in a gauge determined by the Bosonic functional $Y_{(4)}+ Y'_{(4)}$.

Making in $Z_{4|Y_{(4)}}({J}_{(4)})$ an FD $N=4$ BRST  transformation,
${\Phi}_{(4)} \to {\Phi}_{(4)}\tilde{g}(\hat{\lambda})$
and using the relations (\ref{jacobianresN4fd}), (\ref{solcompeq4}) and (\ref{funcdeplafinN4}),
we obtain a $N=4$ \emph{modified Ward} (\emph{Slavnov--Taylor}%
) \emph{identity:}%
\begin{eqnarray}
&& \left\langle \exp\left\{ \frac{i}{\hbar}{J}_{C^t_4}{\Phi}{}^{C^t_4}_{(4)}\left[ \tilde{g}\big(\hat{\lambda}_r\big({\Phi}_{(4)}|\Lambda\big)\big)-1\right]\right\}  \left(  1+\frac{1}{4!}\Lambda(\overleftarrow
{s})^{4}\right)  {}^{-4}\right\rangle _{Y_{(4)},{J}_{(4)}} =1\ . \label{mWIN4}%
\end{eqnarray}
where the  source-dependent average expectation value corresponding to a gauge-fixing
$Y_{(4)}\big({\Phi}_{(4)}\big)$ is determined as in (\ref{aexvN3}) for $N=3$ case.
Due to  $\Lambda\big({\Phi}_{(4)}\big)$, which implies functionally dependent
$\hat{\lambda}_{r}(\Lambda)$, the modified Ward identity depends on a choice
of the gauge Boson $Y_{(4)}\big({\Phi}_{(4)}\big)$ for non-vanishing
${J}_{(4)}$, according to (\ref{solcompeq4}), (\ref{funcdeplafinN4}) with the same as for $N=3$ case interpretation for the  modified Ward identities for the Green functions.
Due to (\ref{mWIN4}) for constant ${\lambda}_{r}$, the usual $G(4)$-quartet
of the Ward identities (\ref{GFGFWIN4}) for $Z_{4|Y_{(4)}}({J}_{(4)})$ follow
from the first order in $\lambda_{r}$.

Then, taking account of (\ref{funcdeplafinN4}),
we find that (\ref{mWIN4}) implies a relation which describes
the gauge dependence of $Z_{4|Y_{(4)}}({J}_{(4)})$ for a finite change
of the gauge, $Y_{(4)}\rightarrow Y_{(4)}+ Y'_{(4)}$:%
\begin{align}
Z_{4|Y_{(4)}+ Y'_{(4)}}({J}_{(4)})  &  =Z_{4|Y_{(4)}}({J}_{(4)})\left\langle\exp\left\{ \frac{i}{\hbar}{J}_{C^t_4}{\Phi}^{C^t_4}_{(4)}\left[ \tilde{g}\big(\hat{\lambda}_r\big({\Phi}_{(4)}|-Y'_{(4)}\big)\big)-1\right]\right\}
\right\rangle _{Y_{(4)},{J}_{(4)}} , \label{GDInewN4}%
\end{align}
so that on the mass-shell for $Z_{4|Y_{(4)}}\big({J}_{(4)}\big)$:  ${J}_{(4)}=0$, the path integral (and therefore
the conventional physical S-matrix) does not depend on the choice
of $Y'_{(4)}\big({\Phi}_{(4)}\big)$.

\subsection{Gauge-independent Gribov-Zwanziger model with local $N=3,4$ BRST symmetries}

\label{NkWIGZ}

Finally, we turn to the Gribov copies problem \cite{Gribov} within the Gribov--Zwanziger model
\cite{Zwanziger}  with a gauge-invariant horizon functional, $H(\mathcal{A}^h)$, recently
proposed to be added to an $N=1$ BRST invariant Yang--Mills quantum action \cite{Pereira}
in Landau gauge with the use of the gauge-invariant (thereby, invariant with respect
to a local $N=1,2$ BRST invariance, as it was shown in \cite{reshmosh}, Eq. (36)--(40))
transverse fields $\mathcal{A}^h_\mu = (\mathcal{A}^{h})^{n}_{\mu} t^n$ \cite{SemenovTyanshan}:
\begin{eqnarray}\label{gitrans}
&\hspace{-1.0em}&  \hspace{-1em} \textstyle \mathcal{A}_{\mu}\hspace{-0.1em}= \hspace{-0.1em}\mathcal{A}^h_{\mu}+  \mathcal{A}^L_{\mu}:  \mathcal{A}^h_{\mu}\hspace{-0.1em}
=\hspace{-0.1em}(\eta_{\mu\nu}- \frac{\partial_\mu\partial_\nu}{\partial^2})\Big(\mathcal{A}^{\nu} -\imath g \big[\frac{\partial \mathcal{A}}{\partial^2},
\mathcal{A}^\nu - \frac{1}{2} \partial^\nu\frac{\partial \mathcal{A}}{\partial^2}\big]  \Big)+ \mathcal{O}(\mathcal{A}^3)\hspace{-0.1em} : \, \mathcal{A}^h_{\mu}\overleftarrow{s}^p=0 ,\\
&\hspace{-1.0em}& \hspace{-1em} \label{FuncM}
H(\mathcal{A}^h)=\gamma^2\int d^dx\big( d^dy f^{mnk}(\mathcal{A}^h)^n_{\mu}(x)(M^{-1})^{ml}(\mathcal{A}^h;x,y)f^{ljk} (\mathcal{A}^h)^{j\mu}(y)\ +\ d(\hat{N}^2{%
-}1)\big).
\end{eqnarray}
Note, that the systematic study for the original   Gribov--Zwanziger model
\cite{Zwanziger} with not BRST-invariant horizon, $H(\mathcal{A})$, within  Lagrangian BRST quantization of gauge theories \cite{bv}, \cite{ht}    from  the viewpoint of so-called soft BRST symmetry breaking was initiated in \cite{llr1}.
 Then, as in the case of $N=1,2$ BRST symmetry, the gauge and $N=1,2$ BRST invariant extension
of the respective quantum Yang--Mills action within the $R_\xi$-family of gauges with a gauge
fermion $\Psi_\xi$ and a boson $Y_\xi$ prescribed by the Gribov--Zwanziger actions are given by
\begin{eqnarray}
\label{brstinvgzN1} &&\textstyle  \hat{S}_{GZ}(\Phi) \hspace{-0.1em}= \hspace{-0.1em}S_0
+ \Psi_\xi\overleftarrow{s} \hspace{-0.1em}+ H(\mathcal{A}^h) ,\ \
\mathrm{for } \ \  \hat{S}_{GZ}(\Phi(1+\overleftarrow{s}\mu))=\hat{S}_{GZ}(\Phi),\\
\label{brstinvgzN2} &&\textstyle  \hat{S}_{GZ}(\Phi_{(2)}) \hspace{-0.1em}= \hspace{-0.1em}S_0
-\frac{1}{2}Y_\xi \overleftarrow{s}^a\overleftarrow{s}_a+ H(\mathcal{A}^h) ,\ \
\mathrm{for } \ \   \hat{S}_{GZ}(\Phi_{(2)}g(\mu_a))=\hat{S}_{GZ}(\Phi_{(2)}),
\end{eqnarray}
with allowance made for (\ref{PintfpLloc}),
(\ref{N1gaugef}) and (\ref{PintfpLlocN2}), (\ref{Y(A,C)})
the same may be done in  $N=3$ and $N=4$  BRST invariant formulations of the respective quantum actions
$ S_{\Psi_{(3)\xi}}\big(\widetilde{\Phi}_{(3)}\big)$ (\ref{qexi3}) and $S_{Y_{(4)\xi}}\big({\Phi}_{(4)}\big)$ (\ref{qexi4}).
Therefore, the $N=3$  and $N=4$  BRST invariant and gauge independent Gribov--Zwanziger
actions within $\Psi_{(3)\xi}$ and respectively within $Y_{(4)\xi}$-family of gauges related to $R_\xi$-gauges
are given by
\begin{eqnarray}\label{brstinvgzN3} &&\textstyle  \hat{S}_{GZ}\big(\widetilde{\Phi}_{(3)}\big) \hspace{-0.1em}= \hspace{-0.1em}S_0
+\frac{1}{3!}\Psi_{(3)\xi} \big(\overleftarrow{s}\big){}^3+ H(\mathcal{A}^h) ,\,
\mathrm{for } \,  \hat{S}_{GZ}\big(\widetilde{\Phi}_{(3)}g(\lambda_p)\big)=\hat{S}_{GZ}\big(\widetilde{\Phi}_{(3)}\big),\\
\label{brstinvgzN4} &&\textstyle  \hat{S}_{GZ}\big(\Phi_{(4)}\big) \hspace{-0.1em}= \hspace{-0.1em}S_0
-\frac{1}{4!}Y_{(4)\xi} \big(\overleftarrow{s}\big){}^{4}+ H(\mathcal{A}^h) ,\,
\mathrm{for } \,  \hat{S}_{GZ}\big(\Phi_{(4)}g(\lambda_r)\big)=\hat{S}_{GZ}\big(\Phi_{(4)}\big).
\end{eqnarray}
As in the case of the $N=1,2$ BRST symmetry, one may expect the unitarity of the theory within
the suggested $N=3$, $N=4$ BRST symmetry generalizations of the Faddeev--Popov quantization rules
\cite{fp}. These problems are under study.

The same results concerning the problems of unitarity and gauge-independence
may be achieved within the local formulations of Gribov--Zwanziger theory \cite{Zwanziger}
when the horizon functional is localized (in the path integral) by means of a quartet of auxiliary
fields $\phi_{\rm {aux}}=\big(\varphi^{mn}_\mu, \bar{\varphi}{}^{mn}_\mu;$
$\omega^{mn}_\mu, \bar{\omega}{}^{mn}_\mu \big)$, having opposite Grassmann parities,
$\epsilon(\varphi, \bar{\varphi})= \epsilon(\omega, \bar{\omega})+1=0$, and being
antisymmetric in $su(\hat{N})$ indices  $m,n$. We  suggest here the only $N=1$ BRST invariant formulation,
\begin{eqnarray}  \label{Sgamma3}
  && \hat{S}_{GZ}\big({\Phi}_{(1)}, \phi_{\rm {aux}}\big)  \ = \  S_0(\mathcal{A})
+\Psi_{\xi}\big({\Phi}_{(1)}\big) \overleftarrow{s}+  S_{\gamma}(\mathcal{A}^h,\phi_{\rm {aux}}). \\
 && S_{\gamma}\ = \ \int \hspace{-0.1em} d^dx\Big\{{\bar \varphi}^{mn}_\mu M^{ml}(\mathcal{A}^h) \varphi^{\mu{} l n}  - {\bar \omega}^{mn}_\mu  M^{m l}(\mathcal{A}^h) \omega^{\mu{} l n} \nonumber\\
 && \ \  + \gamma\,f^{m nl}(\mathcal{A}^h)^{\mu m}(\varphi_\mu^{nl}-\bar{\varphi}%
_\mu^{nl})+\gamma^2d
(\hat{N}{}^2-1)\Big\}\,, \label{Sgamma1}
\end{eqnarray}
with additional non-local  $N=1$ BRST transformations for the fields $\phi_{\rm {aux}}$ with untouched ones for ${\Phi}_{(1)}$ (\ref{N1BRST})
\begin{eqnarray}\label{N1phi}
&& \phi_{\rm {aux}} \overleftarrow{s} = \big(\varphi^{mn}_\mu, \bar{\varphi}{}^{mn}_\mu;
\omega^{mn}_\mu, \bar{\omega}{}^{mn}_\mu \big) \overleftarrow{s} = \big(0,\,\bar{\omega}^{mn}_\mu  ;\,\varphi^{mn}_\mu -\gamma \big(M^{-1}(\mathcal{A}^h)\big)^{mk}f^{knl} (\mathcal{A}^h)_{\mu}^l  ,\,0\big).
\end{eqnarray}%
The part $S_\gamma$ in case of  $N=3$ and $N=4$ BRST formulation for the quantum actions (as well as for the $N=2$ case) should be modified due to another spectra for the auxiliary fields $\phi_{\rm {aux}}$.

Finally, the non-local gauge-invariant transverse fields, $\mathcal{A}^h_{\mu}$, (\ref{gitrans}) can also be localized by using  complex $SU(\hat{N})$-valued auxiliary field, $h(x)$, with non-trivial own gauge  and $N=1$ BRST transformations   \cite{capri}   in order  to reach really localized Gribov-Zwanziger model still $N=1,2,3,4$ BRST invariant without Gribov ambiguity, whose  properties are now under study.

\section{On Feynman diagrammatic technique in $N=3$, $N=4$ BRST quantization}
\label{diagram34}
 \setcounter{equation}{0}

Here, we introduce some  new definitions to develop a Feynman diagrammatic technique for the Yang--Mills theory within suggested $N=3$ and $N=4$ BRST invariant formulations for the  non-renormalized  quantum actions $ S_{\Psi_{(3)\xi}}\big(\widetilde{\Phi}_{(3)}\big)$ given by  (\ref{qexi3})--(\ref{addxi3}), and    $S_{Y_{(4)\xi}}\big({\Phi}_{(4)}\big)$ determined by  (\ref{qexi4})--(\ref{addxi4}). To be complete, we compare the graphs  which contain additional lines  related
to new fictitious fields   to ones with known, i.e. ghost, $C(x)$, antighost, $\overline{C}(x)$, fields in $N=1$ BRST setup and with duplet of ghost-antighost  fields, $C^a(x)$, $a=1,2$ in $N=2$ BRST setup, having in mind that usually the Nakanishi-Lautrup field $B(x)$ is integrated out from the quantum actions.

We present the generating functionals of Green  functions in $R_\xi$-gauges  $
   Z(J)$ (\ref{GFGF}), $Z_Y(J)$ determined with the quantum action $S_{Y_\xi}(\Phi)$ (\ref{SY(A,C)}), ${Z}_{3|\Psi_{\xi}}(\widetilde{J})$ (\ref {GFlocn3}),  $
  {Z}_{4|Y_{\xi}}({J}_{(4)})$ (\ref{GFlocn4}) respectively for $N=1,2,3,4$ BRST symmetry   within the perturbation theory   according to \cite{bookfaddeevslavnov} but for $d$-dimensional space-time
 \begin{eqnarray}
&\hspace{-0.5em}& \hspace{-0.5em}  Z(J) =  \exp\Big\{V\Big(\frac{\hbar}{\imath} \frac{\delta}{\delta J_\mu};\frac{\hbar}{\imath} \frac{\delta}{\delta J},\frac{\hbar}{\imath} \frac{\delta}{\delta \overline{J}} \Big)\Big\}\exp\Big\{\frac{\imath}{2 \hbar} \int d^dx d^dy \,tr\, \Big[ J_{\mu}(x) D^{\mu\nu}(x-y)J_{\nu}(y)
\nonumber \\
&\hspace{-0.5em}& \hspace{-0.5em}  \phantom{Z(J)=} + 2\overline{J}(x)D(x-y)J(y)  \Big] \Big\},\label{GFGFps1}\\
&\hspace{-0.5em}& \hspace{-0.5em}   Z_{Y_{\xi}}(J)= \exp\Big\{V_{Y_\xi}\Big(\frac{\hbar}{\imath} \frac{\delta}{\delta J_\mu};\frac{\hbar}{\imath} \frac{\delta}{\delta J_a} \Big)\Big\}\exp\Big\{\frac{\imath}{2 \hbar} \int d^dx d^dy \,tr\, \Big[ J_{\mu}(x) D^{\mu\nu}(x-y)J_{\nu}(y) \nonumber \\
&\hspace{-0.5em}& \hspace{-0.5em}  \phantom{Z_{Y_{\xi}}(J)=}+ {J}_a(x)D^{ab}(x-y)J_b(y)  \Big] \Big\}, \label{GFGFps2} \\
&\hspace{-0.5em}& \hspace{-0.5em}   Z_{3|\Psi_{\xi}}(\widetilde{J})= \exp\Big\{V_{3|{\xi}}\Big(\frac{\hbar}{\imath} \frac{\delta}{\delta J_\mu};\frac{\hbar}{\imath} \frac{\delta}{\delta J_{(\overline{C})}},\frac{\hbar}{\imath} \frac{\delta}{\delta J_{(\widehat{B})}}; \frac{\hbar}{\imath} \frac{\delta}{\delta J^{(C)}_p},\frac{\hbar}{\imath} \frac{\delta}{\delta J^{(\widehat{B})}_{[p]_2}}; \frac{\hbar}{\imath} \frac{\delta}{\delta J^{(B)}_p},\frac{\hbar}{\imath} \frac{\delta}{\delta J^{({B})}_{[p]_2}} \Big)\Big\} \nonumber \\
&\hspace{-0.5em}& \hspace{-0.5em}  \phantom{Z_{3|\Psi_{\xi}}(\widetilde{J})=} \times \exp\Big\{\frac{\imath}{2 \hbar} \int d^dx d^dy \,tr\, \Big[ J_{\mu}(x) D^{\mu\nu}(x-y)J_{\nu}(y) + 2 J_{(\overline{C})}(x)D_{\overline{C}\widehat{B}}(x-y)J_{(\widehat{B})}(y)  \nonumber \\
&\hspace{-0.5em}& \hspace{-0.5em}  \phantom{Z_{3|\Psi_{\xi}}(\widetilde{J})=} +   J^{(C)}_{p_3}(x)D^{[p]_3}_{C\widehat{B}}(x-y)J^{(\widehat{B})}_{[p]_2}(y)+J^{(B)}_{p_3}(x)D^{[p]_3}_{BB}(x-y)J^{({B})}_{[p]_2}(y)   \Big] \Big\},\label{GFGFps3}\\
&\hspace{-0.5em}& \hspace{-0.5em}  \label{GFGFps4} Z_{4|Y_{\xi}}({J}_{(4)}) = \exp\Big\{V_{4|{\xi}}\Big(\frac{\hbar}{\imath} \frac{\delta}{\delta J_\mu};\frac{\hbar}{\imath} \frac{\delta}{\delta J^{(C)}_r},\frac{\hbar}{\imath} \frac{\delta}{\delta J^{({B})}_{[r]_3}}; \frac{\hbar}{\imath} \frac{\delta}{\delta J^{(B)}_{[r]_2}} \Big)\Big\}\exp\Big\{\frac{\imath}{2 \hbar} \int d^dx d^dy \,tr\, \Big[ J_{\mu}(x)\times \nonumber \\
&\hspace{-0.5em}& \hspace{-0.5em}  \phantom{Z_{4|Y_{\xi}}({J}_{(4)})} \hspace{-0.5em}  D^{\mu\nu}(x-y)J_{\nu}(y) \hspace{-0.1em}+ \frac{1}{3}J^{(C)}_{r_1}(x)D^{[r]_4}_{CB}(x\hspace{-0.1em}-y)J^{({B})}_{[r]_3}(y)\hspace{-0.1em}+\hspace{-0.1em}\frac{1}{4}J^{(B)}_{[r]_2}(x)
D^{[r]_4}_{BB}(x\hspace{-0.1em}-y)J^{({B})}_{r_3r_4}(y)    \hspace{-0.1em} \Big]\hspace{-0.1em} \Big\},
 \end{eqnarray}
for $Sp(2)$-duplet of  sources  $ J_a = \big( J_1, J_2\big) \equiv \big( J, \overline{J}\big)$ to ghost-antighost fields $C^a$, for  $\mathcal{G}(3)$-triplets of Grassmann-odd $ J^{(\widehat{B})}_{[p]_2}$, Grassmann-even $J^{({B})}_{[p]_2}$  and Grassmann-odd singlets $J_{(\overline{C})}, J_{(\widehat{B})}$ of sources .for the  respective fields $\widehat{B}{}^{[p]_2}, {B}{}^{[p]_2}, \overline{C}, \widehat{B}$ mentioned   in the round brackets in the indices and for $\mathcal{G}(4)$-quartets of Grassmann-odd $J^{(C)}_{r_1}, J^{({B})}_{[r]_3}$  and sextet of Grassmann-even $J^{(B)}_{[r]_2}$ sources .for the  fields ${C}^{r_1}$, ${B}^{[r]_3}$, ${B}^{[r]_2}$. The causal Green functions for the vector field $\mathcal{A}_\mu$: $D^{\mu\nu}(x)$ \cite{bookfaddeevslavnov} and for the respective fictitious pair of fields $D(x)$, $D^{ab}(x)$ for the fictitious Grassmann-odd fields in $N=2$;  for $D_{\overline{C}\widehat{B}}(x)$,  $D^{[p]_3}_{C\widehat{B}}(x)$ for Grassmann-odd,  $D^{[p]_3}_{BB}(x)$ for Grassmann-even   fields in $N=3$; $D^{[r]_4}_{CB}(x)$, $D^{[r]_4}_{BB}(x)$ respectively  for Grassmann-odd and Grassmann-even fields in $N=4$ cases are  determined  in terms of the Feynman propagators in momentum representation:
\begin{eqnarray}
  && D^{\mu\nu}(x)\  =\  \frac{1}{\big(2\pi\big)^d} \int d^dp \, e^{-\imath px}D^{\mu\nu}(p), \ \mathrm{for}  \ D^{\mu\nu}(p) =- \Big(\eta_{\mu\nu}-\frac{p_\mu p_\nu(1-\xi)}{p^2+\imath 0}\Big)\frac{1_{su(\hat{N})}}{p^2+\imath 0} , \label{FAprop} \\
  && D(x) \ = \  \frac{1}{\big(2\pi\big)^d} \int d^dp \, e^{-\imath px}D(p), \ \mathrm{for}  \ D(p) =\frac{1_{su(\hat{N})}}{p^2+\imath 0}, \ 1_{su(\hat{N})}\equiv \| \delta^{mn}\| , \label{Fghprop} \\
  && \Big(D^{ab}; D_{\overline{C}\widehat{B}},  D^{[p]_3}_{C\widehat{B}},  D^{[p]_3}_{BB}; D^{[r]_4}_{CB}, D^{[r]_4}_{BB}\Big)(x) \ = \  \Big(\varepsilon^{ab};  1,\, \varepsilon^{[p]_3},\, \varepsilon^{[p]_3};  \varepsilon^{[r]_4},\varepsilon^{[r]_4}\Big)D(x). \label{Fghpropg}
\end{eqnarray}
And  the respective vertexes look as
\begin{eqnarray}
&\hspace{-0.5em}& \hspace{-0.5em}  V\big(\mathcal{A}_\mu; {C}, \overline{C} \big) = \frac{1}{4} \int d^dx \, tr\, \Big\{ 2 \partial^{[\mu}\mathcal{A}^{\nu]}\big[\mathcal{A}_{\mu},\mathcal{A}_{\nu}\big] + \big[\mathcal{A}_{\mu},\mathcal{A}_{\nu}\big]^2 + 4 \overline{C}\partial^{\mu}\big[\mathcal{A}_{\mu}, C\big]\Big\}
,\label{Vps1}\\
&\hspace{-0.5em}& \hspace{-0.5em}   V_{Y_\xi}\big(\mathcal{A}_\mu; {C}^a \big)  =V\big(\mathcal{A}_\mu; {C}, \overline{C} \big)\big|_{({C}, \overline{C})\to C^a} -\frac{\xi }{ 24}  \int d^dx \, tr\,   \big[C^{a},%
C^{c}\big]\big[C^{b},C^{d}\big]\varepsilon_{ab}\varepsilon_{cd} , \label{Vps2} \\
&\hspace{-0.5em}& \hspace{-0.5em}   V_{3|{\xi}}\big(\mathcal{A}_\mu; \overline{C}, \widehat{B}; C^p, \widehat{B}^{[p]_2}; B^p, {B}^{[p]_2} \big) = V\big(\mathcal{A}_\mu; {C}, \overline{C} \big)\big|_{({C}, \overline{C})\to (\widehat{B}, \overline{C})} + \frac{1}{2} \int d^dx \, tr\, \Big\{ B^{p_1}\partial^{\mu}\big[\mathcal{A}_{\mu}, B^{p_2p_3}\big]  \nonumber \\
&\hspace{-0.5em}& \hspace{-0.5em} \phantom{V_{3|{\xi}}\big(\mathcal{A}_\mu; \overline{C}, \widehat{B}; C^p, \widehat{B}^{[p]_2}; B^p, {B}^{[p]_2} \big) =}+ \widehat{B}^{[p]_2}\partial^{\mu}\big[\mathcal{A}_{\mu}, C^{p_3}\big] \Big\}\varepsilon_{[p]_3} +  S_{\mathrm{add}(3)} ,\label{Vps3}\\
&\hspace{-0.5em}& \hspace{-0.5em}  \label{Vps4} V_{4|{\xi}}\big(\mathcal{A}_\mu; C^r, {B}^{[r]_3}; {B}^{[r]_2} \big)= V\big(\mathcal{A}_\mu; 0, 0 \big)+ \int d^dx \, tr\, \Big\{\frac{1}{8}{B}^{[r]_2}\partial^{\mu}\big[\mathcal{A}_{\mu}, B^{r_3r_4}\big] \nonumber \\
&\hspace{-0.5em}& \hspace{-0.5em}\phantom{V_{4|{\xi}}\big(\mathcal{A}_\mu; C^r, {B}^{[r]_3}; {B}^{[r]_2} \big)=} -\frac{1}{3!} C^{r_4}\partial^{\mu}\big[\mathcal{A}_{\mu}, B^{[r]_3}\big]   \Big\}\varepsilon_{[r]_4}+ S_{\mathrm{add}(4)},
 \end{eqnarray}
where each $su(\hat{N})$-commutator implicitly  contains interaction coupling $g$ as multiplier, all the integrations above satisfy to the Feynman boundary conditions  and the respective expressions (\ref{addxi3}), (\ref{addxi4}),   for $S_{\mathrm{add}(3)}$, $S_{\mathrm{add}(4)}$  were used.

The expansion of the functionals (\ref{GFGFps1})--(\ref{GFGFps4}) generates the respective diagrammatic techniques, known for $N=1$ BRST symmetric formulation (\ref{GFGFps1}), e.g. from \cite{bookfaddeevslavnov}. The basic elements for each $N=m, m=1,2,3,4$ we list in the momentum representation, first,
for $N=1$:

\noindent
{\footnotesize\begin{figure}[h]
\begin{picture}(0,2)
\put(4.0,-2.1){\includegraphics[scale=0.25]{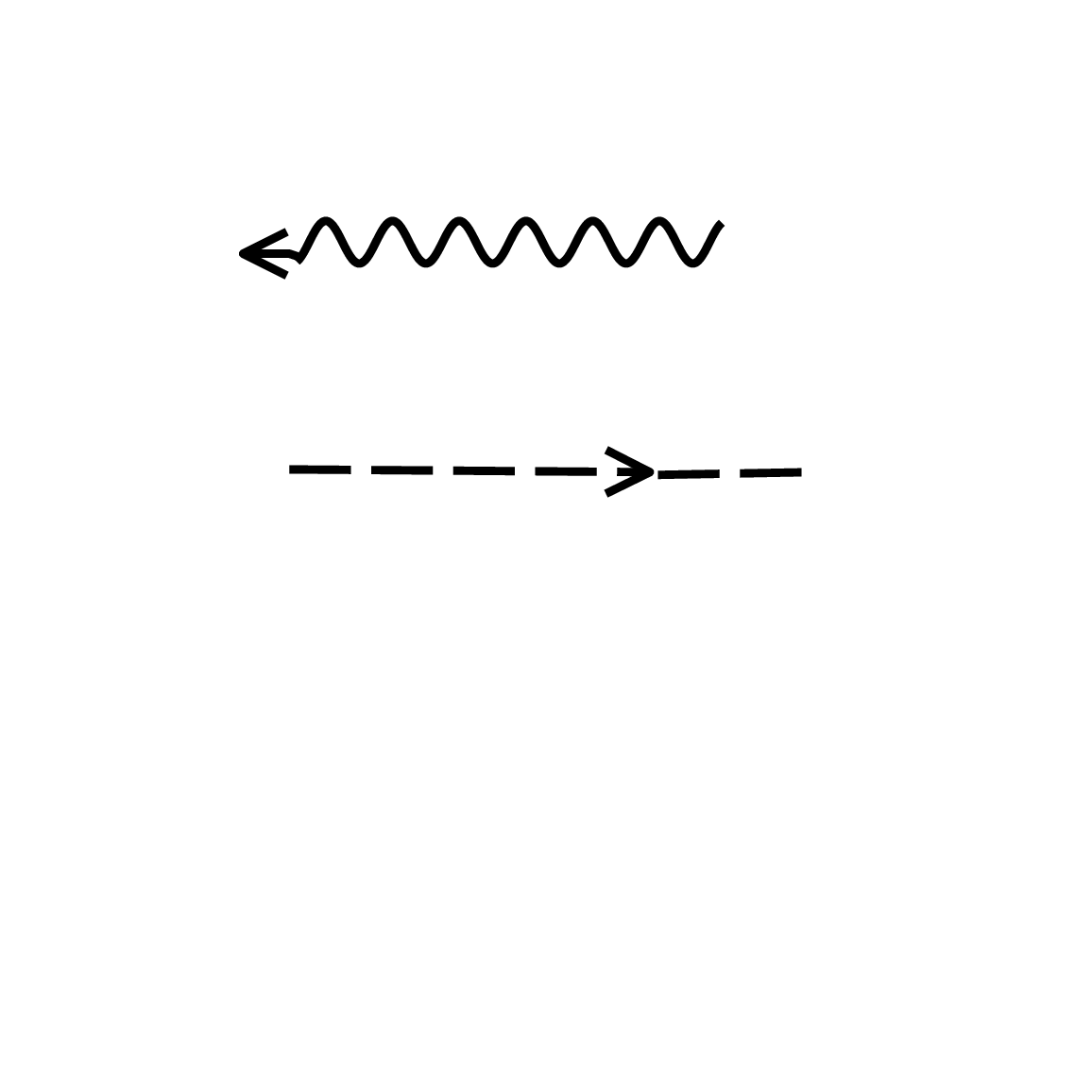}}
\put(2.6,1.6){$D^{\mu\nu}(p)\  \equiv$}
\put(2.8,0.6){$D(p)\ \ \ \equiv$}
\end{picture}
\caption{Propagators for the vector field $\mathcal{A}_\mu$ and for the ghost fields $\overline{C}, C$.}\label{Figure_1_propAcc}
\end{figure}}

\noindent
Second, for $N=2$ case for only different propagator for $Sp(2)$-duplet of ghost-antighost field $C^a$ and quartic in  $C^a, C^b$ ,$C^c$,$C^d$ ($a,b,c,d=1,2$) interaction vertex $V_{(Y_\xi)C^aC^bC^cC^d}(p)$ obtained from
\begin{equation}\label{intN2}
  tr V_{(Y_\xi)C^aC^bC^cC^d}(x)\big[C^{a}%
C^{c}\big]\big[C^{b},C^{d}\big]\hspace{-0.1em} = \hspace{-0.2em}  \frac{1}{\big(2\pi\big)^d} \hspace{-0.2em}\int \hspace{-0.15em}d^dp \, e^{-\imath px} V^{nln_1l_1}_{(Y_\xi)C^aC^bC^cC^d}(p)C^{l_1a}(p)%
C^{n_1c}(p)C^{lb}C^{nd}(p)
\end{equation}

\begin{figure}[h]
{\footnotesize\begin{picture}(0,3.5)
\put(8,-2.0){\includegraphics[scale=0.3]{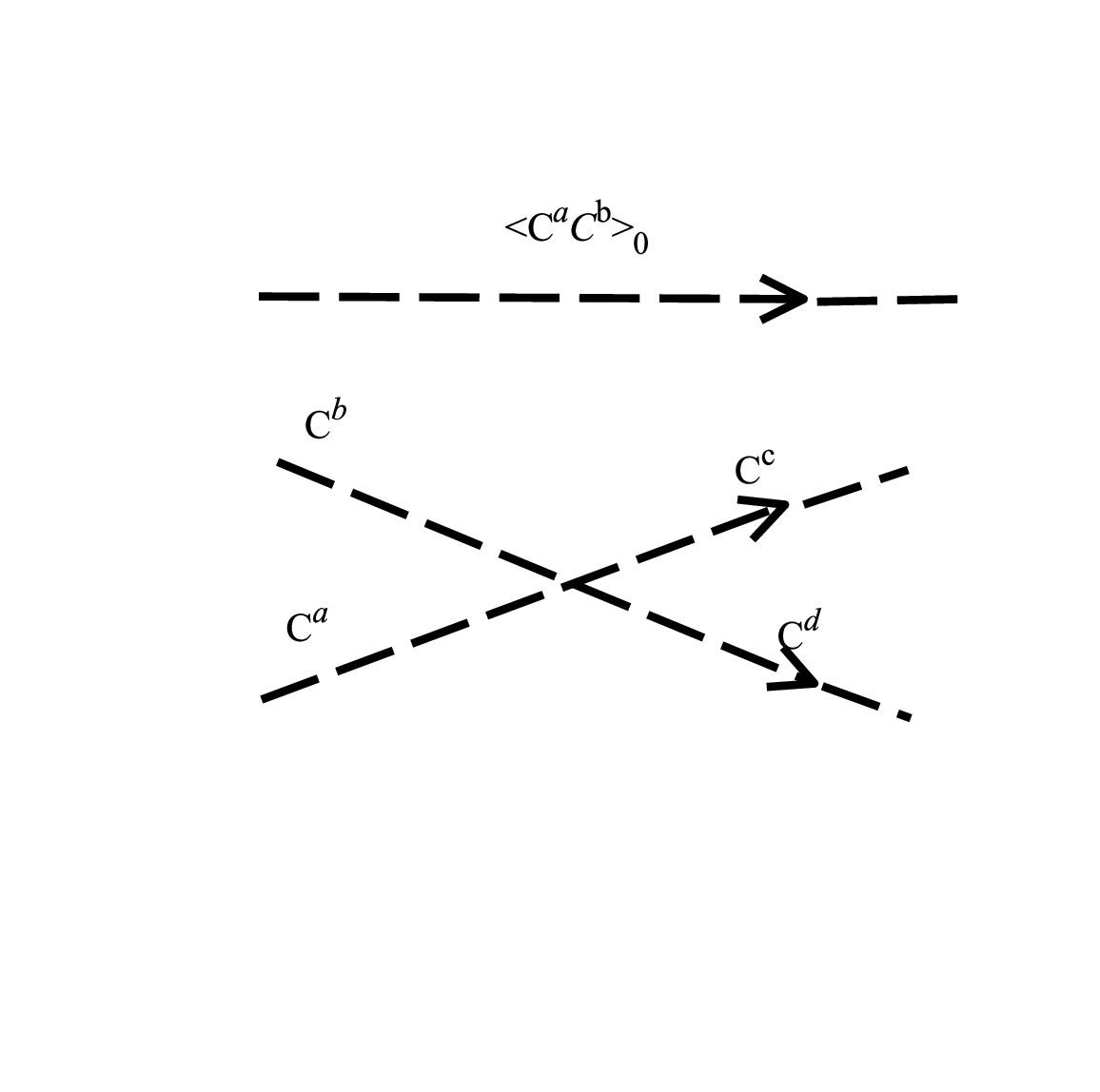}}
\put(4.3,2.2){$D^{ab}(p) = \langle C^a C^b\rangle_0\ \equiv$}
\put(1.5,0.5){$V^{nln_1l_1}_{(Y_\xi)C^aC^bC^cC^d}  = -\displaystyle\frac{\xi }{ 24}  f^{mnl}f^{mn_1l_1}\varepsilon_{ab}\varepsilon_{cd}\ \equiv$}
\end{picture}}
\caption{Propagators for the  fields ${C}^a$ and   self-interaction vertex quartic in the  ghost fields ${C}^a$.}\label{Figure_2_propvertccN2}
\end{figure}

\noindent
Third, for $N=3$ propagators for the fictitious fields   and with account for antisymmetry $(\widehat{B},B)^{qr} = -(\widehat{B},B)^{rq}$
\newpage
\begin{figure}[h]
{\footnotesize\begin{picture}(0,4)
\put(6,-2.0){\includegraphics[scale=0.4]{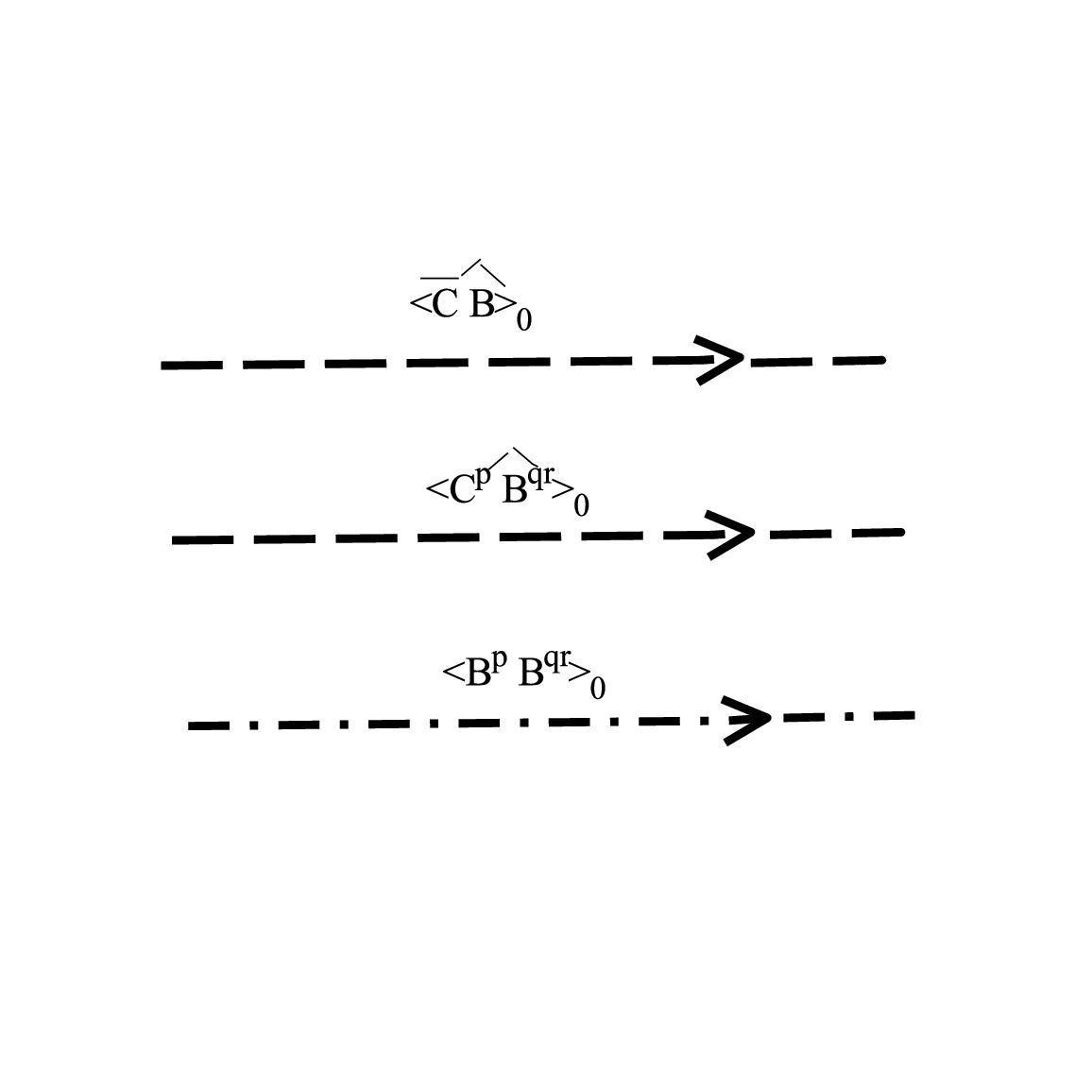}}
\put(3.1,3.1){$D_{\overline{C} \widehat{B}}(p) = \langle \overline{C} \widehat{B}\rangle_0\ \equiv$}
\put(2.8,1.95){$D^{pqr}_{C\widehat{B}}(p) = \langle {C}^p \widehat{B}{}^{qr}\rangle_0 \equiv$}
\put(2.8,0.50){$D^{pqr}_{B{B}}(p) = \langle {B}^p {B}{}^{qr}\rangle_{0} \equiv$}
\end{picture}}
\caption{Propagators for the fictitious Grassman-odd  $\mathcal{G}(3)$ singlets, $\overline{C}, \widehat{B}$, 3 pairs of triplets $C^p, \widehat{B}{}^{qr}$  and 3
pairs  of Grassman-even  triplets ${B}^p$, $B^{qr}$.}\label{Figure_3_PropN3}
\end{figure}

Fourth, for $N=4$ propagators  of the fictitious fields  with account for antisymmetry of $({B}^{r_1r_2r_3},B^{r_1r_2}) = -({B}^{r_2r_1r_3},B^{r_2r_1})$

\begin{figure}[h]
{\footnotesize\begin{picture}(0,3.3)
\put(6,-3.7){\includegraphics[scale=0.4]{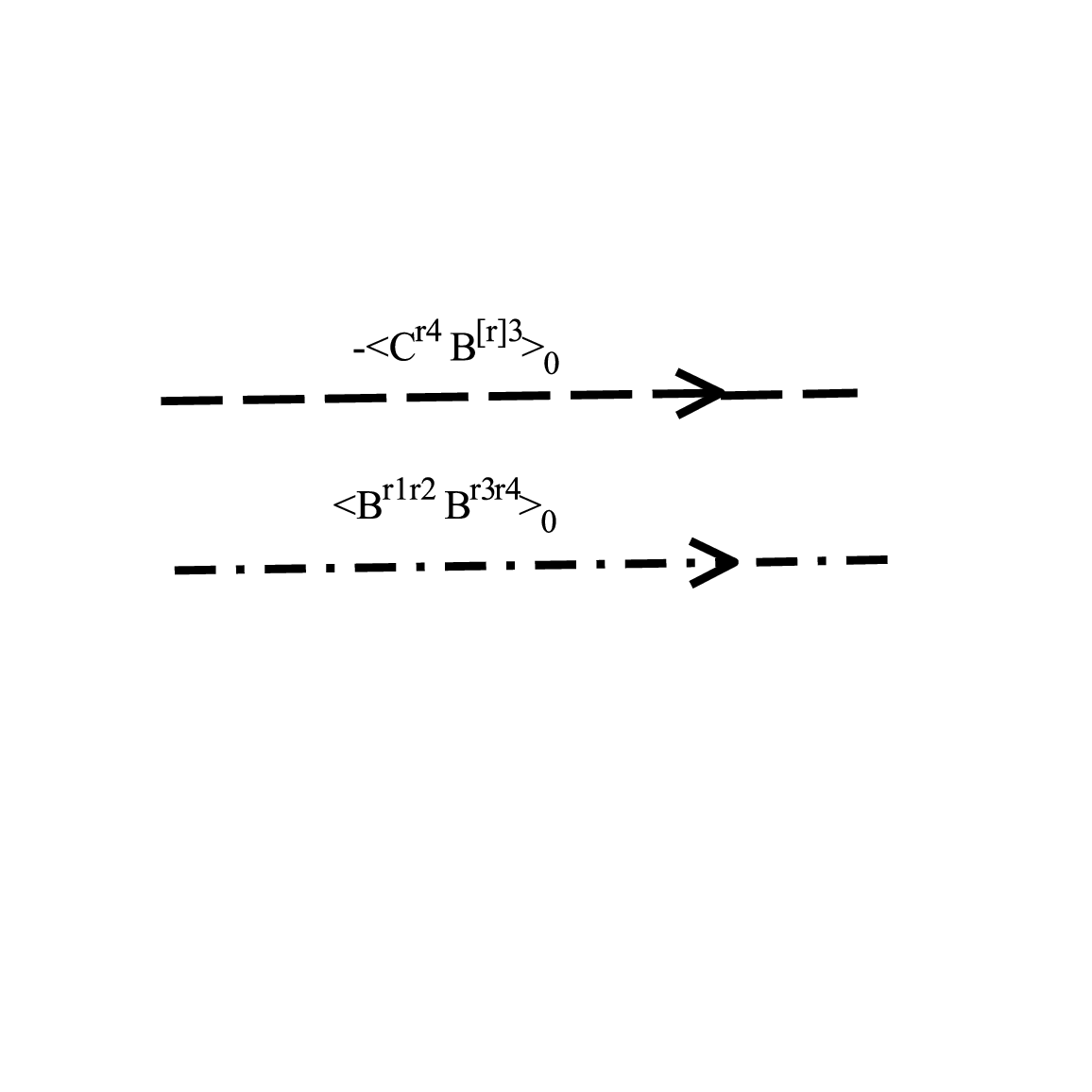}}
\put(2.4,1.20){$D^{[r]_4}_{CB}(p) = -\langle {C}^{r_4} {B}{}^{[r]_3}\rangle_0 \equiv$}
\put(2.3,-0.05){$D^{[r]_4}_{BB}(p) = \langle {B}^{r_1r_2} {B}{}^{r_3r_4}\rangle_0 \equiv$}
\end{picture}}
\caption{Propagators for the fictitious     four  pairs of Grassman-odd $\mathcal{G}(4)$-quartets $C^r, {B}{}^{[r]_3}$  and 3
pairs  of Grassman-even  $\mathcal{G}(4)$-sextet ${B}^{r_1r_2}$, $B^{r_3r_4}$.}\label{Figure_4_PropN4}
\end{figure}

\begin{figure}[h]
{\footnotesize\begin{picture}(0,4)
\put(7, 2.6){\includegraphics[scale=0.28]{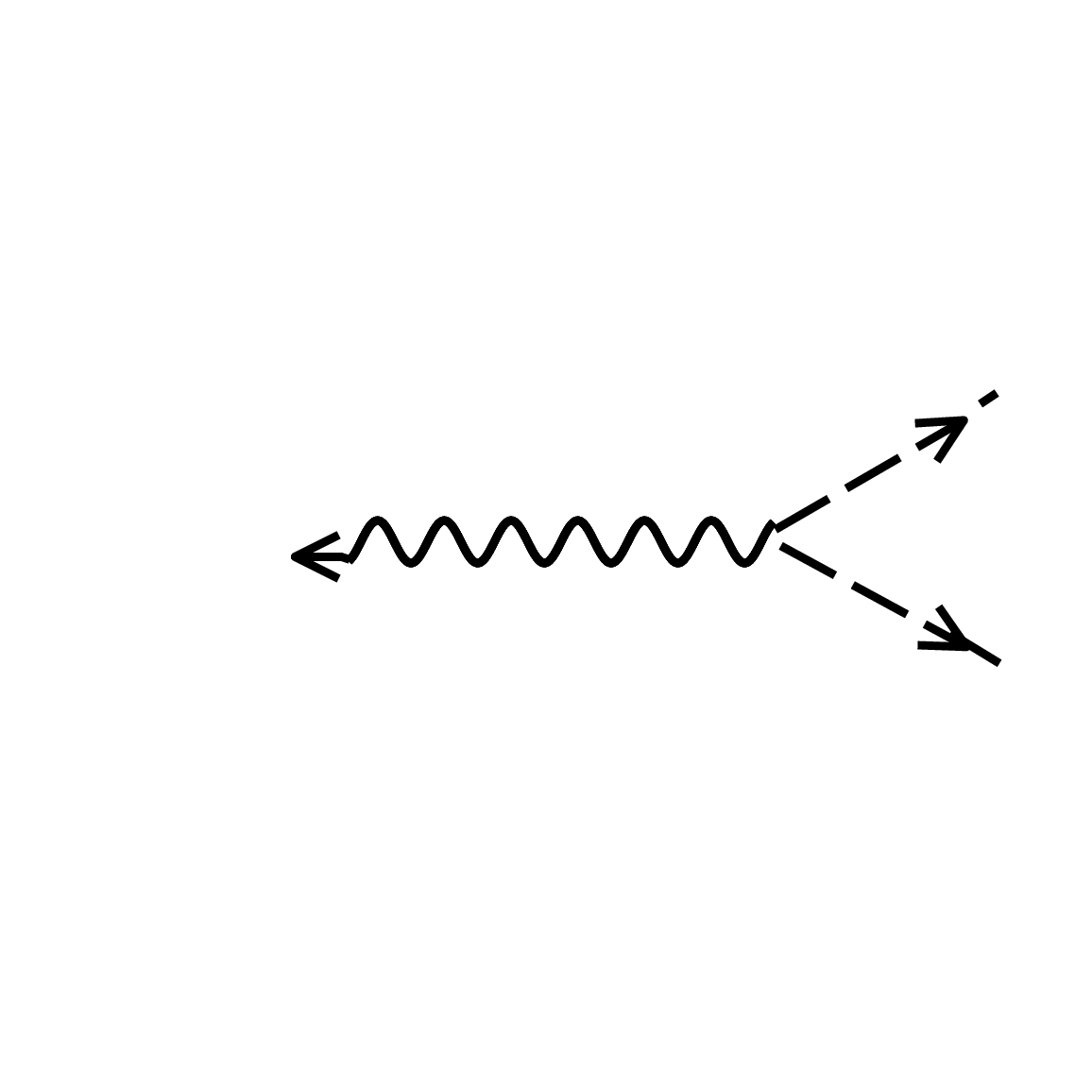}}
\put(7, -0.7){\includegraphics[scale=0.28]{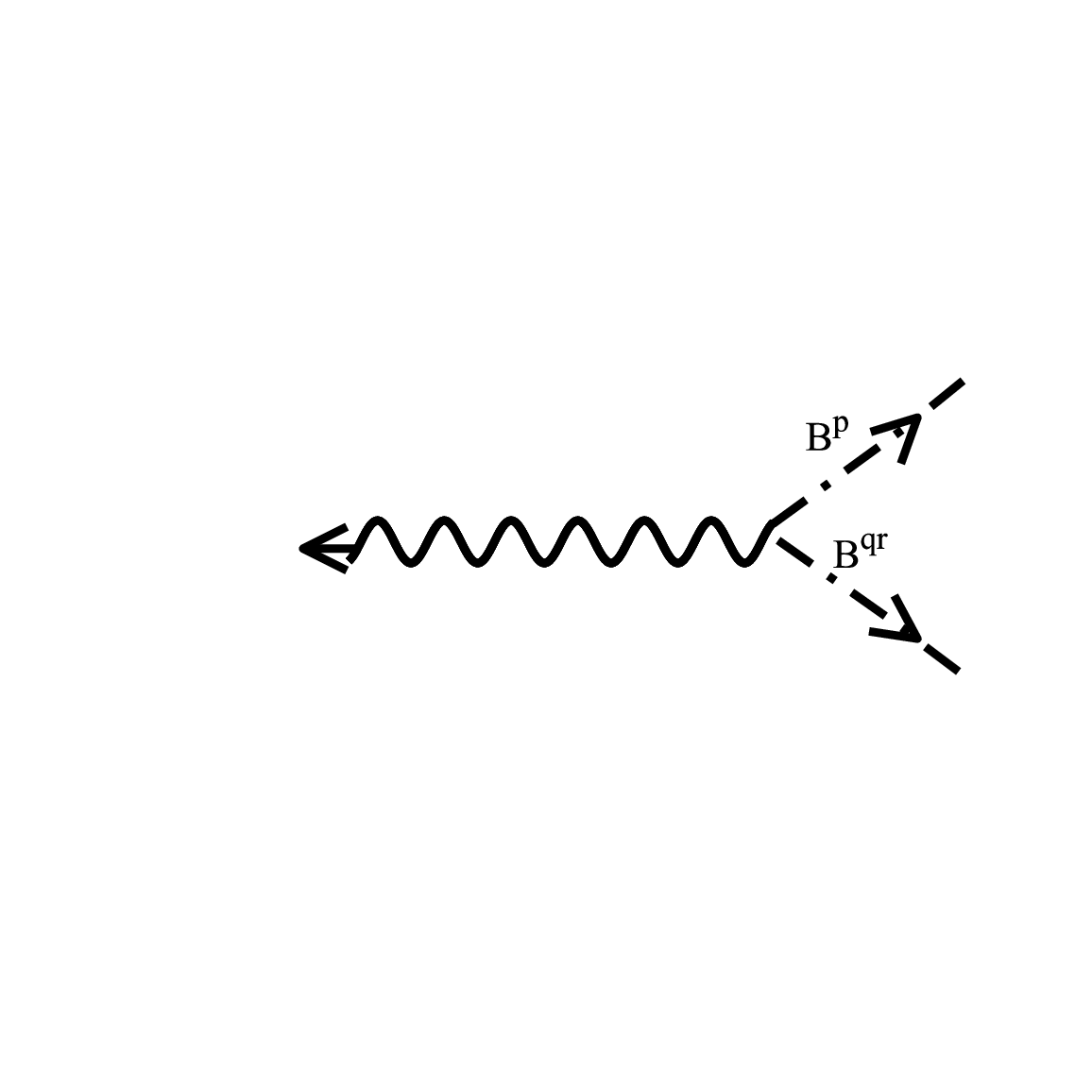}}
\put(3.3,5.15){$V_{\mathcal{A}\overline{C}C}(p) =  f^{mnl}p_{\mu}\ \ \equiv$ }
\put(1.5,1.9){$V_{3|{\xi}|\mathcal{A}B^pB^{[p]_2}}(p) = \displaystyle\frac{1}{2} f^{mnl}p_{\mu}\varepsilon_{[p]_3}\ \ \equiv$ }
\end{picture}}
\vspace{3ex}
\caption{Interaction vertexes of   $\mathcal{A}_\mu$:   with $\overline{C}, C$  fields in $N=1$,  with  any pair from $\mathcal{G}(3)$-triplets of even (Bose) fictitious fields $B^p, {B}{}^{qr}$ in $N=3$  BRST formulations.}\label{Figure_V_N=13 vertexes}
\end{figure}

\begin{figure}[h]
{\footnotesize\begin{picture}(0,3)
\put(7, -1.5){\includegraphics[scale=0.25]{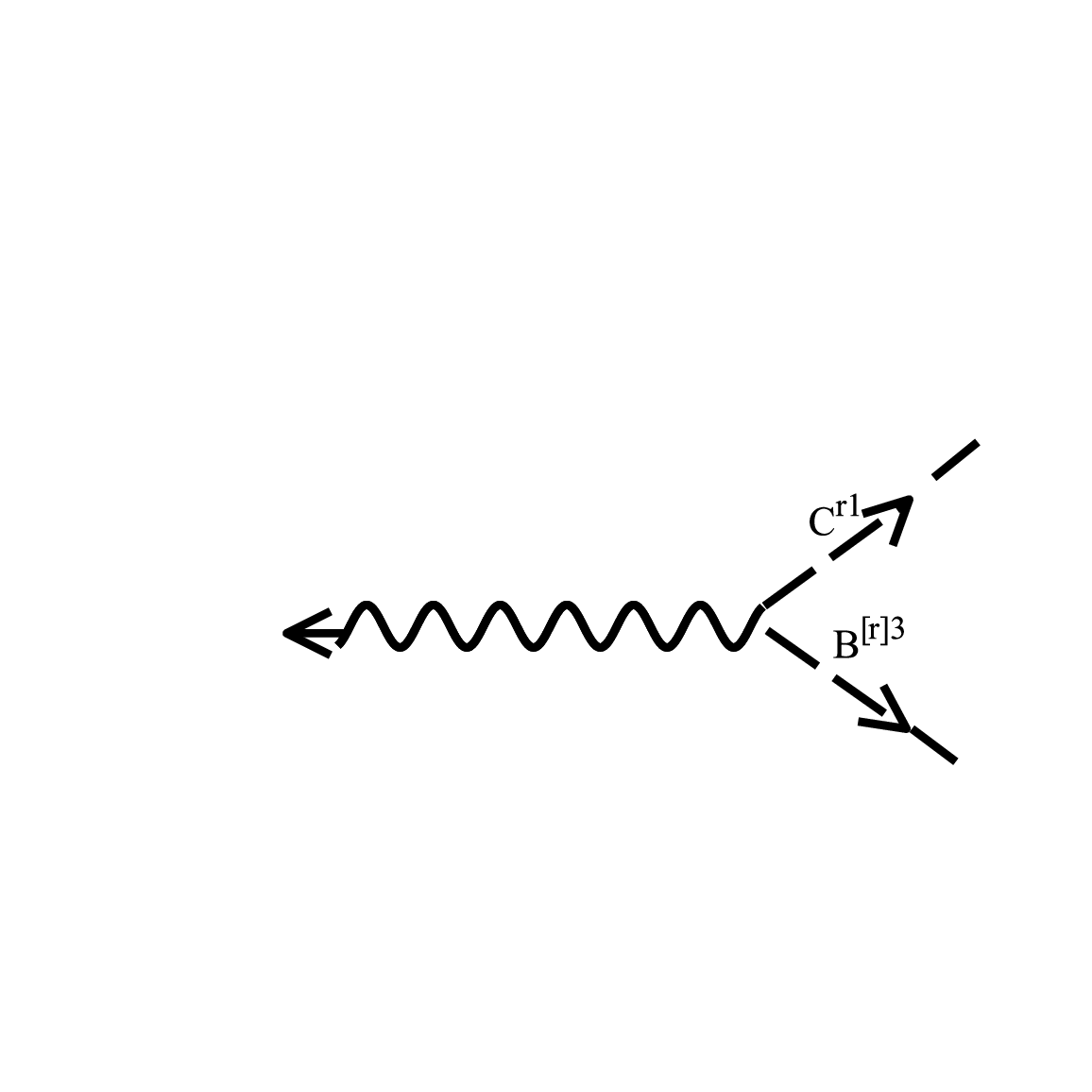}}
\put(1.5, 0.5){$V_{4|{\xi}|\mathcal{A}{C}^{r_4}B^{[r]_3}}(p) =  -\displaystyle\frac{1}{3!}f^{mnl}p_{\mu}\varepsilon_{[r]_4}\ \ \equiv$ }
\end{picture}}
\caption{Interaction vertex of  Yang--Mills field $\mathcal{A}_\mu$   with any pair from $\mathcal{G}(4)$-quartets  of   odd (Fermi)  fictitious fields $C^{r}$,${B}{}^{[r]_3}$ in $N=4$  BRST formulation.}\label{Figure_VI_N=4 vertex}
\end{figure}

And, for some $N=1,3,4$ vertexes  of the gauge vector fields $\mathcal{A}_{\mu}$ with respective fictitious fields (Grassmann-odd for $N=1, 4$ and Grassmann-even for $N=3$ BRST symmetric formulations from the quadratic in the fictitious fields terms with Faddeev-Popov operator $M(\mathcal{A})$ in the momentum representation found as in (\ref{intN2})) in the Figures~\ref{Figure_V_N=13 vertexes},~\ref{Figure_VI_N=4 vertex}

~Note,  starting from the $N=2$ case we have introduced the additional notation of the  respective (being valid for free (quadratic) theory) averaging fields $\langle \ \ \rangle_0$   written under the respective propagator's line   to distinguish different fictitious fields corresponding, in fact,  to the same function, $D(p)$. From the $N=3$ case the propagator's line for the Grassmann-even (Bose) fictitious particle is given by "dash with dot" as compared to the standard "dash" notations  for the Grassmann-odd (Fermi) fictitious particle.  There are 3 independent propagators for Grassmann-odd fields among  $\langle {C}^p \widehat{B}{}^{qr}\rangle_0$
and 3 ones for Grassmann-even from  $\langle {B}^p {B}{}^{qr}\rangle_0$ in the Figure~\ref{Figure_3_PropN3}, which are   $\langle {C}^1 \widehat{B}{}^{23}\rangle_0$, $\langle {C}^2 \widehat{B}{}^{31}\rangle_0$, $\langle {C}^3 \widehat{B}{}^{12}\rangle_0$  and $\langle {B}^1 {B}{}^{23}\rangle_0$, $\langle {B}^2 {B}{}^{31}\rangle_0$, $\langle {B}^3 {B}{}^{12}\rangle_0$, i.e. for $\{p,q,r\}$=$\{(1,2,3);(2,3,1).(3,1,2)\}$. For $N=4$ BRST symmetric case the Figure~\ref{Figure_4_PropN4} contains 4 independent propagators for Grassmann-odd fields  $-\langle {C}^{r_4} {B}{}^{[r]_3}\rangle_0$
and 3 ones for Grassmann-even from  $\langle {B}^{r_1r_2} {B}{}^{r_3r_4}\rangle_0$:   $\langle {C}^{1} {B}{}^{234}\rangle_0$,  $\langle {C}^{2} {B}{}^{314}\rangle_0$, $\langle {C}^{3} {B}{}^{124}\rangle_0$, $\langle {C}^{4} {B}{}^{132}\rangle_0$ and $\langle {B}^{12} {B}{}^{34}\rangle_0$, $\langle {B}^{13} {B}{}^{42}\rangle_0$, $\langle {B}^{14} {B}{}^{23}\rangle_0$.

There exist more additional vertexes from (\ref{Vps2}), (\ref{Vps3}), (\ref{Vps4}) which can be analogously represented as in the Figures~\ref{Figure_V_N=13 vertexes},~\ref{Figure_VI_N=4 vertex}.

\section{Conclusion}

In the present work  a generalization of the Faddeev--Popov proposal presenting the Lagrangian
path integral for the Yang--Mills theory in Landau and Feynman gauges \cite{fp}, \cite{fp-ghosts} is proposed  for non-local  form by inserting the special unity, ${\det}^{k} M(\mathcal{A}) \,  {\det}^{-k} M(\mathcal{A}) $, depending on non-negative  integer $k$     in  (\ref{Pintgen}), (\ref{PintgenF}) and for local form in (\ref{PintfpLlock}),
with numbers of fictitious Grassmann-odd and Grassmann-even fields (with the same number of physical degrees of freedom
as compared to the case of space-time extended  SUSY gauge theories) in the spectrum
of the total configuration spaces larger than those for $N=1,2$ BRST symmetry cases.
It is shown in the Statement 1, that to realize the $N=m$ BRST symmetry transformations with more than two Grassmann-odd parameters, $\lambda_p$, $p=1,2,...,m$  (in substituting instead of the infinitesimal gauge parameters $\xi = C^p \lambda_p $  the $m$-plet of Grassmann-odd ghost fields)  when formulating the corresponding quantum actions, $S^L_{(N(k))}(\Phi_{(N(k))})$ (\ref{qe}) with the  gauge-fixing terms
(respecting $N(k)=m$ BRST invariance) to be added to the classical Yang--Mills action the spectrum of $k=k(N)$ should obey to the relation (\ref{kN}), whereas to perform the gauge-fixing procedure without using an odd non-degenerate transformation changing the Grassmann parities for some fictitious fields   its spectrum  $k=k_u(N)$ is described  by (\ref{kNu}).

An irreducible representation space, $\mathcal{M}^{(3)}_{min}$, for the $3$-parametric abelian superalgebra
$\mathcal{G}(3)$ of anticommuting  generators $\overleftarrow{s}{}^p$ with triplet of Grassmann-odd constant parameters, $\lambda_p$,  with its  action on the local coordinates, fields $\Phi_{(3)}$,  has been explicitly
constructed by Eqs. (\ref{minN3}).    To formulate a local quantum action
with appropriate gauge-fixing procedure, we have followed two ways. First, that proves the condition (\ref{kN}) of the Statement 1,  is based on the original using of Grassmann-odd non-degenerate operator $\Pi$ which changes the Grassmann parities and  acts  on $\mathcal{G}(3)$-irreducible space of initial Yang--Mills fields $\mathcal{A}^\mu$, triplets $C^p$, $B^{pq}$, odd singlet $\widehat{B}$,   in such a way, that the respective change of variables in the subspace of part of fictitious fields, $\phi^M$ (\ref{calN}) has permitted to  make possible to pass to a new
basis of fictitious fields in which the local quantum action (\ref{finSminN3}) and path integral (\ref{PintfpLloc2f}) in Landau gauge with a new form of $N=3$ BRST symmetry transformations, (\ref{N3Afin})--(\ref{N3Bfin}),
have been constructed.  Second, the non-minimal sector of the fields $\overline{\Phi}_{(3)}$ containing antighost field (as  a connection) $\overline{C}$ to incorporate the usual gauge condition, $\chi(\mathcal{A},B)$, into a gauge  fermionic functional $\Psi_{(3)}$  (\ref{N3gaugef})   has been
introduced, on which a new $N=3$ representation of   $\mathcal{G}(3)$-superalgebra is explicitly realized (\ref{nminN3}). The sector contains two $\mathcal{G}(3)$-singlets, $\overline{C}, B$ with usual Nakanishi-Lautrup field and two $\mathcal{G}(3)$-triplets, $B^p, \widehat{B}{}^{pq}$ and together with the fields, $\Phi_{(3)}$ composes the fields $\widetilde{\Phi}_{(3)}$ of  reducible  $\mathcal{G}(3)$-superalgebra representation  parameterizing  the total configuration space  $\mathcal{M}^{(3)}_{tot}$, on which the quantum action $ S_{\Psi_{(3)\xi}}\big(\widetilde{\Phi}_{(3)}\big)$ (\ref{qexi3})--(\ref{addxi3}) and path integral ${Z}_{3|\Psi}(0)$  (\ref{Pintlocn3}) in $R_\xi$-like gauges, determined by the gauge fermionic functional $\Psi_{(3)\xi}\big(\widetilde{\Phi}_{(3)}\big)$ (\ref{N3gaugef}),  have been explicitly constructed. The set of the transformations in $\mathcal{M}^{(3)}_{tot}$,  $\delta_\lambda \widetilde{\Phi}_{(3)} = \widetilde{\Phi}_{(3)} \overleftarrow{s}{}^{p} \lambda_p $, determined  by (\ref{minN3}), (\ref{nminN3}),  leaving both the quantum action and the integrand of the respective path integral by invariant,we call  $N=3$ BRST transformations.
The quantum (non-renormalized) action $S_{\Psi_{(3)\xi}}$ contains the terms quadratic in fictitious fields leading to the same one-loop contribution for the effective action as one for the quantum actions constructed according to $N=1$ and $N=2$ BRST symmetry principles in smaller configuration space, whereas for more than quadratic in powers of ghost fields terms in $S_{\Psi_{(3)\xi}}$,  described by the $ S_{\mathrm{add}(3)}$ (\ref{addxi3}) which generates the   ghost vertexes to be different than ones derived from the former actions.

We have established with the help of   $\mathcal{G}(1)$-superalgebra with nilpotent generator $\overleftarrow{\bar{s}}$ and parameter $\bar{\lambda}$ being additional  to $\mathcal{G}(3)$-superalgebra, but acting on  the fields of $\mathcal{M}^{(3)}_{tot}$ by the rule (\ref{aN1}), the fact that  the $\mathcal{G}(1)$- invariant path integral ${Z}_{1|\Psi}(0)$ (\ref{Pintlocn1}) with the quantum action $S_{\Psi_{(1)}}$  and, at least for special quadratic  gauge fermionic functional $\Psi_{(1)} $ (\ref{aN1gaugef})  given on $\mathcal{M}^{(3)}_{tot}$ is equivalent to the $N=1$ antiBRST invariant  path integral (\ref{Pintlocn1fin}) with the quantum action $S_{\Psi}$, (\ref{eqZpsi0}) constructed by the standard Faddeev-Popov method  with use of $N=1$ antiBRST symmetry transformations acting in the standard  configuration space, $\mathcal{M}_{tot}$  of fields $\mathcal{A}_\mu, C, \overline{C}, B$. We call the transformations  (\ref{abrstrans}) with parameter  $\bar{\lambda}$ which led to the $\mathcal{G}(1)$- invariance of  $S_{\Psi_{(1)}}$ and integrand of ${Z}_{1|\Psi}(0)$ by $N=1$ antiBRST symmetry transformations in  $\mathcal{M}^{(3)}_{tot}$.

It was shown the Grassmann-odd parameters: $\mathcal{G}(3)$-triplet $\lambda_p$ and $\mathcal{G}(1)$-singlet  $\bar{\lambda}$,  of  $\mathcal{G}(3)$ and $\mathcal{G}(1)$ superalgebras  acting on  the space $\mathcal{M}^{(3)}_{tot}$, are uniquely combined  within quartet of parameters  $\lambda_r = \big(\lambda_p, \bar{\lambda}\big)$, as well as the quartet of the generators $\overleftarrow{s}{}^r = \big(\overleftarrow{s}{}^p, \overleftarrow{\bar{s}}\big)$  to form   a $\mathcal{G}(4)$-superalgebra  whose irreducible representation contains the same fields as reducible one for $\mathcal{G}(3)$-superalgebra in $\mathcal{M}^{(3)}_{tot}$ but  organized in $\mathcal{G}(4)$-antisymmetric tensors, $\Phi_{(4)} = \big(\mathcal{A}^\mu,C^r, B^{r_1r_2}, B^{r_1r_2r_3},B\big)$ according to the rule (\ref{corg3g4}), which  parameterize   $N=4$ total configuration space $\mathcal{M}^{(4)}_{tot}$. The explicit action of the generators $\overleftarrow{s}{}^r$  on each component from $\Phi_{(4)}$ was constructed by Eqs. (\ref{minN4})  with preservation of the $\mathcal{G}(4)$-superalgebra: $\{\overleftarrow{s}{}^{r_1},\,\overleftarrow{s}{}^{r_2}\}=0$.  The respective $N=4$ SUSY transformations,  $\delta_\lambda \Phi_{(4)} = \Phi_{(4)} \overleftarrow{s}{}^{r} \lambda_r $ have appeared, according to their definition,   by $N=4$ BRST transformations for the quantum action $S_{Y_{(4)}}$ and local path integral  ${Z}_{4|Y}(0)$ (\ref{Pintlocn4})  constructed with help of addition to the classical action of the $N=4$ BRST exact term generated  by the quartic powers in $\overleftarrow{s}{}^{r}$ applied to the gauge Bosonic functional, $Y_{(4)}\big(\Phi_{(4)}\big)$ (\ref{N4gaugeb}).  For $R_\xi$-like family of gauges determined by the functional $Y^0_{(4)\xi}({\Phi}_{(4)})$ (\ref{gBn4xi}) the quantum action $
  S_{Y_{(4)\xi}}\big({\Phi}_{(4)}\big)$  (\ref{qexi4}) was exactly calculated for the Landau gauge ($\xi=0$), whereas for the Feynman gauge ($\xi=1$)  the additional summand $S_{\mathrm{add}(4)}$ (\ref{addxi4}) to the standard gauge-fixed and quadratic [in 4 Grassmann-odd $(C^1, B^{234})$, $(C^2, B^{134})$, $(C^3, B^{124})$, $(C^4, B^{123})$  and 3 Grassmann-even $(B^{12}, B^{34})$, $(B^{13}, B^{24})$, $(B^{14}, B^{23})$  pairs  of ghost fields] parts $S_{\mathrm{gf}(4)}$, $S_{\mathrm{gh}(4)}$ of the quantum action contains the $8$-th  powers in odd $C^r$  and $4$-th powers  in even  $B^{r_1r_2}$ fields. For any $\xi$  classical action and  the functionals $S_{\mathrm{gf}(4)}$, $S_{\mathrm{gh}(4)}$ lead to the same  contribution into one-loop effective action as  those for the known and above quantum actions
  constructed according to the $N=1$, $N=2$ and both $N=3$ BRST symmetry recipes.
It was explicitly  shown on the level of the non-renormalized  path integrals the equivalence among $N=3$ BRST invariant path integral evaluated in the $R_\xi$-like gauges and usual $N=1$ BRST invariant path integral in the $R_\xi$ -gauges in (\ref{N3N1eqv}). For $N=4$ BRST invariant path integral its equivalence with $N=1$ BRST invariant path integral was found in case of Landau gauge in (\ref{N4N1eqv}).

For both $N=3$ and $N=4$ BRST invariant formulations of the quantum actions the generating functionals of Greens functions, including effective actions were determined and Ward identities (\ref{GFGFWIN3}), (\ref{GFGFWIN4}) for them, which  follow from the respective algebraic $N=m$, $m=3,4$ BRST invariance,  were derived as well as the independence on the choice of the gauge condition for the respective path integral under the corresponding small variation of the gauge: $\Psi_{(3)}\to \Psi_{(3)}+\delta\Psi_{(3)}$ and  $Y_{(4)}\to Y_{(4)}+\delta Y_{(4)}$ were established by means of infinitesimal FD $N=m$ BRST transformations.

The finite $N=3$ and $N=4$ BRST transformations were restored to form respectively the Abelian supergroups $G(m) = \exp\{\overleftarrow{s}{}^p \lambda_p\}$, $p=1,2,...,m$  acting on the respective configuration space  $\mathcal{M}^{(m)}_{tot}$ by means of two ways: first, by continuation of the  invariance of any regular functional  under algebraic $N=m$, $m=3,4$ BRST transformations to full invariance under finite transformations, second by means of resolution of the Lie equations.
The sets $\tilde{G}(m)$ (\ref{tildeG3}), (\ref{tildeG4}) of finite FD $N=m$ BRST transformations  were introduced and the respective Jacobians of the change of variables in $\mathcal{M}^{(m)}_{tot}$ generated by these transformations were calculated in (\ref{jacobianresN3}),  (\ref{jacobianresN4}). For  functionally-dependent Grassman-odd parameters, $\hat{\lambda}_{p_1} = -(-1)^m\frac{1}{(m-1)!}\Lambda_{(m)} \big({\Phi}_{(m)}\big)\varepsilon
_{[p]_m}\overleftarrow{s}{}^{p_2}
...\overleftarrow{s}{}^{p_m}$ with a some potential functional $\Lambda_{(m)}$ Grassmann-odd(even) for $m=3$ ($m=4$)    (\ref{fdepparam}), (\ref{fdepparam4}) the Jacobians above are transformed  to the respective $N=m$ BRST exact terms (\ref{jacobianresN3fd}), (\ref{jacobianresN4fd}). The latter Jacobians were applied, first, to the establishing of the  independence upon the choice of  the gauge condition for finite variation of the respective  path integral, $Z_{3|\Psi_{(3)}}(0)=Z_{3|\Psi_{(3)}+\Psi'_{(3)}}(0)$, ${Z}_{4|Y}(0)={Z}_{4|Y+Y'}(0)$, from the solutions of the corresponding compensation equations (\ref{eqexplN3}), (\ref{eqexplN4}) relating   the parameters $\hat{\lambda}_{p_1}$ with respective change of the gauge condition $\Psi'_{(3)}, Y'_{(4)}$ in (\ref{funcdeplafinN3})  and (\ref{funcdeplafinN4}). Second, they were used  to derive new  modified   Ward identities (\ref{mWIN3}), (\ref{mWIN4}) for the generating functionals of Green functions $Z_{3|\Psi_{(3)}}(\widetilde{J}_{(3)})$, $Z_{4|Y_{(4)}}({J}_{(4)})$   depending on the functionally-dependent FD parameters $\hat{\lambda}_{p_1}$, $p_1=1,2..,m$, and therefore on the finite variation of the gauge $\Psi'_{(3)}, Y'_{(4)}$  respectively. Third, they have permitted to establish gauge independence of  $Z_{3|\Psi_{(3)}}(\widetilde{J}_{(3)})$, $Z_{4|Y_{(4)}}({J}_{(4)})$ upon the respective choice of the gauge condition $\Psi_{(3)}\to \Psi_{(3)}+\Psi'_{(3)}$ and  $Y_{(4)}\to Y_{(4)}+Y'_{(4)}$  on the corresponding mass-shell: $\widetilde{J}_{(3)}=0$, ${J}_{(4)}=0$.

The new  Ward identities (\ref{meq3})
for the extended (by means of sources $K_{p},  K_{pq}, \overline{K}$ to the $N=3$ BRST variations $\widetilde{\Phi}_{(3)} \overleftarrow{s}{}^p$,
$\widetilde{\Phi}{}_{(3)} \overleftarrow{s}{}^p\overleftarrow{s}{}^q$, and $\widetilde{\Phi}{}_{(3)} (\overleftarrow{s})^3$)   generating functional of vertex Green's functions, $\Gamma\big(\langle\widetilde{\Phi}_{(3)}\rangle, K_{p},  K_{pq}, \overline{K}\big)$ (\ref{avfields1N3}) obtained from the part of extended $N=3$ BRST transformations (\ref{extBRSTN3}) in the space of $\widetilde{\Phi}_{(3)}, K_{p},  K_{pq}, \overline{K}$ for constant $\lambda_p$,  reproduced the new
differential-geometric objects. i.e., $G(3)$-triplets of antibrackets: $(\bullet,\bullet)^p$
and odd-valued first-order differential operators  $V^p$ (\ref{eN3br}).

The gauge-independent Gribov-Zwanziger model of Yang--Mills fields without residual Gribov ambiguity in the infrared region of the field $\mathcal{A}^\mu$ configurations described by gauge-invariant, and therefore $N=m$ BRST invariant, for $m=3,4$, horizon functional $H(\mathcal{A}^h)$ (\ref{FuncM}) in terms of gauge-invariant transverse fields $\mathcal{A}^h_\mu$ (\ref{gitrans}) \cite{SemenovTyanshan}, firstly proposed in \cite{Pereira} within $N=1$ BRST symmetry realization but with non-local BRST transformations was suggested in non-local form but with local $N=3, N=4$ BRST invariance by the Eqs. (\ref{brstinvgzN3}), (\ref{brstinvgzN4}). The partially local, (in view of residual presence of non-local vector  field $\mathcal{A}^h_\mu$) Gribov-Zwanziger model was proposed with non-local $N=1$ BRST symmetry (\ref{N1BRST}), (\ref{N1phi}), due to inverse gauge-invariant Faddeev-Popov matrix $\big(M^{-1}\big)(\mathcal{A}^h)$ presence for auxiliary fields in (\ref{N1phi}).

The extension of the basics for the diagrammatic Feynman technique within perturbation theory for the $N=3$ and $N=4$ BRST invariant quantum actions for the Yang--Mills theory were proposed due to the presence of additional  both Grassmann-odd and Geassmann-even fictitious fields.

Concluding, let us present  the  spectrum of irreducible representations for a $\mathcal{G}(l)$ Abelian superalgebra
with $l=0$ (non-gauge theories), $l=1$ (BRST symmetry algebra),
$l=2$ (BRST-antiBRST symmetry algebra), $l=3$ (superalgebra with $3$ BRST symmetries),
and so on according to the chain (\ref{chain1})--(\ref{chain2}),  by a numeric pyramid partially similar to the Pascal triangle
(\ref{Pastriangle}), which contains in its left-hand side the symbol "$d|\mathcal{A}$" relating to number of degrees of freedom  of the classical
Yang--Mills fields $\mathcal{A}_\mu$ with suppressed $su(\hat{N})$ indices:
\begin{table}
\begin{center}
{\footnotesize\begin{tabular}{ccccccccc ccccccccc}
     $N=0$:    &&&& && & $d|\mathcal{A}$ & & &&&&&\\
$N=1$: &&& & &&$d|\mathcal{A}$&\hspace{-0.5em}$\stackrel{s}{\to}$ & $1|C$&& &&&&&\\
$N=2$: & & && &\hspace{-0.5em}$d|\mathcal{A}$&$\stackrel{s{}^a}{\to}$& \hspace{-0.5em}$2|C^{a}$ &$\stackrel{s{}^a}{\to}$&$1|B$&&&\\
$N=3$: & && &$d|\mathcal{A}$&\hspace{-0.5em}$\stackrel{s{}^p}{\to}$ &\hspace{-0.5em}$3|C^{[p]_1}$& $\stackrel{s{}^p}{\to}$ &$3|B^{[p]_2}$&$\stackrel{s{}^p}{\to}$&\hspace{-0.5em}$1|\widehat{B}$&& \\
$N=4$:    &&&$d|\mathcal{A}$& $\stackrel{s{}^r}{\to}$ &$4|C^{[r]_1}$&\hspace{-0.5em}$\stackrel{s{}^r}{\to}$&$6|B^{[r]_2}$&$\stackrel{s{}^r}{\to}$&\hspace{-0.5em}$4|B^{[r]_3}$&$\hspace{-0.5em}\stackrel{s{}^r}{\to}$&\hspace{-0.8em}$1|B$ &\\
...&... & ...&... & ... &...
&...&...&...&...&...&...&&\\
$N=2K$:&\hspace{-0.5em}$d|\mathcal{A}$& \hspace{-0.5em}$\stackrel{s{}^r}{\to}$ &\hspace{-0.5em} $2K|C^{[r]_1}$&\hspace{-0.5em}$\stackrel{s{}^r}{\to}$& \hspace{-0.5em} $C^{2K}_2|B_{(2K)}^{[r]_2}$ &&....&&\hspace{-0.7em}$\stackrel{s{}^r}{\to}$ &\hspace{-0.7em}${2K}|B^{[r]_{2K-1}}_{(2K)} $&\hspace{-0.5em}$\stackrel{s{}^r}{\to}$&\hspace{-0.5em}$1|B$
\end{tabular}}
\end{center}
 \caption{Numbers of fictitious fields in addition to $\mathcal{A}_\mu$ for each $N=0,1,2,...,2K$} \label{Pastriangle}
\end{table}
where the  $l$-th row, corresponding to the field content $\Phi_{(l)}$ of an irrep space
for the $\mathcal{G}(l)$ superalgebra, is constructed from the symbols
of $d|\mathcal{A}$, $l|C^{[r]_1}$, $C^l_2|B^{[r]_2}$..., $1|B_{(l)}$ ($C^k_l= k!/(l!(k-l)!)$), corresponding
to the degrees of freedom (modulo the dimension of $su(\hat{N})$) for $\mathcal{A}_\mu$,
$C^{p_l}$,  $B^{p_lq_l}$, ..., ${B}_{(l)}$, $p_l=1,2,...,l$, whose sum is equal
to  $(2^l+d-1)$. The symbols related by an arrow: $d|\mathcal{A}\stackrel{s{}^{r_l}}{\to} l|C^{[r]_1}$ meaning the part of the chain generated  by the $N=l$-BRST generator $\overleftarrow{s}{}^{r_l}$, $r_l=1,2,...,l$ by the rule: $\mathcal{A}_\mu \overleftarrow{s}{}^{r_l} = D_\mu C^{r_l}$  with omitting the arrow over $\overleftarrow{s}{}^{r_l}$ for the readability in the Table~\ref{Pastriangle}. From the second row ($N=2$), the rule of filling the triangle starts
to work, whereas for $N=0$ there is no fictitious fields, and
in the case of $N=1$ it is only $\mathcal{A}_\mu$ and ghost field $C$ that compose
an irrep space of the $N=1$ BRST algebra without an additional trivial BRST doublet,
$\bar{C}, B$ necessary to construct quantum action and local path integral which as the fields from the non-minimal sector, answering for the reducible representation of $\mathcal{G}(1)$-superalgebra,  selected into another Table~\ref{Pastrianglenmin}.  Notice, that the second left-hand side only contains the numbers
$1, 2,3,...,2K$ of Grassmann-odd fictitious fields, $C, C^{p_2}, C^{p_3},...,C^{p_{2K}}$;
the third left-hand side (starting from $N=2$) only contains the numbers
$1, 3, 6,...,C_2^{2K}$ of Grassmann-even fictitious fields
$B,  B^{p_3q_3},B^{p_4q_4},...,B^{p_{2K}q_{2K}}$, etc. The final right-hand side
of the triangle (\ref{Pastriangle}) is composed of the Nakanishi--Lautrup
$\mathcal{G}(l)$-singlet fields $B\equiv \widehat{B}_2,
\widehat{B}_3\equiv \widehat{B}, \widehat{B}_4,...,\widehat{B}_l$,
with alternating Grassmann parity, $\epsilon(\widehat{B}_l)= l$,
respectively for  $l=2,3,...,{2K}$.

In turn,  for the reducible representation space of $\mathcal{G}(2K-1)$-superalgebra, for integer $K$ determining  the non-minimal sector of fields to be necessary to provide gauge-fixing procedure without odd supermatrix, the spectrum of additional fields is described by the Table~\ref{Pastrianglenmin} corresponding to the exact sequence (\ref{exseqnm}).
\begin{table}[b]
\begin{center}
{\footnotesize\begin{tabular}{ccrcc ccccc cc lccc}
$N=1$: &&& & && $1|\overline{C}$ &$\stackrel{{s}}{\to}$&$1|B$ &&&&&&\\
$N=3$:& &  && $1|\overline{C}$&$\stackrel{{s}{}^p}{\to}$&$3|B^{[p]_1}$&$\stackrel{{s}{}^p}{\to}$&$3|\widehat{B}^{[p]_2}$ &$\stackrel{{s}{}^p}{\to}$& $1|B$&&&&\\
$N=5$: && \hspace{-0.7em}$1|\overline{C}$&\hspace{-0.7em}$\stackrel{{s}{}^r}{\to}$& \hspace{-0.5em} $5|B_{(5)}^{r}$&$\stackrel{s{}^r}{\to}$&$10|\widehat{B}_{(5)}^{[r]_2}$&$\stackrel{s{}^r}{\to}$&$10|B_{(5)}^{[r]_3}$&$\stackrel{s{}^r}{\to}$&
$5|\widehat{B}_{(5)}^{[r]_4}$&\hspace{-0.7em}$\stackrel{s{}^r}{\to}$&\hspace{-0.9em}$1|B$&&\\
...&  & ... &\hspace{-0.5em} ... &\hspace{-0.5em} ...
&...&  ...&...& ... & ...&...&\hspace{-0.5em} ...&\hspace{-0.5em} ...&\hspace{-0.5em}&\\
$N=2K-1$:& $1|\overline{C}$ &$\stackrel{s{}^r}{\to}$&\hspace{-0.5em}$N|B_{(N)}^{r}$&\hspace{-0.5em}$\stackrel{s{}^r}{\to}$& ...&...&...&...&...&$\stackrel{s{}^r}{\to}$&\hspace{-0.5em}$N|\widehat{B}_{(N)}^{[r]_{N-1}}$ &\hspace{-0.5em}$\stackrel{s{}^r}{\to}$&\hspace{-0.5em}$1|B$&
\end{tabular}}
\end{center}
 \caption{Numbers of fictitious fields from the non-minimal sectors for each odd $N=1,3,5,...,2K-1$} \label{Pastrianglenmin}
\end{table}
In particular, from Table~\ref{Pastriangle} it follows that, for
odd numbers $N=2K-1$ of parameters in the $\mathcal{G}(N)$ superalgebra,
the generalized Faddeev--Popov rules must be described by odd non-degenerate transformation,
$\Pi$, intended to present the path integral with the Grassmann-even
Nakanishi--Lautrup field $B_{(2K-1)}=  \Pi\widehat{B}_{2k-1}$ exponentiating
the standard gauge condition, added to the classical action using
an $N=(2K-1)$ BRST-exact form.

It follows from the both Tables that the generalization of the Faddeev-Popov quantizations for the case of $N=2K-1$ BRST invariance without using
of an odd non-degenerate transformation, when formulating the local quantum action and  path integral leads to the dimension of the total configuration space  $\mathcal{M}^{(2K-1)}_{tot}$ to coinciding with the one for $\mathcal{M}^{(2K)}_{tot}$ realizing $N=2K$ BRST symmetry for the same purpose.

There are various directions to extend the results of the present study.
Let us mention some of them. First, to develop the case of $N=3, 4$ BRST symmetries
transformations in a Yang-Mills theory as a dynamical system with first-class
constraints in the generalized canonical formalism \cite{bfv}, \cite{bfv1}, \cite{Henneaux1}.
Second: to develop the case of $N=3, 4$ BRST symmetry transformations
for irreducible general gauge theories in Lagrangian formalism \cite{bv},
including theories with a closed algebra of rank 1. Third: to generalize,
in a manifest way, the Faddeev--Popov rules in Yang--Mills theories to the case
of $N=2K-1$ and $N=2K$, $K>2$ BRST symmetry transformations in Lagrangian formalism
and in generalized canonical formalism. Then, it is intended to examine
the case of irreducible dynamical systems subject to $N=2K-1$, $N=2K$, $K>2$
BRST symmetry transformations  and to compare the
results with superfield formulations with $N$ BRST charges in
\cite{Nsupercharge}. Next, it is planned to consider an
irreducible general gauge theory subject to $N=2K-1$, $N=2K$ $K>2$
 BRST symmetry transformations
in the Lagrangian formalism. The problem  of  study of the  renormalizability  for the suggested $N=3, 4$ BRST invariant formulations of the quantum actions  so as to have
completely renormalized respective effective actions  remains a
very important question, as well as adopting of the $N=m$,
$m=2,3,4$ BRST invariance to the renormalizability of
$\mathcal{N}=1$ space-time super Yang--Mills theory in terms of
$\mathcal{N}=1$ superfields considered for $N=1$ BRST symmetry in
\cite{Piguet}, \cite{Stepanyantz} on the basis of preserving the
gauge-invariance, and, hence, the $N=m$ BRST symmetry,
regularization by higher-derivatives \cite{Slavnov:1972sq},
recently developed for $\mathcal{N}=2$ superfield formulation of
Abelian and super Yang--Mills theories
\cite{BuchbinderStepanyantz} on a basis of $\mathcal{N}=2$
harmonic superspace approach \cite{harmonic}.  We intend to study these problems in forthcoming works.

\vspace{1ex} \noindent \textbf{Acknowledgments \ } The author is thankful to J.L~ Buchbinder for illuminating discussion,  V.P, Spiridonov for the comments. He is  grateful to P.Yu. Moshin  and  K.V. Stepanyantz for the comments and advises  within
  many useful  discussions, as well as to B.P. Mandal and S. Upadhyay for participation at the initial
research stage for the problem.

\appendix
\section*{Appendix}
\section{On $N=3$ BRST invariant gauge-fixing in $N=3$ irreducible superspace}\label{app1}

 \renewcommand{\theequation}{\Alph{section}.\arabic{equation}} \setcounter{equation}{0}

Here, we will prove that it is impossible to perform  $N=3$ BRST invariant gauge-fixing  procedure within the  set  of fields $\Phi^{A_3}_{(3)}$ parameterizing  the superspace of  irreducible  representation  of $\mathcal{G}(3)$-superalgebra without using of non-degenerate odd-valued change of variables among  the components of  $\Phi_{(3)}$ to explicitly construct such a gauge-fixing.

Indeed, it is easy to see that in the basis of additional to $\mathcal{A}^\mu$ fields in  $\Phi^{A_3}_{(3)}$ = $(\mathcal{A}^\mu, C^p, B^{pq}, \widehat{B})$ composing the irreducible  representation space of $\mathcal{G}(3)$-superalgebra,  on which due to Lemma 1 the $N=3$ SUSY transformations is realized  (\ref{minN3}), there are  no enough coordinates  to reach a non-local
Faddeev--Popov path integral  (\ref{Pintgen}) with preservation of the  symmetry above. The terms
in the functional $S^L_{(3)}(\Phi_{(3)})$ (\ref{qe}) for $N(k)=3$, $k=1$, with the fermionic  gauge-fixing functional, $\frac{1}{3!} F_{(3)\xi}\big(\Phi_{(3)}\big)\prod\overleftarrow{s}{}^{p^1_1}\overleftarrow{s}{}^{p^2_1}\overleftarrow{s}{}^{p^{3}_1}\varepsilon_{p^1_1p^2_1 p^3_1}$ are calculated following to the rules (\ref{auxsaK})--(\ref{s3(AB)}) similar to the $N=2$ BRST symmetry case
(\ref{Y(A,C)}) for $\xi=0$, when $F_{(3)0}\big(\Phi_{(3)}\big)=F_{(3)0}\big(\mathcal{A}\big)$:
\begin{eqnarray}
&& \label{auxsaKapp}   \hspace{-0.9em} F_{(3)0}\big(\mathcal{A}\big) \overleftarrow{s}{}^{p} = \int d^dx \,tr\, \frac{\delta F_{(3)0}}{\delta \mathcal{A}_{\mu
}}D_\mu C^p = - \int d^dx \,tr\,  D_\mu \Big(\frac{\delta F_{(3)0}}{\delta \mathcal{A}_{\mu
}}\Big) C^p  = - \int d^dx \,tr\, \chi^F(\mathcal{A}) C^p,\\
&&  \hspace{-0.9em} F_{(3)0}\big(\mathcal{A}\big) \overleftarrow{s}{}^{p}\overleftarrow{s}{}^{q}\varepsilon_{pqr} =   \int d^dx \, tr\, \left(\int d^dy  C^p(x)M^F(\mathcal{A},x;y) C^q(y)    - \chi^F(\mathcal{A})  B^{pq}\right)\varepsilon_{pqr} , \label{auxapp} \\
 && \ \ \mathrm{for} \ M^F(\mathcal{A},x;y)  =  \frac{\delta \chi^F(\mathcal{A},x)}{\delta \mathcal{A}_{\mu
}(y)}D_\mu(y), \label{auxsaK1app}  \\
  && \hspace{-0.9em}  F_{(3)0}\big(\mathcal{A}\big) (\overleftarrow{s})^{3}=   \int d^dx \, tr\, \Bigg\{\int d^dy \Big(  2B^{qp}(x)M^F(\mathcal{A},x;y) C^r(y)   +C^{p}(x)M^F(\mathcal{A},x;y) B^{qr}(y)  \nonumber \\
&&  \hspace{-0.9em} \ \  - \int\hspace{-0.2em} d^dz C^p(x) \frac{\delta M^F(\mathcal{A},x;y)}{\delta \mathcal{A}_{\mu
}(z)}D_\mu(z) C^r(z)  C^{q}(y)\Big)\varepsilon_{pqr} -  \chi^F(\mathcal{A}) \Big(3!\widehat{B}  + \frac{1}{2} \big[B^{pq}, C^r\big] \varepsilon_{pqr} \Big)\hspace{-0.2em}\Bigg\}.\label{s3(AB)app}%
\end{eqnarray}
Hence,
\begin{eqnarray}\label{F(A)o}
\hspace{-0.5em} &\hspace{-0.9em}&\hspace{-0.9em}
S_{F_{(3)0}}(\Phi_{(3)} ) \equiv S^L_{(3)}(\Phi_{(3)}) = S_0+ \textstyle\frac{1}{3!}%
F_{(3)0}(\mathcal{A}) (\overleftarrow{s})^3 = S_0+ S_{F_{(3)0}|\mathrm{gf}} +  S_{F_{(3)0}|\mathrm{gh}}+  S_{F_{(3)0}|\mathrm{add}}  , \end{eqnarray}
\vspace{-1ex}
\begin{eqnarray}
\hspace{-0.5em} &\hspace{-0.9em}&\hspace{-0.9em}
S_{F_{(3)0}|\mathrm{gf}} +  S_{F_{(3)0}|\mathrm{gh}}   = \int d^d x\, tr\,  \Big\{ \widehat{B} \chi^F(\mathcal{A})  +  \frac{1}{3!}\int d^d y   \Big(  2B^{qp}(x)M^F(\mathcal{A},x;y) C^r(y)  \nonumber \\
\hspace{-0.5em} &\hspace{-0.9em}&\hspace{-0.9em}\phantom{S_{F_{(3)0}|\mathrm{gf}} +  S_{F_{(3)0}|\mathrm{gh}}  } +C^{p}(x)M^F(\mathcal{A},x;y) B^{qr}(y) \Big)\varepsilon_{pqr}\Big\}
 ,\label{Sgh3o}\\
\hspace{-0.9em} &\hspace{-0.9em}&\hspace{-0.9em}
S_{F_{(3)0}|\mathrm{add}}   = -\frac{\epsilon_{pqr}}{3!}\hspace{-0.2em}\int \hspace{-0.2em}d^{d}x \hspace{-0.1em} \Big\{\frac{1}{2}\chi^F(\mathcal{A}) \big[B^{pq}, C^r\big] +\hspace{-0.15em}\int \hspace{-0.15em}d^{d}y d^dz C^p(x) \frac{\delta M^F(\mathcal{A},x;y)}{\delta \mathcal{A}_{\mu
}(z)}D_\mu(z) C^r(z)  C^{q}(y) \hspace{-0.1em}\Big\}\hspace{-0.1em} ,  \label{Sadd3o}
\end{eqnarray}
where $\chi^F(\mathcal{A})$ may be interpreted as a Grassmann-odd analog of gauge conditions (\ref{N1gaugef}), (\ref{N2lqact})
used in the $N=1, 2$ BRST symmetry realizations for the quantum action, and therefore
$M^F(\mathcal{A},x;y)$ should be considered as a Grassmann-odd analog of
the Faddeev--Popov matrix (\ref{operatorFP}).

\subsection{Non-degenerate odd-valued change of fictitious fields}\label{N3oddfict}

To provide a satisfactory description, we must deal neither with the appearance
in $Z^L_0$ of the $\delta$-function $\delta(\chi^F)$ from odd-valued functions,
nor with the superdeterminant $\mathrm{sdet}M^F(\mathcal{A})$ from an odd-valued
matrix $M^F(\mathcal{A})$\footnote{If one attempts to exponentiate
the non-local path integral  (\ref{Pintgen}) over $\mathcal{M}^{(3)}= \{\Phi^{A_3}_{(3)}\}$
in the basis of, first, the auxiliary fields $\{C^p, B^{pq}, \widehat{B}\}$ by means of
the one Lie-group $G$-valued  field $B^{12}$ from the triplet of Grassmann-even
fields $B^{pq}$, to exponentiate $\delta(\chi)$, second, the pair, $C^{1},  C^{2}$
from the triplet of Grassmann-odd fields $C^{p}$, to exponentiate  $\det M$,
third, the pair $B^{13}, B^{23}$ from Grassmann-even fields $B^{pq}$,
and the remaining pair of Grassmann-odd fields, $C^{ {3}}, \widehat{B} $,  to exponentiate,
respectively, $\mathrm{det}^{-1}{}M$ and $\mathrm{det}{}M$, we get: $$ Z^L_0 = \int dA \delta(\chi)\,
\mathrm{det}^2 M(\mathcal{A}) \, \mathrm{det}^{-1} M(\mathcal{A})
\exp\Big\{\frac{\imath}{\hbar}S_0(\mathcal{A})\Big\} \ = \int d\Phi _{(3)}
\exp\Big\{\frac{\imath}{\hbar}\widetilde{S}_L(\Phi_{(3)})\Big\}, $$
 $$ \ \mathrm{for} \ \widetilde{S}_L(\Phi_{(3)}) = S_0(A) +
 \int d^dx \,tr\, \Big(\chi(\mathcal{A}) {B}^{12} + C^{1}M(\mathcal{A}) C^{2}
 + B^{23}M(\mathcal{A}) B^{13} +C^{3}M(\mathcal{A})\widehat{B}\Big).
$$ However, to provide  $N=3$ BRST invariance of the local action
$\widetilde{S}_L(\Phi_{(3)})$ for Yang--Mills theory one must impose  additional
requirement:   $\delta_\lambda  B^{12}  = 0 $, being rather restrictive one.},
we may pass to another basis of auxiliary fields, $\widehat{\Phi}_{(3)}$,
in the representation space $\mathcal{M}^{(3)}$, $\mathcal{M}^{(3)} =\mathcal{M}^{(3)}_{\mathrm{min}}$,  of the $N=3$  superalgebra
${\mathcal{G}}(3)$ with the same number of Grassmann-odd and Grassmann-even fields.
To this end, we introduce a non-degenerate transformation in $\mathcal{M}^{(3)}$:
${\Phi}_{(3)} \to \widehat{\Phi}_{(3)}=\Xi {\Phi}_{(3)}$, with
unaffected Yang--Mills fields $\mathcal{A}_\mu$, ghost fields $C^1, C^3$, bosonic fields
$B^{13}$, and to be transformed fictitious fields
$\phi^M = (B^{23}, B^{12}, C^{ 2}, \widehat{B}{})$,
 by introducing a Grassmann-odd non-degenerate matrix
$\mathcal{N}=\|\mathcal{N}_{MN}\|$ (analogous to the odd supermatrix
$\omega = \|\omega^{\mathbf{AB}}\|$ = $\|(\Gamma^{\mathbf{A}},\Gamma^{\mathbf{B}})\| $,
$\epsilon(\omega)=1$, resulting from the odd Poisson bracket, $ (\bullet,\bullet)$,   calculations
with respect to the field-antifield variables $\Gamma^\mathbf{A}$ in the field-antifield
formalism \cite{bv}, \cite{ht}), composed from the unit matrices $1_{(\hat{N}{}^2-1)}$ with suppressed $su(\hat{N})$-indices,
  as follows:
\begin{equation}\label{calN}
  \phi^M \to \widehat{\phi}{}^M = \mathcal{N}^{MN}\phi^N : \  \left(\begin{array}{c}
    \overline{B}{}_2 \\
     {B} \\
     \overline{C}{}^{1}\\
     \overline{C}{}^{3}
  \end{array}\right)= \left(\begin{array}{cccc}
                              0 & 0 & \Pi & 0 \\
                              0 & 0 & 0 & \Pi\\
                              \Pi & 0 & 0 & 0 \\
                              0 & \Pi & 0 & 0
                            \end{array}
  \right) \left(\begin{array}{c}
    {B}{}^{23} \\
     {B}^{12} \\
     {C}{}^{2}\\
      \widehat{B}{}
  \end{array}\right),
\end{equation}
with the  odd non-degenerate supermatrix $\Pi$, which turns the only fields of definite parity into new fields with the same properties but with opposite parity:
$\Pi \big({B}{}^{23},   {B}^{12},   {C}{}^{2},  \widehat{B}\big)  = \big(  \overline{C}{}^{1},     \overline{C}{}^{3},   \overline{B}{}_2,  {B}\big)$,
  so that by definition,
the  property  to be idempotent for $\Pi$ holds: $\Pi^2=1$.
Notice that the separation of the (un)transformed fields in ${\Phi}_{(3)}$ is not unique
for unaffected $\mathcal{A}^\mu$. Note, that in the usual sense \cite{berezin}, \cite{books3}  $\mathrm{sdet} \mathcal{N}=0$.

The supermatrix $\mathcal{N}$ plays the role of an inverse for itself, which make it possible
to express the initial fictitious fields $\phi^N $ from (\ref{calN})
as functions of new fictitious fields $\widehat{\phi}^N $:
\begin{equation}\label{invtrans}
  \phi^M =   \mathcal{N}^{MN}\widehat{\phi}^N,\  \epsilon(\mathcal{N})=1,\   \mathrm{because\ of} \   \mathcal{N}^2= \mathbf{1}_{4(\hat{N}{}^2-1)} .
\end{equation}

\subsection{$N=3$ BRST-invariance and path integral in new fictitious fields}\label{N3FPtricklocnew}

The following step is based on a definition of the gauge fermion $F_{(3)0}(\mathcal{A})$
with help of  the odd matrix $\Pi$ in quadratic form consistent for the Landau gauge:
\begin{eqnarray}
&& F^L_{(3)0}(\mathcal{A})\ =  -\frac{1}{2}\int d^d x \, tr\,   \mathcal{A}^{\mu}\Pi \mathcal{A}_\mu = - \frac{1}{2}\int d^d x \,   \mathcal{A}^{\mu m}\Pi^{mn} \mathcal{A}^n_\mu , \  \ \mathrm{for }\ \ \epsilon(F^L_{(3)0})=1. \label{N3gaugefap}
\end{eqnarray}
Because  the  map $\Pi$ acts  linearly, turning the points (with coordinates) in a fiber of the respective bundle into the same  points (with coordinates) in  a fiber of  another bundle, but with opposite parity, then the respective infinitesimal  gauge for $\mathcal{A}^\mu$ and $N(3)$ SUSY transformations for $\Phi^{A_3}_{(3)}$ make by natural the properties:
\begin{eqnarray}
  &&  \Pi \delta  \mathcal{A}^\mu =    \delta  (\Pi \mathcal{A}^\mu)  \Rightarrow \Pi \delta_{\lambda}  \mathcal{A}^\mu =    \delta_\lambda  (\Pi \mathcal{A}^\mu) =\Pi D^\mu (\mathcal{A})C^p\lambda_p, \label{piamu}   \\
&& \Pi \partial_\mu =  \partial_\mu\Pi, \  \ \Pi D^\mu (\mathcal{A})C^p =  D^\mu (\mathcal{A}) \Pi C^p \  \Leftrightarrow \  \Pi \big[\mathcal{A}, C^p\big] =  \big[\mathcal{A}, \Pi C^p\big], \label{piamu2}
\end{eqnarray}
where the last relation maybe considered as the continuation of the commutativity property of $\Pi$  with partial derivative $\partial_\mu$.

Now,
we can write the path integral related to (\ref{Pintgen}) in a local form,
 (\ref{PintfpLlock}) for $k=1$  with the action $S_{F^L_{(3)0}}$ (\ref{F(A)o}),  fermionic functional $F^L_{(3)0}$, in terms of
a new basis of $\{\widehat{\Phi}^{A_3}_{(3)}\}$  for the representation space
of the $\mathcal{G}(3) $ superalgebra, as follows:
\begin{eqnarray}
\label{PintfpLloc2f}
\hspace{-0.5em}  &\hspace{-1em}&\hspace{-0.5em} {Z}_{{F}^L_{(3)0}} =  \int  d \widehat{\Phi}_{(3)}   \exp \Big\{\frac{\imath}{\hbar}S_{F^L_{(3)0}}\big(\widehat{\Phi}_{(3)}\big)\Big\}, \ \mathrm{with}  \ S_{F^L_{(3)0}}  = S_0(\mathcal{A}) +\frac{1}{3!}%
F^L_{(3)0}(\mathcal{A}) \big(\overleftarrow{s}\big){}^3\Big|_{\big({\phi}^M \to \widehat{\phi}^M\big)}, \\
  \hspace{-0.5em}&\hspace{-1em}&\hspace{-0.5em} S_{F^L_{(3)0}|\mathrm{gf}} +  S_{F^L_{(3)0}|\mathrm{gh}}   = \frac{1}{2}\int d^d x\, tr\,     \left\{\Big(  \big(\Pi \chi(\mathcal{A})\big)+\chi(\mathcal{A})\Pi \Big)\widehat{B}     -  \frac{1}{2}\Big(  {M}(\mathcal{A}) \big(\Pi C^q \big)\right. \nonumber\\
\hspace{-0.5em}&\hspace{-0.5em}&\hspace{-0.5em}  \qquad \left.
- {M}(\mathcal{A})C^q\Pi \Big) \varepsilon_{pqr} B^{pr}\right\}\Big|_{\big({\phi}^M \to \widehat{\phi}^M\big)}, \label{Sgh3fin}\\
\hspace{-0.5em} &\hspace{-0.5em}&\hspace{-0.5em}
 S_{F^L_{(3)0}|\mathrm{add}}   =  \frac{\varepsilon_{pqr}}{2\cdot 3!}\int d^dx \,tr\, \Bigg\{\frac{1}{2}\Big(  \big(\Pi \chi(\mathcal{A})\big)+\chi(\mathcal{A})\Pi \Big)  \big[B^{pq}, C^r\big] +\Big\{   \big[{M}(\mathcal{A})C^r, \Pi C^{q}\big]
\nonumber\\
\hspace{-0.5em}&\hspace{-0.5em}&\hspace{-0.5em} \qquad
 + \big[D^\mu(\mathcal{A})C^r, \partial_\mu \Pi C^{q}\big]   -\Big(\big[{M}(\mathcal{A})C^r, C^{q}\big]+ \big[D^\mu(\mathcal{A})C^r, \partial_\mu C^{q}\big]\Big)\Pi \Big\}  C^{p}\Bigg\}\Big|_{\big({\phi}^M \to \widehat{\phi}^M\big)} ,  \label{Sadd3fin}
\end{eqnarray}
with usual Faddeev-Popov matrix, ${M}={M}(\mathcal{A})$ and with  taken account for the relations
\begin{eqnarray}
\hspace{-0.5em}&\hspace{-0.5em}& \hspace{-0.5em} \label{PauxsaKapp}    F^L_{(3)0}\big(\mathcal{A}\big) \overleftarrow{s}{}^{p} = \frac{1}{2}\hspace{-0.15em}\int\hspace{-0.15em} d^dx \,tr\, \Big\{D_\mu C^p \Pi\mathcal{A}^{\mu
}-\mathcal{A}^{\mu
}\Pi D_\mu C^p\Big\}\hspace{-0.15em}   = \frac{1}{2} \hspace{-0.15em}\int \hspace{-0.15em}d^dx \,tr\, \Big\{  \big(\Pi \chi(\mathcal{A})\big)+\chi(\mathcal{A})\Pi \hspace{-0.15em}\Big\}  C^p,\\
\hspace{-0.5em}&\hspace{-0.5em}& \hspace{-0.5em} F^L_{(3)0}\big(\mathcal{A}\big) \overleftarrow{s}{}^{p}\overleftarrow{s}{}^{q}\varepsilon_{pqr} =
\frac{1}{2} \int d^dx \,tr\, \left\{\Big(  \big(\Pi \chi(\mathcal{A})\big)+\chi(\mathcal{A})\Pi \Big)  B^{pq}-  \Big(  \big( {M}(\mathcal{A}) \Pi C^q \big) \right. \nonumber\\
\hspace{-0.5em}&\hspace{-0.5em}& \hspace{-0.5em}\phantom{F^L_{(3)0}\big(\mathcal{A}\big) \overleftarrow{s}{}^{p}\overleftarrow{s}{}^{q}\varepsilon_{pqr}} \left. -{M}(\mathcal{A})C^q\Pi \Big)  C^{p}\right\}\varepsilon_{pqr}
 , \label{Pauxapp} \\
\hspace{-0.5em}&\hspace{-0.5em}& \hspace{-0.5em} F_{(3)0}\big(\mathcal{A}\big) (\overleftarrow{s})^{3}=
\frac{1}{2} \int d^dx \,tr\, \Bigg\{\Big(  \big(\Pi \chi(\mathcal{A})\big)+\chi(\mathcal{A})\Pi \Big)  \Big(3!\widehat{B} +  \frac{1}{2} \big[B^{pq}, C^r\big] \varepsilon_{pqr} \Big) \nonumber\\
\hspace{-0.5em}&\hspace{-0.5em}& \hspace{-0.5em} \quad  -  \Big\{3\Big(   \big( {M}(\mathcal{A}) \Pi C^q \big) -{M}(\mathcal{A})C^q\Pi \Big)  B^{pr} +\Big( \Big(\big[{M}(\mathcal{A})C^r, C^{q}\big]
\nonumber\\
\hspace{-0.5em}&\hspace{-0.5em}& \hspace{-0.5em} \quad+ \big[D^\mu(\mathcal{A})C^r, \partial_\mu C^{q}\big]\Big)\Pi -  \big[{M}(\mathcal{A})C^r, \Pi C^{q}\big]
  -\big[D^\mu(\mathcal{A})C^r, \partial_\mu \Pi C^{q}\big]   \Big)  C^{p}\Big\}\varepsilon_{pqr}\Bigg\}.\label{Ps3(AB)app}%
\end{eqnarray}
Here the relations (\ref{auxsaK}), (\ref{auxsaK1}),  (\ref{piamu}), (\ref{piamu2}) and (\ref{idDmu})  for Landau gauge $\chi(\mathcal{A})=0$ were used as well as the vanishing of the terms, $\big[C^{p_1},C^{p_2}\big] \varepsilon_{[p]_3} \equiv 0$.

Note, first, the  terms proportional to the $\Pi\chi(\mathcal{A})$  in (\ref{Ps3(AB)app}) maybe easily elaborated by the rule
\begin{eqnarray}\label{addrule1}
&& tr\,\big(\Pi \chi(\mathcal{A})\big)\widehat{B} =  \Pi^{mn} \chi^n(\mathcal{A}) \widehat{B}^m = \chi^n(\mathcal{A})\Pi^{mn} \widehat{B}^m =  tr\, \chi(\mathcal{A})\Pi\widehat{B},\\
  \label{addrule2}
&& tr\,\big(\Pi \chi(\mathcal{A})\big)  \big[B^{pq}, C^r\big] = tr\, \chi(\mathcal{A})  \Pi\big[B^{pq}, C^r\big] ,
\end{eqnarray}
by virtue of the properties (\ref{piamu}), (\ref{piamu2}). Second,  the  quadratic in the fictitious  fields with Faddeev--Popov matrix summands,  we can present due to the same properties as follows:
\begin{eqnarray}\label{addrule3}
&& tr\, \big( {M}(\mathcal{A}) \Pi C^q \big)  B^{pr}  = tr\, B^{pr} {M}(\mathcal{A}) \Pi C^q = tr\, B^{pr} \Pi {M}(\mathcal{A})  C^q =tr\, (\Pi B^{pr}) {M}(\mathcal{A})  C^q.
\end{eqnarray}
Expressing the fields $\phi^M$ in terms of $\widehat{\phi}^M$, according to the change of variables (\ref{calN}) in $\mathcal{M}^{(3)}$, we get for the action $S_{F^L_{(3)0}}$ (\ref{PintfpLloc2f})--(\ref{Sadd3fin}) with use of (\ref{addrule1})--(\ref{addrule3})  and with use of dual field $B_2 =  - B^{13}=\varepsilon_{132}B^{13}$:
\begin{eqnarray}
  && S_{F^L_{(3)0}}\big(\widehat{\phi}\big)  = S_0(\mathcal{A}) + \int d^d x\, tr\,     \Big(  \chi(\mathcal{A}){B}     +
    \overline{C}{}^3 {M}(\mathcal{A})  C^3 +\overline{C}{}^1 {M}(\mathcal{A})  C^1+ \overline{B}_2{M}(\mathcal{A}) B_2 \Big) \nonumber\\
&&  \qquad
+ \frac{1}{3!}\int d^dx \,tr\, \Bigg\{   \chi(\mathcal{A})\Pi \Big( \big[B_2,\,\Pi \overline{B}_2\big] +\sum_{k=0}^1\big[\Pi \overline{C}{}^{2k+1},\, C^{2k+1}\big] \Big) \nonumber\\
&& \qquad+\frac{\varepsilon_{pqr}}{2}\Big\{   \big[{M}(\mathcal{A})C^r, \Pi C^{q}\big]  C^{p}+ \big[D^\mu(\mathcal{A})C^r, \partial_\mu \Pi C^{q}\big]  C^{p}
\nonumber\\
&& \qquad
  -\Big(\big[{M}(\mathcal{A})C^r, C^{q}\big]+ \big[D^\mu(\mathcal{A})C^r, \partial_\mu C^{q}\big]\Big)\Pi  C^{p}\Big\} \Big|_{\big[(\Pi C^2, C^2) \to ( \overline{B}_2, \Pi \overline{B}_2)\big]} \Bigg\}. \label{finSminN3}%
\end{eqnarray}
Here, the role of Faddeev--Popov ghosts is a mixed one, in comparison
with the initial basis of fictitious fields $C^p, B^{pq}, \widehat{B}$.
For example, in the first  row of (\ref{finSminN3}) for the fields $C, \overline{C}$,
used within the original Faddeev--Popov  quantization  as ghost and antighost fields, we have,
respectively, $C^{1}, \Pi{B}^{ 23}$ and $C^{3}, \Pi{B}^{12}$.

Therefore, as far as the last condition in (\ref{piamu2}) holds true, the functional
$S_{F^L_{(3)0}}\big(\widehat{\phi}\big) $, with the gauge functional (\ref{N3gaugefap}) which
determines the path integral $
 {Z}_{{F}^L_{(3)0}}$ (\ref{PintfpLloc2f}) in the Landau gauge with a local quantum
action solving the problem of generalization of the Faddeev--Popov rules in the case
of the irreducible representation $N=3$-parametric $\mathcal{G}(3)$ superalgebra.

The latter local action (as well as the measure $d \widehat{\Phi}_{(3)}$) corresponding
to the Landau gauge is invariant under $N=3$   (therefore called as $N=3$ BRST)  transformations, which, at the  algebraic
level in a new basis of fields,  $\widehat{\Phi}^{A_3}_{(3)}$, are written with allowance for
(\ref{minN3}), (\ref{calN}), (\ref{invtrans}), as follows:
 \begin{eqnarray}
\hspace{-0.5em}&\hspace{-0.5em}& \hspace{-0.5em}  \mathcal{A}_\mu  \overleftarrow{s}{}^p   \ = \ D_\mu \big(C^1 \delta^{1p}+ \Pi \overline{B}_2  \delta^{2p}+ C^3\delta^{3p}\big), \label{N3Afin}\\
\hspace{-0.5em}&\hspace{-0.5em}& \hspace{-0.5em}  C^1 \overleftarrow{s}{}^p   \ = \ \textstyle\frac{1}{2}\big[C^{1}, C^{1}\big]\delta^{1p} + \big\{\Pi \overline{C}{}^{3}+\frac{1}{2}\big[C^{1}, \Pi \overline{B}_{2}\big] \big\}\delta^{2p}+\big\{ -B_2+\textstyle\frac{1}{2}\big[C^{1}, C^{3}\big]  \big\}\delta^{3p}  %
 , \nonumber\\
  \hspace{-0.5em}&\hspace{-0.5em}& \hspace{-0.5em}  \overline{B}_2 \overleftarrow{s}{}^p   \ = \ \big\{-\overline{C}{}^3+\textstyle\frac{1}{2}\Pi\big[\Pi \overline{B}{}_2, C^{1}\big]\big\} \delta^{1p} + \frac{1}{2}\Pi\big[\Pi \overline{B}_2, \Pi \overline{B}_{2}\big] \delta^{2p} +\big\{ \overline{C}{}^1+\frac{1}{2}\Pi\big[\Pi \overline{B}_2, C^{3}\big]  \big\}\delta^{3p}  %
 , \label{N3Cfin}\\
       \hspace{-0.5em}&\hspace{-0.5em}& \hspace{-0.5em} C^3 \overleftarrow{s}{}^p   \ = \ \big\{ B_2 +\textstyle\frac{1}{2}\big[C^{3}, C^{1}\big]\big\}\delta^{1p} + \big\{-\Pi \overline{C}{}^{1}+\frac{1}{2}\big[C^{3}, \Pi \overline{B}_{2}\big] \big\}\delta^{2p}+\frac{1}{2}\big[C^{3}, C^{3}\big] \delta^{3p}  %
 , \nonumber\\
     \hspace{-0.5em}&\hspace{-0.5em}& \hspace{-0.5em} \overline{C}{}^3 \overleftarrow{s}{}^p =   \textstyle\frac{1}{2}\Pi\Big(
\big[\Pi \overline{C}{}^3, C^{1}\big]-\frac{1}{6}\big[C^{[1},\,\big[\Pi \overline{B}{}^{2]},\,C^{1}\big]\big]\Big)\delta^{1p}+ \textstyle\frac{1}{2}\Pi\Big(
\big[\Pi \overline{C}{}^3, \Pi \overline{B}_{2}\big]-\frac{1}{6}\big[C^{[1},\,\big[\Pi \overline{B}{}^{2]},\,\Pi \overline{B}_{2}\big]\big]\Big)\delta^{2p}\nonumber \\
 \hspace{-0.5em}&\hspace{-0.5em}& \hspace{-0.5em} \phantom{\overline{C}{}^3 \overleftarrow{s}{}^r}+\Big\{{B} +\textstyle\frac{1}{2} \Pi\Big(
\big[\Pi \overline{C}{}^3,\, C^{3}\big]-\frac{1}{6}\big[C^{[1},\,\big[\Pi \overline{B}{}^{2]},\,C^{3}\big]\big]\Big)\Big\}\delta^{3p} ,  \label{N3algBRST1} \\
  &&   B_2  \overleftarrow{s}{}^p =  \textstyle\frac{1}{2}\Big(
\big[B_2, C^{1}\big]+\frac{1}{6}\big[C^{[1},\big[C^{3]},C^{1}\big]\big]\Big)\delta^{1p} + \Big\{\Pi{B} +\textstyle\frac{1}{2}\Big(
\big[B_2,\, \Pi\overline{B}_2\big]+\frac{1}{6}\big[C^{[1},\big[{C}^{3]},\, \Pi\overline{B}_{2}\big]\big]\Big)\Big\}\delta^{2p}\nonumber \\
\hspace{-0.5em}&\hspace{-0.5em}& \hspace{-0.5em} \phantom{B_2  \overleftarrow{s}{}^r}   + \textstyle\frac{1}{2}\Big(
\big[B_2, C^{3}\big]+\frac{1}{6}\big[C^{[1},\,\big[{C}^{3]},\,C^{3}\big]\big]\Big)\delta^{3p},  \label{N3algBRST2}
\end{eqnarray}
\vspace{-1ex} \begin{eqnarray}
     \hspace{-0.5em}&\hspace{-0.5em}& \hspace{-0.5em} \overline{C}{}^1  \overleftarrow{s}{}^p =  \Big\{{B} +\textstyle\frac{1}{2}\Pi\Big(
\big[\Pi\overline{C}{}^1, \, C^{1}\big]-\frac{1}{6}\big[\Pi\overline{B}{}^{[2},\big[C^{3]},\,C^{1}\big]\big]\Big)\Big\}\delta^{1p}
 +\textstyle\frac{1}{2}\Pi\Big(
\big[\Pi\overline{C}{}^1, C^{3}\big]
\nonumber \\
\hspace{-0.5em}&\hspace{-0.5em}& \hspace{-0.5em} \phantom{\overline{C}{}^1  \overleftarrow{s}{}^r} -\textstyle\frac{1}{6}\big[\Pi\overline{B}{}^{[2},\,\big[C^{3]},\,C^{3}\big]\big]\Big)\delta^{3p} +
\textstyle\frac{1}{2} \Pi\Big(
\big[\Pi\overline{C}{}^1, \, \Pi\overline{B}_{2}\big]-\textstyle\frac{1}{6}\big[\Pi\overline{B}{}^{[2},\,\big[C^{3]},\,\Pi \overline{B}_{2}\big]\big]\Big)\delta^{2p}
     ,  \label{N3algBRST3} \\
   \hspace{-0.5em}&\hspace{-0.5em}& \hspace{-0.5em}   {B}\overleftarrow{s}{}^p  \ =\   \textstyle\frac{1}{2}\Pi\Bigg(\big[\Pi{B},\,C^1\delta^{1p}+\, \Pi \overline{B}_2\delta^{2p}+\,C^3\delta^{3p}\big] - \Big\{\frac{1}{2}\Big[\Big(\big[\Pi \overline{C}{}^{3},\,C^3\big] +\big[B_{2},\, \Pi \overline{B}_2\big]\nonumber\\
    \hspace{-0.5em}&\hspace{-0.5em}& \hspace{-0.5em} \phantom{{B}\overleftarrow{s}{}^p\ } +\big[\Pi \overline{C}{}^{1},\,C^1\big]\Big),\,\big\{C^1\delta^{1p}+\, \Pi\overline{B}_2\delta^{2p}+\,C^3\delta^{3p}\big\}\Big]+\textstyle\frac{1}{3} \big[\big[\big\{\Pi \overline{C}{}^{3}\delta^{2p} - B_2\delta^{3p}\big\},\,C^{[2}\big],\,C^{3]}\big]
     \nonumber\\
    \hspace{-0.5em}&\hspace{-0.5em}& \hspace{-0.5em} \phantom{{B}\overleftarrow{s}{}^p\ }
     +\textstyle\textstyle\frac{1}{3} \big[\big[\big\{-\Pi \overline{C}{}^{3}\delta^{1p} + \Pi \overline{C}{}^1\delta^{3p}\big\},\,C^{[3}\big],\,C^{1]}\big]
     +\frac{1}{3} \big[\big[\big\{B_{2}\delta^{1p} - \Pi \overline{C}{}^1\delta^{2p}\big\},\,C^{[1}\big],\,\Pi \overline{B}{}^{2]}\big]
     \Big\} \Bigg) ,\label{N3Bfin}
 \end{eqnarray}
where we introduced  the formal identification $\overline{B}_{2}=\overline{B}{}^{2}$ to use the  antisymmetry  of: $C^{[1} \Pi\overline{B}{}^{2]} = C^{1}\Pi\overline{B}{}^{2} - \Pi\overline{B}{}^{2}C^{1}$ being inherited from one for  $C^{[1} C^{2]}$\footnote{The action of the Grassmann-odd  operator $\Pi$ may be determined on the $su(\hat{N})$ commutator $\big[A,\,B \big]$ of any Grassmann- homogeneous quantities  $A, B$  as $\Pi\big[A,\,B \big]=\big[\Pi A,\,B \big] =(-1)^{\epsilon(A)}\big[A,\,\Pi B \big]$ in such a way that $\Pi $ should act only on the  fields $\widehat{\phi}{}^M$ (\ref{calN}) and $\Pi\widehat{\phi}{}^M$. E.g.
$\Pi\big[\Pi{B},\,C^1\big] = \big[{B},\,C^1\big]$ and $\Pi\big[\Pi{B},\, \Pi \overline{B}_2\big]=\big[{B},\, \Pi \overline{B}_2\big]=-\big[\Pi{B},\,  \overline{B}_2\big]$.}.

Thus, we see, that the preservation of the explicit $N=3$ BRST symmetry for the quantum action $S^L_{(3)}(\Phi_{(3)})$ in the space of $\mathcal{G}(3)$-irreducible representation $\mathcal{M}^{(3)}$ requires the introduction of odd non-degenerate supermatrix $\mathcal{N}$ with destroying of $\mathcal{G}(3)$-covariance of the fields $\Phi_{(3)}$  to get local  path integral (\ref{PintfpLloc2f}) with  $N=3$ BRST invariance (\ref{N3Afin})-(\ref{N3Bfin}).

This fact proves the validity condition (\ref{kN}) of the Statement~1 concerning gauge-fixing procedure for odd $N$.

\section{$N=4$ BRST Invariant Yang--Mills Action in $R_{\xi}$-like Gauges}

\label{AppB} \renewcommand{\theequation}{\Alph{section}.\arabic{equation}} \setcounter{equation}{0}

In this Appendix, we present the details of calculations used in
Section~\ref{N4gf} to find $N=4$ BRST invariant quantum  action  (\ref{qexi4})--(\ref{addxi4})  and
establish a correspondence between the gauge-fixing
procedures in the Yang--Mills theory described by a gauge-fixing function
$\chi(\mathcal{A},B) = 0$ from the class of $R_{\xi}$-gauges in  $N=1$ BRST formulation and by a gauge-fixing functional $Y^0_{(4)\xi}$ in the suggested $N=4$ BRST
quantization.

To calculate $ S_{Y_{(4)\xi}}\big({\Phi}_{(4)}\big)$ we have used the results of applications  (\ref{auxsaK})--(\ref{s3(AB)}),  (\ref{varAs}), (\ref{varABs}),  (\ref{sABcom})  adapted for $N=4$ case, as well as the property (\ref{s4(AB)}) for differentiation of the product and commutator of any two functions by products of the generators $\overleftarrow{s}{}^p$ up to $4$-th order.

Thus,  for the quadratic gauge bosonic functional, $Y^0_{(4)\xi}(\Phi_{(4)}) = Y^0_{(4)}(\mathcal{A})+Y^B_{(4)\xi}(B^{q_1q_2})$, (\ref{gBn4xi}) we need  the preliminary calculations with action of  the first and second powers of  $\overleftarrow{s}{}^{r_1}$  on $Y^0_{(4)}(\mathcal{A})$ with use of the notation  for the compact writing, $\varepsilon_{r_1r_2r_3r_4} \equiv \varepsilon_{[r]_4}$:
\begin{eqnarray}
&& Y^0_{(4)}(\mathcal{A}) \overleftarrow{s}{}^{r_1} =  \int
d^{d}x\ tr \mathcal{ A}_{\mu } D^\mu(\mathcal{A})C^{r_1} = -\int
d^{d}x\ tr  (\partial^\mu\mathcal{ A}_{\mu }) C^{r_1}   ,\label{AAmu1}\\
  && \Big((\partial^\mu\mathcal{ A}_{\mu }) C^{r_1}\Big) \overleftarrow{s}{}^{r_1}\overleftarrow{s}{}^{r_2}\varepsilon_{[r]_4} = \Big(C^{r_1}M(\mathcal{A})  C^{r_2}  + (\partial^\mu\mathcal{ A}_{\mu })  B^{r_1r_2} \Big)\varepsilon_{[r]_4}, \label{AAmu12}
\end{eqnarray}
of the third powers, with account for the identities (\ref{idDmu})  below and equalities $\int
d^{d}x  \,tr C^{r_1}M(\mathcal{A}) B^{r_2r_3}  = \int
d^{d}x  \,tr B^{r_2r_3}\big\{M(\mathcal{A})  -[(\partial^\mu\mathcal{ A}_{\mu }), \ ]\big\}C^{r_1}$, obtained with help of  the integration by parts:
\begin{eqnarray}   \hspace{-0.5em}&\hspace{-0.5em}& \hspace{-0.5em}Y^0_{(4)}(\mathcal{A}) \prod_{k=1}^{3} \overleftarrow{s}{}^{r_k}\varepsilon_{[r]_4}=  - \int
d^{d}x  \,tr \Big(-B^{r_1r_3}M(\mathcal{A})  C^{r_2}+ C^{r_1}M(\mathcal{A})  B^{r_2r_3}-C^{r_1}\big\{\big[M(\mathcal{A})C^{r_3},\,  C^{r_2}\big]\nonumber\\
   \hspace{-0.5em}&\hspace{-0.5em}& \hspace{-0.5em}\quad   +  \big[D^\mu(\mathcal{A})C^{r_3},\,  \partial_\mu C^{r_2}\big]\big\} + B^{r_1r_2}M(\mathcal{A})C^{r_3} + (\partial^\mu\mathcal{ A}_{\mu }) \Big\{ B^{r_1r_2r_3} + \textstyle\frac{1}{2}
\big[B^{r_1r_2}, C^{r_3}\big]\Big\} \Big)\varepsilon_{[r]_4}\nonumber \\
\hspace{-0.5em}&\hspace{-0.5em}& \hspace{-0.5em}  = - \int
\hspace{-0.15em}d^{d}x  \,tr \Big(\hspace{-0.1em} 3 B^{r_1r_2}M(\mathcal{A}) C^{r_3} \hspace{-0.15em}+ C^{r_1}\partial_\mu\big[D^\mu C^{r_2},  C^{r_3}\big]\hspace{-0.15em}+ \Big\{\hspace{-0.1em} B^{r_1r_2r_3}\hspace{-0.15em} + \textstyle\frac{3}{2}
\big[B^{r_1r_2}, C^{r_3}\big]\hspace{-0.1em}\Big\}\partial^\mu\mathcal{ A}_{\mu }\hspace{-0.1em} \Big)\varepsilon_{[r]_4} ,\label{AAmu123}
\end{eqnarray}
and of the fourth power:
\begin{eqnarray}    && Y^0_{(4)}(\mathcal{A}) \prod_{k=1}^{4} \overleftarrow{s}{}^{r_k}\varepsilon_{[r]_4} = -\int
d^{d}x  \,tr  \Big( -3\Big\{ B^{r_1r_2r_4} + \textstyle\frac{1}{2}
\big[B^{r_1r_2}, C^{r_4}\big]\Big\}M(\mathcal{A})C^{r_3}
  + 3 B^{r_1r_4}M(\mathcal{A})  B^{r_2r_3} \nonumber \\
&& \quad
 + (\partial^\mu\mathcal{ A}_{\mu }) \Big\{ \varepsilon^{[r]_4}{B}  +\textstyle\frac{1}{2}\Big(\hspace{-0.1em}\big[B^{r_1r_2r_3},C^{r_4}\big]
  - \displaystyle\sum_{P}\hspace{-0.15em}(-1)^{P(r_1,r_2,r_3)}\hspace{-0.1em}\Big\{\hspace{-0.1em}\textstyle\frac{1}{4}\big[\big[B^{r_1r_2},C^{r_3}\big],C^{r_4}\big] \nonumber \\
&& \quad +\textstyle\frac{1}{3} \big[\big[B^{r_1r_4},C^{r_2}\big],C^{r_3}\big] \hspace{-0.1em} \Big\} \hspace{-0.1em}\Big) - \textstyle\frac{3}{2}
\big[B^{r_1r_2r_4}, C^{r_3}\big] -  \textstyle\frac{3}{4}
\big[\big[B^{r_1r_2}, C^{r_4}\big], C^{r_3}\big]\Big\} \nonumber \\
&& \quad  +\Big\{ B^{r_1r_2r_3} + \textstyle\frac{3}{2}
\big[B^{r_1r_2}, C^{r_3}\big]\Big\} M(\mathcal{A})  C^{r_4}     +
 C^{r_1}\partial_\mu \Big\{ \big[D^\mu C^{r_2},\,    B^{r_3r_4}\big] -\big[D^\mu B^{r_2r_4},\,   C^{r_3} \big]\nonumber \\
&& \quad +\big[\big[D^\mu C^{r_4},\,    C^{r_2}\big],\,    C^{r_3}\big]\Big\}- 3 B^{r_1r_2}\partial_\mu \big[D^\mu C^{r_4},\,    C^{r_3}\big]  + B^{r_1r_4}\partial_\mu \big[D^\mu C^{r_2},\,    C^{r_3}\big]\Big)\varepsilon_{[r]_4}\label{ghgf4} \\
&& \   = -\int
d^{d}x  \,tr    \Big(\Big\{4 B^{r_1r_2r_3}M(\mathcal{A}) C^{r_4}  + 3 B^{r_1r_2}M(\mathcal{A})  B^{r_3r_4}+  (\partial^\mu\mathcal{ A}_{\mu })  {B}\varepsilon^{[r]_4}\Big\}  \nonumber \\
&& \quad  + (\partial^\mu\mathcal{ A}_{\mu })  \Big\{2\big[B^{r_1r_2r_3}, C^{r_4}\big]  - \big[\big[B^{r_1r_2}, C^{r_3}\big], C^{r_4}\big]\Big\} -  B^{r_1r_2}\Big\{\big[ C^{r_3},  M(\mathcal{A})C^{r_4}\big]+4\big[ \partial_\mu C^{r_3},  D^\mu C^{r_4}\big]\Big\}\nonumber \\
&& \quad   +
 C^{r_1}\partial_\mu \Big\{ \big[D^\mu C^{r_2},\,    B^{r_3r_4}\big] -\big[C^{r_2},\, D^\mu B^{r_3r_4}   \big] + \big[\big[D^\mu C^{r_2},\, C^{r_3}\big],  C^{r_4}\big]\Big\}\Big)\varepsilon_{[r]_4}.
   \label{AAmu1234}
\end{eqnarray}
Here, we have used that, $\big[C^{r_1},C^{r_2}\big]\varepsilon_{r_1r_2r_3r_4}\equiv 0$, definition of the Faddeev-Popov operator (\ref{operatorFP}), integration by parts, relations (\ref{auxsaK}),  (\ref{auxsaK1}) its analog, $\big(M(\mathcal{A})B^{r_1r_2}\big)\overleftarrow{s}{}^{r_3}$,    and easily checked  Leibnitz rule of the commutator differentiation for covariant derivative, $D_\mu(\mathcal{A})$:
\begin{eqnarray}
&& \big(M(\mathcal{A})B^{r_1r_2}\big)\overleftarrow{s}{}^{r_3}  \ =\    \partial_\mu \big[D^\mu(\mathcal{A}) C^{r_3},\,B^{r_1r_2}\big] + M(\mathcal{A})\big(B^{r_1r_2}\overleftarrow{s}{}^{r_3}\big) \nonumber \\
 && \phantom{\big(M(\mathcal{A})B^{r_1r_2}\big)\overleftarrow{s}{}^{p}}\ = \ \big[M(\mathcal{A}) C^{r_3},\,B^{r_1r_2}\big] +\big[D^\mu(\mathcal{A}) C^{r_3},\, \partial_\mu B^{r_1r_2}\big] \nonumber \\
 && \phantom{\big(M(\mathcal{A})B^{r_1r_2}\big)\overleftarrow{s}{}^{p}}\  + M(\mathcal{A})\Big\{{B}^{r_1r_2r_3}+ \textstyle\frac{1}{2}
\big[B^{r_1r_2}, C^{r_3}\big]-\frac{1}{12}\big[C^{[r_1},\big[C^{r_2]},C^{r_3}\big]\big]\Big\}
, \label{auxsaK2} \\
\label{dercovcomm}
  && D_\mu(\mathcal{A})\big[B^{r_1r_2}, C^{r_3}\big] \ = \ \big[\big(D_\mu(\mathcal{A})B^{r_1r_2}, C^{r_3}\big]+\big[B^{r_1r_2}, D_\mu(\mathcal{A})C^{r_3}\big],
\end{eqnarray}
as well as the relations, first, for the terms with permutation, $P(r_1,r_2,r_3)$, and second,  for $su(\hat{N})$-valued functions $F, G$:
\begin{eqnarray}
&&   -\Big(\displaystyle\sum_{P}\hspace{-0.15em}(-1)^{P(r_1,r_2,r_3)}\hspace{-0.1em}\textstyle\frac{1}{2}\Big\{\textstyle\frac{1}{4}\big[\big[B^{r_1r_2},C^{r_3}\big],C^{r_4}\big] +\frac{1}{3} \big[\big[B^{r_1r_4},C^{r_2}\big],C^{r_3}\big] \hspace{-0.1em} \Big\}  + \textstyle\frac{3}{4}
\big[\big[B^{r_1r_2}, C^{r_4}\big], C^{r_3}\big] \Big)\varepsilon_{[r]_4}\nonumber \\
 && \quad = -\Big( \textstyle\frac{3}{4}\big[\big[B^{r_1r_2},C^{r_3}\big],C^{r_4}\big]  + \big[\big[B^{r_1r_4},C^{r_2}\big],C^{r_3}\big] -  \textstyle\frac{3}{4}
\big[\big[B^{r_1r_2}, C^{r_3}\big], C^{r_4}\big] \Big)\varepsilon_{[r]_4}\nonumber \\
 && \quad =   -\big[\big[B^{r_1r_2}, C^{r_3}\big], C^{r_4}\big]\varepsilon_{[r]_4}, \label{permBr1234}\\
&&  D_{\mu}A^{\mu}=  \partial_{\mu}A^{\mu}\ ,\ \ \ \int d^{d}x\ tr \left(
D_{\mu}F\right)  G=-\int d^{d}x\ tr F D_{\mu}G\ .
\label{idDmu}%
\end{eqnarray}
  In turn, the input from the gauge boson part $Y^B_{(4)\xi}(B^{q_1q_2})$, (\ref{gBn4xi}) into the quantum action (\ref{qexi4})  may be presented as:
  \begin{eqnarray}\label{yb4pres}
    Y^B_{(4)\xi} \prod_{k=1}^4\overleftarrow{s}{}^{r_k}\varepsilon_{[r]_4}& =& - 2 \frac{\xi g^2}{4!} \int d^dx \ tr \Big\{B^{q_1q_2}\prod_{k=1}^4\overleftarrow{s}{}^{r_k}B^{q_3q_4}+ 4B^{q_1q_2}\prod_{k=1}^3\overleftarrow{s}{}^{r_k}\big(B^{q_3q_4} \overleftarrow{s}{}^{r_4}\big) \nonumber\\
    &&  +3B^{q_1q_2}\overleftarrow{s}{}^{r_1}\overleftarrow{s}{}^{r_2}\big(B^{q_3q_4} \overleftarrow{s}{}^{r_3}\overleftarrow{s}{}^{r_4}\big)\Big\}\varepsilon_{[r]_4}\varepsilon_{[q]_4},
  \end{eqnarray}
  so that to derive the quadratic in the fields $B$  terms, which should determine  the gauge-fixed action for the Feynman-like gauge it is sufficient to calculate the last  summand above, because of, $B^{q_1q_2}\overleftarrow{s}{}^{r_1}\overleftarrow{s}{}^{r_2} = \varepsilon^{q_1q_2r_1r_2}B + o (B,C)$, according to $N=4$ BRST transformations (\ref{minN4}).

  Let us find the action of the operators  $\overleftarrow{s}{}^{r_1}\varepsilon_{[r]_4}$ and $\overleftarrow{s}{}^{r_1}\overleftarrow{s}{}^{r_2}\varepsilon_{[r]_4}$  on $B^{q_1q_2}\varepsilon_{[q]_4}$:
\begin{eqnarray}
&& B^{q_1q_2} \overleftarrow{s}{}^{r_1}\varepsilon_{[r]_4}\varepsilon_{[q]_4} =    \Big\{B^{q_1q_2r_1}+ \frac{1}{2}\big[\big[B^{q_1q_2},C^{r_1}\big] -\textstyle\frac{1}{6}\big[C^{q_1},\big[C^{q_2},\,C^{r_1}\big]\big]  \Big\}\varepsilon_{[r]_4}\varepsilon_{[q]_4}, \label{Bqqs1}\\
&& B^{q_1q_2} \overleftarrow{s}{}^{r_1}\overleftarrow{s}{}^{r_2}\varepsilon_{[r]_4}\varepsilon_{[q]_4} =    \Big\{\varepsilon^{q_1q_2r_1r_2}{B}+\big[B^{q_1q_2r_1},C^{r_2}\big]+\big[\big[B^{q_1r_1},C^{q_2}\big],C^{r_2}\big]\hspace{-0.1em} \nonumber \\
&& \quad  -\textstyle\frac{1}{3}\big[\big[B^{q_1r_1},C^{r_2}\big],C^{q_2}\big]- \textstyle\frac{1}{6}\big[\big[B^{r_1r_2},C^{q_1}\big],C^{q_2}\big]
+\textstyle\frac{1}{6}\big[C^{q_1},\big[\big[C^{q_2},\,C^{r_2}\big] , C^{r_1}\big]\big]\nonumber \\
&& \quad  +
 \textstyle\frac{1}{2}
\big[B^{q_1q_2}, B^{r_1r_2}\big] \Big\}\varepsilon_{[r]_4}\varepsilon_{[q]_4}. \label{Bqqs12}
\end{eqnarray}
Then, for the  last term in (\ref{yb4pres}) we have
 \begin{eqnarray}\label{yb4presfin}
    && -  \frac{\xi g^2}{4} \int d^dx \ tr  B^{q_1q_2}\overleftarrow{s}{}^{r_1}\overleftarrow{s}{}^{r_2}\big(B^{q_3q_4} \overleftarrow{s}{}^{r_3}\overleftarrow{s}{}^{r_4}\big)\varepsilon_{[r]_4}\varepsilon_{[q]_4} = -  \frac{\xi g^2}{4} \int d^dx \ tr  \Big\{\varepsilon^{q_1q_2r_1r_2}{B}\nonumber \\
&& \quad +\big[B^{q_1q_2r_1},C^{r_2}\big]+\big[\big[B^{q_1r_1},C^{q_2}\big],C^{r_2}\big]\hspace{-0.1em}   -\textstyle\frac{1}{3}\big[\big[B^{q_1r_1},C^{r_2}\big],C^{q_2}\big]- \textstyle\frac{1}{6}\big[\big[B^{r_1r_2},C^{q_1}\big],C^{q_2}\big]\nonumber \\
&& \quad
+\textstyle\frac{1}{6}\big[C^{q_1},\big[\big[C^{q_2},\,C^{r_2}\big] , C^{r_1}\big]\big]  +
 \textstyle\frac{1}{2}
\big[B^{q_1q_2}, B^{r_1r_2}\big] \Big\}\times \Big\{\varepsilon^{q_3q_4r_3r_4}{B}+\big[B^{q_3q_4r_3},C^{r_4}\big]\nonumber \\
&& \quad +\big[\big[B^{q_3r_3},C^{q_4}\big],C^{r_4}\big]\hspace{-0.1em}   -\textstyle\frac{1}{3}\big[\big[B^{q_3r_3},C^{r_4}\big],C^{q_4}\big]- \textstyle\frac{1}{6}\big[\big[B^{r_3r_4},C^{q_3}\big],C^{q_4}\big]\nonumber \\
&& \quad
+\textstyle\frac{1}{6}\big[C^{q_3},\big[\big[C^{q_4},\,C^{r_4}\big] , C^{r_3}\big]\big]  +
 \textstyle\frac{1}{2}
\big[B^{q_3q_4}, B^{r_3r_4}\big] \Big\}\varepsilon_{[r]_4}\varepsilon_{[q]_4}  \nonumber \\
&& \ =   -  {\xi g^2} \int d^dx \ tr  \Bigg\{4! {B}^2  + 2 B \Big(\big[B^{q_1q_2r_1},C^{r_2}\big]+\big[\big[B^{q_1r_1},C^{q_2}\big],C^{r_2}\big]\hspace{-0.1em}
-\textstyle\frac{1}{3}\big[\big[B^{q_1r_1},C^{r_2}\big],C^{q_2}\big]\nonumber \\
&& \quad   - \textstyle\frac{1}{6}\big[\big[B^{r_1r_2},C^{q_1}\big],C^{q_2}\big]+\textstyle\frac{1}{6}\big[C^{q_1},\big[\big[C^{q_2},\,C^{r_2}\big] , C^{r_1}\big]\big]  +
 \textstyle\frac{1}{2}
\big[B^{q_1q_2}, B^{r_1r_2}\big] \Big)\varepsilon_{q_1q_2r_1r_2} \nonumber \\
&& \quad
 +  \textstyle\frac{1}{4}\Big\{\big[B^{q_1q_2r_1},C^{r_2}\big] +\big[\big[B^{q_1r_1},C^{q_2}\big],C^{r_2}\big]\hspace{-0.1em}   -\textstyle\frac{1}{3}\big[\big[B^{q_1r_1},C^{r_2}\big],C^{q_2}\big]- \textstyle\frac{1}{6}\big[\big[B^{r_1r_2},C^{q_1}\big],C^{q_2}\big]\nonumber \\
&& \quad
+\textstyle\frac{1}{6}\big[C^{q_1},\big[\big[C^{q_2},\,C^{r_2}\big] , C^{r_1}\big]\big]  +
 \textstyle\frac{1}{2}
\big[B^{q_1q_2}, B^{r_1r_2}\big] \Big\}\times \Big\{\big[B^{q_3q_4r_3},C^{r_4}\big]+\big[\big[B^{q_3r_3},C^{q_4}\big],C^{r_4}\big]\hspace{-0.1em} \nonumber \\
&& \quad   -\textstyle\frac{1}{3}\big[\big[B^{q_3r_3},C^{r_4}\big],C^{q_4}\big]- \textstyle\frac{1}{6}\big[\big[B^{r_3r_4},C^{q_3}\big],C^{q_4}\big]+\textstyle\frac{1}{6}\big[C^{q_3},\big[\big[C^{q_4},\,C^{r_4}\big] , C^{r_3}\big]\big]
\nonumber \\
&& \quad +
 \textstyle\frac{1}{2}
\big[B^{q_3q_4}, B^{r_3r_4}\big] \Big\}\varepsilon_{[r]_4}\varepsilon_{[q]_4} \Bigg\}, \label{yb4presfin1}
  \end{eqnarray}
  where we have used the Fierz-like identities for the products of Levi-Civita tensors:
  \begin{equation}\label{Firtz}
\varepsilon^{q_1q_2r_1r_2}\varepsilon_{[r]_4}\varepsilon^{q_3q_4r_3r_4} = 4 \varepsilon^{[q]_4}, \ \mathrm{and} \ \varepsilon^{q_1q_2r_1r_2}\varepsilon_{[r]_4}\varepsilon^{q_3q_4r_3r_4}\varepsilon_{[q]_4}= 4 \cdot 4!,
  \end{equation}
  and its normalization (\ref{lchevn}), (\ref{eprts}).

Now, we are waiting that the first and second terms  in (\ref{yb4pres}) of the  third and fourth orders in $\overleftarrow{s}{}^r$ when acting on $B^{q_1q_2}$  will not  produce new summands to the gauge-fixed  and quadratic in the fictitious fields parts of the action  (\ref{qexi4}). Their role concerns only to exclude non-diagonal terms from the   last quantity in (\ref{yb4pres}) given explicitly  in (\ref{yb4presfin1}).
To justify the proposal let us show, that the terms linear in $B$ in   (\ref{yb4presfin1}) are absent in $S_{Y_{(4)\xi}}\big({\Phi}_{(4)}\big)$ (\ref{qexi4}).
To do so we need  the product of three antisymmetrized generators, $\overleftarrow{s}{}^{r_1}\overleftarrow{s}{}^{r_2}\overleftarrow{s}{}^{r_3}\varepsilon_{[r]_4}$ applied to $B^{q_1q_2}$:
\begin{eqnarray}
&& B^{q_1q_2}\prod_{k=1}^{3} \overleftarrow{s}{}^{r_k}\varepsilon_{[r]_4}\varepsilon_{[q]_4} = \Bigg\{\varepsilon^{q_1q_2r_1r_2}\Big( \textstyle\frac{1}{2}\big[{B},\,C^{r_3}\big] - \frac{1}{4!}\big[\big[B^{s_1s_2s_3},\,C^{s_4}\big],\,C^{r_3}\big] \varepsilon_{s_1s_2s_3s_4}\Big) \nonumber \\
&& \quad +\big[B^{q_1q_2r_1},\,B^{r_2r_3}\big]- \Big[\Big(
\varepsilon^{q_1q_2r_1r_3}{B}+\textstyle\frac{1}{2}\hspace{-0.1em}\big[B^{q_1q_2r_1},C^{r_3}\big] - \displaystyle\sum_{P}\hspace{-0.15em}(-1)^{P(q_1,q_2,r_1)}\hspace{-0.1em}\Big\{\hspace{-0.1em}\textstyle\frac{1}{8}\big[\big[B^{q_1q_2},C^{r_1}\big],C^{r_3}\big] \nonumber \\
&& \quad +\textstyle\frac{1}{6} \big[\big[B^{q_1r_3},C^{q_2}\big],C^{r_1}\big] \hspace{-0.1em} \Big\} \hspace{-0.1em} \Big),C^{r_2}\Big]
+\big[\big[B^{q_1r_1},C^{q_2}\big],B^{r_2r_3}\big] -\big[\big[B^{q_1r_1}, \big\{B^{q_2r_3}+\frac{1}{2}\big[C^{q_2},\,C^{r_3}\big]\big\}\big],C^{r_2}\big] \nonumber \\
&& \quad +\Big[\Big[\Big(
  {B}^{q_1r_1r_3}+ \textstyle\frac{1}{2}
\big[B^{q_1r_1}, C^{r_3}\big]+\frac{1}{12}\big[C^{r_1},\big[C^{q_1},C^{r_3}\big]\big]\Big)
 ,\,C^{q_2}\Big],C^{r_2}\Big]
\hspace{-0.1em}  \nonumber \\
&& \quad   -\textstyle\frac{1}{3}\big[\big[B^{q_1r_1},C^{r_2}\big], \big\{B^{q_2r_3}+\frac{1}{2}\big[C^{q_2},\,C^{r_3}\big]\big\}\big]  +\textstyle\frac{1}{3}\big[\big[B^{q_1r_1},B^{r_2r_3}\big],C^{q_2}\big] -\textstyle\frac{1}{3}\Big[\Big[\Big(
  {B}^{q_1r_1r_3}+ \textstyle\frac{1}{2}
\big[B^{q_1r_1}, C^{r_3}\big]
\nonumber \\
&& \quad   +\textstyle\frac{1}{12}\big[C^{r_1},\big[C^{q_1},C^{r_3}\big]\big]\Big)
,C^{r_2}\Big],C^{q_2}\Big]
- \textstyle\frac{1}{6}\big[\big[B^{r_1r_2},C^{q_1}\big],
\big\{B^{q_2r_3}+\frac{1}{2}\big[C^{q_2},\,C^{r_3}\big]\big\}
\big]  \nonumber \\
&& \quad +\textstyle\frac{1}{6}\big[\big[B^{r_1r_2}, \big\{B^{q_1r_3}+\frac{1}{2}\big[C^{q_1},\,C^{r_3}\big]\big\}\big],C^{q_2}\big] - \textstyle\frac{1}{6}\Big[\Big[\Big(
 {B}^{r_1r_2r_3}+ \textstyle\frac{1}{2}
\big[B^{r_1r_2}, C^{r_3}\big] \Big)
,C^{q_1}\Big],C^{q_2}\Big] \nonumber \\
&& \quad+\textstyle\frac{1}{6}\big[C^{q_1},\big[\big[C^{q_2},\,C^{r_2}\big] , B^{r_1r_3}\big]\big]-\textstyle\frac{1}{6}\big[C^{q_1},\big[\big[C^{q_2},\,B^{r_2r_3}\big] , C^{r_1}\big]\big]
 +\textstyle\frac{1}{6}\big[C^{q_1},\big[\big[
 \big\{B^{q_2r_3}  \nonumber \\
&& \quad
 +\textstyle\frac{1}{2}\big[C^{q_2},\,C^{r_3}\big]\big\},\,C^{r_2}\big] , C^{r_1}\big]\big]
-\textstyle\frac{1}{6}\big[\big\{B^{q_1r_3}+\frac{1}{2}\big[C^{q_1},\,C^{r_3}\big]\big\},\big[\big[C^{q_2},\,C^{r_2}\big] , C^{r_1}\big]\big]  +
 \textstyle\frac{1}{2}
\Big[B^{q_1q_2},  \Big(
 {B}^{r_1r_2r_3} \nonumber \\
&& \quad+ \textstyle\frac{1}{2}
\big[B^{r_1r_2}, C^{r_3}\big] \Big)\Big] +
 \textstyle\frac{1}{2}
\big[\Big(
 {B}^{q_1q_2r_3}+ \textstyle\frac{1}{2}
\big[B^{q_1q_2}, C^{r_3}\big] -\frac{1}{6}\big[C^{q_1},\big[C^{q_2},C^{r_3}\big]\big]\Big), B^{r_1r_2}\big]\Bigg\}\varepsilon_{[r]_4}\varepsilon_{[q]_4}. \label{Bqqs123}
\end{eqnarray}
Consider,  e.g.  the summand, $B \big[B^{q_1q_2r_1},C^{r_2}\big]$ in (\ref{yb4presfin1}). The only  second  term in   (\ref{yb4pres}) gives similar contribution from (\ref{Bqqs123}), so that their sum is equal to:
 \begin{eqnarray}\label{BBqqrC}
   && \int d^dx \ tr   \Big( 4 \cdot 3 \cdot 2  B \big[B^{q_1q_2r_1},C^{r_2}\big] \varepsilon_{q_1q_2r_1r_2}  +  4 \cdot \textstyle\frac{3}{2} \varepsilon^{q_1q_2r_1r_2} \big[B,\,C^{r_3}\big]B^{q_3q_4r_4} \varepsilon_{[r]_4}\varepsilon_{[q]_4}\Big)   \nonumber \\
   && \quad = 4! \int d^dx \ tr   \Big(   B \big[B^{q_1q_2r_1},C^{r_2}\big] \varepsilon_{q_1q_2r_1r_2}   +    \big[B,\,C^{r_3}\big]B^{q_3q_4r_4} \varepsilon_{q_3q_4r_3r_4}\Big)  \\
   && \quad  = 4! \int d^dx \ tr   \Big(   B \big[B^{q_1q_2r_1},C^{r_2}\big]    +    B\big[B^{q_1q_2r_2},\,C^{r_1}\big] \Big) \varepsilon_{q_1q_2r_1r_2} \equiv  0, \nonumber
 \end{eqnarray}
due to the antisymmetry in  $r_1,r_2$ of, $ \big[B^{q_1q_2r_2},C^{r_1}\big]\varepsilon_{q_1q_2r_1r_2} = -\big[B^{q_1q_2r_1},C^{r_2}\big]\varepsilon_{q_1q_2r_1r_2}$ and the property for $su(\hat{N})$-valued functions with definite Grassmann parities:
\begin{equation}\label{suNtr}
  tr \,  F \big[G,\,H\big] = F^m\,  f^{mnl}\, G^n\,H^l  =  tr \,   \big[F,\,G\big]H =  -  tr \,   G \big[F,\,H\big](-1)^{\epsilon(F)\epsilon(G)}.
\end{equation}
The checking that the remaining  terms linear in $B$ in (\ref{yb4presfin1}) do not contribute in the quantum action (\ref{qexi4})  may be fulfilled analogously, but we leave out of the paper scope the proof of this  fact.

 The   $\xi$-dependent  part of  $N=4$ BRST invariant  quantum action  (\ref{qexi4})  take the form:
\begin{align}
  \xi\frac{\partial}{\partial\xi} S_{Y_{(4)\xi}}\big({\Phi}_{(4)}\big) &  =  {\xi g^2} \int d^{d}x\ tr \Bigg\{  B^2+\frac{1}{4!}\Big(\frac{1}{4^2}\big[B^{q_1q_2}, B^{r_1r_2}\big]
\big[B^{q_3q_4}, B^{r_3r_4}\big]  \label{addxi4app} \\
 \phantom{ S_{add(4)}} &   + \frac{1}{4!3!}\big[C^{q_1},\big[\big[C^{q_2},\,C^{r_2}\big] , C^{r_1}\big]\big]  \big[C^{q_3},\big[\big[C^{q_4},\,C^{r_4}\big] , C^{r_3}\big]\big]\Big)\varepsilon_{[r]_4}\varepsilon_{[q]_4} \Bigg\}+ \widetilde{S}_\xi , \nonumber
\end{align}
without terms linear in $B$ in $\widetilde{S}_\xi$, which should be determined from (\ref{yb4pres})--(\ref{yb4presfin1}), (\ref{Bqqs123}) and the  results of the product of four antisymmetrized generators, $\overleftarrow{s}{}^{r_1}\overleftarrow{s}{}^{r_2}\overleftarrow{s}{}^{r_3}\overleftarrow{s}{}^{r_4}\varepsilon_{[r]_4}$ applied to $B^{q_1q_2}$.

Therefore, combining (\ref{ghgf4}),  (\ref{addxi4app}) we have
\begin{align}
  S_{Y_{(4)\xi}}\big({\Phi}_{(4)}\big) & = S_0 +  \int d^{d}x\ tr \Bigg\{  \Big[    \partial^{\mu}\mathcal{A}_{\mu
}  + {\xi g^2} B\Big]  B +  \Big\{\frac{1}{3!} B^{r_1r_2r_3}M(\mathcal{A})  C^{r_4} + \frac{1}{8} B^{r_1r_2}M(\mathcal{A})  B^{r_3r_4} \Big\}\varepsilon_{[r]_4} \nonumber \\
 \phantom{S_{Y_{(4)\xi}}\big({\Phi}_{(4)}\big)} & + \frac{1}{4!}\Big\{(\partial^\mu\mathcal{ A}_{\mu })  \Big(2\big[B^{r_1r_2r_3}, C^{r_4}\big]  - \big[\big[B^{r_1r_2}, C^{r_3}\big], C^{r_4}\big]\Big) -  B^{r_1r_2}\Big(\big[ C^{r_3},  M(\mathcal{A})C^{r_4}\big]\nonumber \\
\phantom{S_{Y_{(4)\xi}}}&    +4\big[ \partial_\mu C^{r_3},  D^\mu C^{r_4}\big]\Big)+
 C^{r_1}\partial_\mu \Big( \big[D^\mu C^{r_2},\,    B^{r_3r_4}\big] -\big[C^{r_2},\, D^\mu B^{r_3r_4}   \big] + \big[\big[D^\mu C^{r_2},\, C^{r_3}\big],  C^{r_4}\big]\Big)\Big\}\varepsilon_{[r]_4}\nonumber \\
\phantom{S_{Y_{(4)\xi}}\big({\Phi}_{(4)}\big)} &+ \frac{{\xi g^2}}{4!}\Big(\frac{1}{4^2}\big[B^{q_1q_2}, B^{r_1r_2}\big]
\big[B^{q_3q_4}, B^{r_3r_4}\big]   + \frac{1}{4!3!}\big[C^{q_1},\big[\big[C^{q_2},\,C^{r_2}\big] , C^{r_1}\big]\big]\times \nonumber\\
 \phantom{S_{Y_{(4)\xi}}\big({\Phi}_{(4)}\big)} &    \times \big[C^{q_3},\big[\big[C^{q_4},\,C^{r_4}\big] , C^{r_3}\big]\big]\Big)\varepsilon_{[r]_4}\varepsilon_{[q]_4} \Bigg\}+ \widetilde{S}_\xi,
\label{qexi4app}
\end{align}
for $\widetilde{S}_\xi\big|_{(\xi=0)} =0$, that proves the representation (\ref{qexi4})--(\ref{addxi4}) for the quantum action.
Determining the dual  fields (with lower $\mathcal{G}(4)$-indices) for Grassmann-even  $B^{r_1r_2}$ and Grassmann-odd  $B^{r_1r_2r_3}$ fields:
\begin{eqnarray}\label{dual42}
  && B_{r_1r_2} =  \frac{1}{2}B^{r_3r_4}\varepsilon_{r_1r_2r_3r_4} = \big(B^{34}, -B^{24}, B^{23}, B^{14}, -B^{13}, B^{12}\big) ,\\
   \label{dual43}
  && \quad   C_{r_4}= \frac{1}{3!}B^{r_1r_2r_3}\varepsilon_{r_1r_2r_3r_4} = \big(-B^{234}, B^{134},-B^{124}, B^{123}\big)
\end{eqnarray}
the action (\ref{qexi4app}) can be equivalently presented as follows
\begin{align}
  S_{Y_{(4)\xi}} & = S_0 +  \int d^{d}x\ tr \Bigg\{  \Big[    \partial^{\mu}\mathcal{A}_{\mu
}  + {\xi g^2} B\Big]  B +  \Big\{  C_{r_1}M(\mathcal{A})  C^{r_1} + \sum_{1\leq r_1<r_2 \leq 3} B^{r_1r_2}M(\mathcal{A})  B_{r_1r_2} \Big\} \nonumber \\
 \phantom{S_{Y_{(4)\xi}}} &
 + \frac{1}{4!}\Big\{2(\partial^\mu\mathcal{ A}_{\mu })  \Big(3!\big[C_{r_1}, C^{r_1}\big]  - \big[\big[B_{r_1r_2}, C^{r_1}\big], C^{r_2}\big]\Big) -  2B_{r_1r_2}\Big(\big[ C^{r_1},  M(\mathcal{A})C^{r_2}\big]\nonumber \\
\phantom{S_{Y_{(4)\xi}}}&    +4\big[ \partial_\mu C^{r_1},  D^\mu C^{r_2}\big]\Big)+
 2C^{r_1}\partial_\mu \Big( \big[D^\mu C^{r_2},\,    B_{r_1r_2}\big] -\big[C^{r_2},\, D^\mu B_{r_1r_2}   \big] + \frac{1}{2}\big[\big[D^\mu C^{r_2},\, C^{r_3}\big],  C^{r_4}\big]\varepsilon_{[r]_4}\Big)\Big\}
 \nonumber \\
\phantom{S_{Y_{(4)\xi}}} &+\frac{{\xi g^2}}{4\cdot4!}\big[B^{q_1q_2}, B^{r_1r_2}\big]
\big[B_{q_1q_2}, B_{r_1r_2}\big]   + \frac{{\xi g^2}}{3!(4!)^2}\big[C^{q_1},\big[\big[C^{q_2},\,C^{r_1}\big] , C^{r_2}\big]\big]\times \nonumber\\
 \phantom{S_{Y_{(4)\xi}}} &    \times \big[C^{q_3},\big[\big[C^{q_4},\,C^{r_3}\big] , C^{r_4}\big]\big]\varepsilon_{[r]_4}\varepsilon_{[q]_4} \Bigg\}+\widetilde{S}_\xi,
\label{qexi4appgen}
\end{align}
   by virtue of easily checked   identity, $\sum_{1\leq r_1<r_2 \leq 3} B^{r_1r_2}M(\mathcal{A})  B_{r_1r_2} \equiv \frac{1}{4} B^{r_1r_2}M(\mathcal{A})  B_{r_1r_2}$,  justifying the representation (\ref{SintfpLlock}) for the quantum action for the case  $N=4$ ($k(4)$=3) in the Landau gauge ($\xi=0$) with the identification:
   \begin{equation}\label{identifgh4}
   \big(C^0, C^{[3]}; \overline{C}{}^0 , \overline{C}{}^{[3]};  B^{[3]}, \overline{B}{}^{[3]}\big) =  \big(  C^{r};  {C}_{r} ;  {B}_{r_1r_2}, {B}^{r_1r_2}\big) \ \mathrm{for} \   r=1,...,4; \ 1\leq r_1<r_2 \leq 3.
 \end{equation}

\end{document}